\documentstyle[12pt]{book}

\begin{document}

\sloppy
\raggedbottom

\begin{titlepage}

{\Huge\bf\begin{center}\begin{tabular}{c}
THE LIMITS OF\\
MATHEMATICS\\
\end{tabular}\end{center}}\vfill

{\Large\bf\begin{center}\begin{tabular}{c}
G. J. Chaitin\\
IBM, P O Box 704\\
Yorktown Heights, NY 10598\\
{\it chaitin@watson.ibm.com}
\end{tabular}\end{center}}\vfill

{\Large\bf\begin{center}
Draft July 19, 1994 
\end{center}}

\end{titlepage}

\begin{titlepage}

\hfill
\vfill

\end{titlepage}

\begin{titlepage}

\hfill
\vfill

\begin{center}
{\it To Fran\c{c}oise}
\end{center}

\vfill
\hfill

\end{titlepage}

\begin{titlepage}

\hfill

\end{titlepage}

\pagenumbering{roman}

\chapter*{Preface}

\setcounter{page}{5}

In a remarkable development, I have constructed a new definition for a
self-delimiting universal Turing machine (UTM) that is easy to program
and runs very quickly.  This provides a new foundation for algorithmic
information theory (AIT), which is the theory of the size in bits of
programs for self-delimiting UTM's.  Previously, AIT had an abstract
mathematical quality.  Now it is possible to write down executable
programs that embody the constructions in the proofs of theorems.  So
AIT goes from dealing with remote idealized mythical objects to being
a theory about practical down-to-earth gadgets that one can actually
play with and use.

This new self-delimiting UTM is implemented via an extremely fast LISP
interpreter written in C. The universal Turing machine $U$ is written in
this LISP.  This LISP also serves as a very high-level assembler to
put together binary programs for $U.$ The programs that go into $U,$ and
whose size we measure, are bit strings.  The output from $U,$ on the
other hand, consists of a single LISP S-expression, in the case of finite
computations, and of an infinite set of these S-expressions, in the case
of infinite computations.

The LISP used here is based on the version
of LISP that I used in my book {\it Algorithmic Information Theory,}
Cambridge University Press, 1987.  The difference is that a) I have
found a self-delimiting way to give binary data to LISP programs,
b) I have found a natural way to handle unending computations,
which is what formal axiomatic systems are, in LISP, and c) I have
found a way to eliminate more parentheses from M-expressions.

Using this new software, as well as the latest theoretical ideas, it
is now possible to give a self-contained ``hands on'' course
presenting very concretely my latest proofs of my two fundamental
information-theoretic incompleteness theorems.  The first of these
theorems states that an $N$-bit formal axiomatic system cannot enable
one to exhibit any specific object with program-size complexity
greater than $N+c$.  The second of these theorems states that an
$N$-bit formal axiomatic system cannot enable one to determine more
than $N+c'$ scattered bits of the halting probability $\Omega$.

Most people believe that anything that is true is true for a reason.
These theorems show that some things are true for no reason at all,
i.e., accidentally, or at random.

The latest and I believe the deepest proofs of these two theorems were
originally presented in my paper ``Information-theoretic
incompleteness,'' {\it Applied Mathematics and Computation\/} {\bf 52}
(1992), pp.\ 83--101.  This paper is reprinted in my book {\it
Information-Theoretic Incompleteness,} World Scientific, 1992.

As is shown in this course, the algorithms considered in the proofs of
these two theorems are now easy to program and run, and by looking at
the size in bits of these programs one can actually, for the first
time, determine exact values for the constants $c$ and $c'$.
Indeed, $c = 994$ bits and $c' = 3192$ bits.

This approach and software were used in an intense short course on
the limits of mathematics that I gave at the University of Maine in
Orono in June 1994.  I wish to thank Prof.\ George Markowsky of the
University of Maine for his stimulating invitation to give this
course, for all his hard work organizing the course and porting
my software to PC's,
for many helpful suggestions on presenting this material, and
in particular for the crucial suggestion that led me
to a vastly improved algorithm for computing lower bounds on the
halting probability $\Omega$ (see {\tt omega.l} below).

I thank the dozen participants
in the short course, who were mostly professors of mathematics
and computer science, for their enthusiasm and determination and
extremely useful feedback.
I also thank Prof.\ Cristian Calude of the University of
Auckland, Prof.\ John Casti of the Santa Fe Institute, and Prof.\
Walter Meyerstein of the University of Barcelona for stimulating
discussions at the delicate initial phase of this project.

I am grateful to IBM for its enthusiastic and unwavering support of my
research for a quarter of a century, and to my current management
chain at the IBM Research Division, Jeff Jaffe, Eric Kronstadt, and
Marty Hopkins.  Finally I thank the RISC project group, of which I am
a member, for designing the marvelous IBM RISC System/6000
workstations that I have used for all these calculations, and for
providing me with the latest models of this spectacular computing
equipment.

All enquires, comments and suggestions regarding this software should
be sent via e-mail to {\tt chaitin} at {\tt watson.ibm.com}.

\begin{flushright}
{\sc Gregory Chaitin}
\end{flushright}

\tableofcontents

\markboth{}{}

\newcommand
{\chap}[1]{\chapter*{#1}\markboth{The Limits of Mathematics}{#1}
\addcontentsline{toc}{chapter}{#1}}

\newcommand
{\Size}{\small}   

\chap{Responses to ``Theoretical Mathematics'' (excerpt)}

\pagenumbering{arabic}

\setcounter{page}{1}

\section*{AMS Bulletin 30 (1994), pp.\ 181--182}

\section*{}

One normally thinks that everything that is true is true for a reason.
I've found mathematical truths that are true for no reason at all.
These mathematical truths are beyond the power of mathematical
reasoning because they are accidental and random.

Using software written in {\sl Mathematica\/} that runs on an IBM
RS/6000 workstation [5,7], I constructed a perverse 200-page exponential
diophantine equation with a parameter $N$ and 17,000 unknowns:
\begin{center}
        Left-Hand-Side($N$) = Right-Hand-Side($N$).
\end{center}
For each nonnegative value of the parameter $N$, ask whether this
equation has a finite or an infinite number of nonnegative solutions.
The answers escape the power of mathematical reason because they are
completely random and accidental.

This work is part of a new field that I call algorithmic information
theory [2,3,4].

What does this have to with Jaffe and Quinn [1]?

The result presented above is an example of how my
information-theoretic approach to incompleteness makes incompleteness
appear pervasive and natural.  This is because algorithmic information
theory sometimes enables one to measure the information content of a
set of axioms and of a theorem and to deduce that the theorem cannot
be obtained from the axioms because it contains too much information.

This suggests to me that sometimes to prove more one must assume more,
in other words, that sometimes one must put more in to get more out.
I therefore believe that elementary number theory should be pursued
somewhat more in the spirit of experimental science.  Euclid declared
that an axiom is a self-evident truth, but physicists are willing to
assume new principles like the Schr\"odinger equation that are not
self-evident because they are extremely useful.  Perhaps number
theorists, even when they are doing elementary number theory, should
behave a little more like physicists do and should sometimes adopt new
axioms.  I have argued this at greater length in a lecture [6,8] that I
gave at Cambridge University and at the University of New Mexico.

\section*{References}

\begin{itemize}
\item[{[1]}] A. Jaffe and F. Quinn, {\it Theoretical Mathematics: Toward
a cultural synthesis of mathematics and theoretical physics,}
Bull.\ Amer.\ Math.\ Soc.\ {\bf 29} (1993), 1--13.
\item[{[2]}] G. J. Chaitin, {\it Algorithmic information theory,} revised third
printing,
Cambridge Univ.\ Press, Cambridge, 1990.
\item[{[3]}] G. J. Chaitin, {\it Information, randomness \& incompleteness---%
Papers on algorithmic information theory,} second edition,
World Scientific, Singapore, 1990.
\item[{[4]}] G. J. Chaitin, {\it Information-theoretic incompleteness,}
World Scientific, Singapore, 1992.
\item[{[5]}] G. J. Chaitin,
{\it Exhibiting randomness in arithmetic using Mathematica and  C,}
IBM Research Report RC-18946, June 1993.
\item[{[6]}] G. J. Chaitin,
{\it Randomness in arithmetic and the decline and fall of reductionism in
pure mathematics,} Bull.\ European Assoc.\ for Theoret.\
Comput.\ Sci., no.\ 50 (June 1993), 314--328.
\item[{[7]}] G. J. Chaitin,
{\it The limits of mathematics---Course outline and software,}
IBM Research Report RC-19324, December 1993.
\item[{[8]}] G. J. Chaitin, {\it Randomness and complexity in pure
mathematics,}
Internat.\ J.\  Bifur.\ Chaos {\bf 4} (1994), 3--15.
\end{itemize}

\chap{The Berry Paradox}

{\it
Lecture given Wednesday 27 October 1993 at a Physics -- Computer
Science Colloquium at the University of New Mexico.  The lecture was
videotaped; this is an edited transcript.  It also incorporates
remarks made at the Limits to Scientific Knowledge meeting held at the
Santa Fe Institute 24--26 May 1994.
}

In early 1974, I was visiting the Watson Research Center and I got the
idea of calling G\"odel on the phone.  I picked up the phone and
called and G\"odel answered the phone.  I said, ``Professor G\"odel,
I'm fascinated by your incompleteness theorem.  I have a new proof
based on the Berry paradox that I'd like to tell you about.''  G\"odel
said, ``It doesn't matter which paradox you use.''  He had used a
paradox called the liar paradox.  I said, ``Yes, but this suggests to
me an information-theoretic view of incompleteness that I would very
much like to tell you about and get your reaction.''  So G\"odel said,
``Send me one of your papers.  I'll take a look at it.  Call me again
in a few weeks and I'll see if I give you an appointment.''

I had had this idea in 1970, and it was 1974.  So far I had only
published brief abstracts.  Fortunately I had just gotten the galley
proofs of my first substantial paper on this subject.  I put these in
an envelope and mailed them to G\"odel.

I called G\"odel again and he gave me an appointment!  As you can
imagine I was delighted.  I figured out how to go to Princeton by
train.  The day arrived and it had snowed and there were a few inches
of snow everywhere.  This was certainly not going to stop me from
visiting G\"odel!  I was about to leave for the train and the phone
rang and it was G\"odel's secretary.  She said that G\"odel was very
careful about his health and because of the snow he wasn't coming to
the Institute that day and therefore my appointment was canceled.

And that's how I had two phone conversations with G\"odel but never
met him.  I never tried again.

I'd like to tell you what I would have told G\"odel.  What I wanted to
tell G\"odel is the difference between what you get when you study the
limits of mathematics the way G\"odel did using the paradox of the
liar, and what I get using the Berry paradox instead.

What is the paradox of the liar?  Well, the paradox of the liar is
\[
\begin{array}{l}
   \mbox{``This statement is false!''}
\end{array}
\]
Why is this a paradox?  What does ``false'' mean?  Well, ``false''
means ``does not correspond to reality.''  This statement says that it
is false.  If that doesn't correspond to reality, it must mean that
the statement is true, right?  On the other hand, if the statement is
true it means that what it says corresponds to reality.  But it says
that it is false.  Therefore the statement must be false.  So whether
you assume that it's true or false you then conclude the opposite!  So
this is the paradox of the liar.

Now let's look at the Berry paradox.  First of all, why ``Berry''?
Well it has nothing to do with fruit!  This paradox was published at
the beginning of this century by Bertrand Russell.  Now there's a
famous paradox which is called Russell's paradox and this is not it!
This is another paradox that he published.  I guess people felt that
if you just said the Russell paradox and there were two of them it
would be confusing.  And Bertrand Russell when he published this
paradox had a footnote saying that it was suggested to him by Mr G. G.
Berry.  So it ended up being called the Berry paradox even though it
was published by Russell.

Here is a version of the Berry paradox:
\[
\begin{array}{l}
   \mbox{``the first positive integer that cannot} \\
   \mbox{be specified in less than a billion words''.}
\end{array}
\]
This is a phrase in English that specifies a particular positive
integer.  Which positive integer?  Well, there are an infinity of
positive integers, but there are only a finite number of words in
English.  Therefore, if you have a billion words, there's only
going to be a finite number of possibilities.  But there's an infinite
number of positive integers.  Therefore most positive integers require
more than a billion words, and let's just take the first one.  But
wait a second.  By definition this integer is supposed to take a
billion words to specify, but I just specified it using much less than
a billion words!  That's the Berry paradox.

What does one do with these paradoxes?  Let's take a look again at the
liar paradox:
\[
\begin{array}{l}
   \mbox{``This statement is false!''}
\end{array}
\]
The first thing that G\"odel does is to change it from ``This
statement is false'' to ``This statement is unprovable'':
\[
\begin{array}{l}
   \mbox{``This statement is unprovable!''}
\end{array}
\]
What do we mean by ``unprovable''?

In order to be able to show that mathematical reasoning has limits
you've got to say very precisely what the axioms and methods of
reasoning are that you have in mind.  In other words, you have to
specify how mathematics is done with mathematical precision so that it
becomes a clear-cut question.  Hilbert put it this way: The rules
should be so clear, that if somebody gives you what they claim is a
proof, there is a mechanical procedure that will check whether the
proof is correct or not, whether it obeys the rules or not.  This
proof-checking algorithm is the heart of this notion of a completely
formal axiomatic system.

So ``This statement is unprovable'' doesn't mean unprovable in a vague
way.  It means unprovable when you have in mind a specific formal
axiomatic system {\sl FAS\/} with its mechanical proof-checking
algorithm.  So there is a subscript:
\[
\begin{array}{l}
   \mbox{``This statement is unprovable${}_{\it FAS}$!''}
\end{array}
\]

And the particular formal axiomatic system that G\"odel was interested
in dealt with 1, 2, 3, 4, 5, and addition and multiplication, that was
what it was about.  Now what happens with ``This statement is
unprovable''?  Remember the liar paradox:
\[
\begin{array}{l}
   \mbox{``This statement is false!''} \\
\end{array}
\]
But here
\[
\begin{array}{l}
   \mbox{``This statement is unprovable${}_{\it FAS}$!''}
\end{array}
\]
the paradox disappears and we get a theorem.  We get incompleteness,
in fact.  Why?

Consider ``This statement is unprovable''.  There are two
possibilities: either it's provable or it's unprovable.

If ``This statement is unprovable'' turns out to be unprovable within
the formal axiomatic system, that means that the formal axiomatic
system is incomplete.  Because if ``This statement is unprovable'' is
unprovable, then it's a true statement.  Then there's something true
that's unprovable which means that the system is incomplete.  So that
would be bad.

What about the other possibility?  What if
\[
\begin{array}{l}
   \mbox{``This statement is unprovable${}_{\it FAS}$!''}
\end{array}
\]
is provable?  That's even worse.  Because if this statement is
provable
\[
\begin{array}{l}
   \mbox{``This statement is unprovable${}_{\it FAS}$!''}
\end{array}
\]
and it says of itself that it's unprovable, then we're proving
something that's false.

So G\"odel's incompleteness result is that if you assume that only
true theorems are provable, then this
\[
\begin{array}{l}
   \mbox{``This statement is unprovable${}_{\it FAS}$!''}
\end{array}
\]
is an example of a statement that is true but unprovable.

But wait a second, how can a statement deny that it is provable?  In
what branch of mathematics does one encounter such statements?
G\"odel cleverly converts this
\[
\begin{array}{l}
   \mbox{``This statement is unprovable${}_{\it FAS}$!''}
\end{array}
\]
into an arithmetical statement, a statement that only involves 1, 2,
3, 4, 5, and addition and multiplication.  How does he do this?

The idea is called g\"odel numbering.  We all know that a string of
characters can also be thought of as a number.  Characters are either
8 or 16 bits in binary.  Therefore, a string of $N$ characters is
either $8N$ or $16N$ bits, and it is also the base-two notation for a
large positive integer.  Thus every mathematical statement in this
formal axiomatic system
\[
\begin{array}{l}
   \mbox{``This statement is unprovable${}_{FAS\longleftarrow}$!''}
\end{array}
\]
is also a number.  And a proof, which is a sequence of steps, is also
a long character string, and therefore is also a number.  Then you can
define this very funny numerical relationship between two numbers $X$
and $Y$ which is that $X$ is the g\"odel number of a proof of the
statement whose g\"odel number is $Y$.  Thus
\[
\begin{array}{l}
   \mbox{``This statement is unprovable${}_{\it FAS}$!''}
\end{array}
\]
ends up looking like a very complicated numerical statement.

There is another serious difficulty.  How can this statement refer to
itself?  Well you can't directly put the g\"odel number of this
statement inside this statement, it's too big to fit!  But you can do
it indirectly.  This is how G\"odel does it: The statement doesn't
refer to itself directly.  It says that if you perform a certain
procedure to calculate a number, this is the g\"odel number of a
statement which cannot be proved.  And it turns out that the number
you calculate is precisely the g\"odel number of the entire statement
\[
\begin{array}{l}
   \mbox{``This statement is unprovable${}_{\it FAS}$!''}
\end{array}
\]
That is how G\"odel proves his incompleteness theorem.

What happens if you start with this
\[
\begin{array}{l}
   \mbox{``the first positive integer that cannot} \\
   \mbox{be specified in less than a billion words''}
\end{array}
\]
instead?  Everything has a rather different flavor.  Let's see why.

The first problem we've got here is what does it mean to specify a
number using words in English?---this is very vague.  So instead let's
use a computer.  Pick a standard general-purpose computer, in other
words, pick a universal Turing machine ({\sl UTM\/}).  Now the way you specify
a number is with a computer program.  When you run this computer
program on your {\sl UTM} it prints out this number and halts.  So a
program is said to specify a number, a positive integer, if you start
the program running on your standard {\sl UTM,} and after a finite amount of
time it prints out one and only one great big positive integer and it
says ``I'm finished'' and halts.

Now it's not English text measured in words, it's computer programs
measured in bits.  This is what we get.  It's
\[
\begin{array}{l}
   \mbox{``the first positive integer that cannot} \\
   \mbox{be specified${}_{\it UTM}$ by a computer program} \\
   \mbox{with less than a billion bits''.}
\end{array}
\]
By the way the computer program must be self-contained.  If it has
any data, the data is included in the program as a constant.

Next we have to do what G\"odel did when he changed ``This statement
is false'' into ``This statement is unprovable.''  So now it's
\[
\begin{array}{l}
   \mbox{``the first positive integer that can be proved${}_{\it FAS}$} \\
   \mbox{to have the property that it cannot} \\
   \mbox{be specified${}_{\it UTM}$ by a computer program} \\
   \mbox{with less than a billion bits''.}
\end{array}
\]
And to make things clearer let's replace ``a billion bits'' by ``$N$ bits''.
So we get:
\[
\begin{array}{l}
   \mbox{``the first positive integer that can be proved${}_{\it FAS}$} \\
   \mbox{to have the property that it cannot} \\
   \mbox{be specified${}_{\it UTM}$ by a computer program} \\
   \mbox{with less than $N$ bits''.}
\end{array}
\]

The interesting fact is that there is a computer program
\[
    \log_2 N + c_{\it FAS}
\]
bits long for calculating this number that supposedly cannot be calculated
by any program that is less than $N$ bits long. And
\[
    \log_2 N + c_{\it FAS}
\]
is much much smaller than $N$ for all sufficiently large $N$!  Thus
for such $N$ our {\sl FAS\/} cannot enable us to exhibit any numbers
that require programs more than $N$ bits long.  This is my
information-theoretic incompleteness result that I wanted to discuss
with G\"odel.

Why is there a program that is
\[
    \log_2 N + c_{\it FAS}
\]
bits long for calculating
\[
\begin{array}{l}
   \mbox{``the first positive integer that can be proved${}_{\it FAS}$} \\
   \mbox{to have the property that it cannot} \\
   \mbox{be specified${}_{\it UTM}$ by a computer program} \\
   \mbox{with less than $N$ bits'' ?}
\end{array}
\]
Well here is how you do it.

You start running through all possible proofs in the formal axiomatic
system in size order.  You apply the proof-checking algorithm to each
proof.  And after filtering out all the invalid proofs, you search for
the first proof that a particular positive integer requires at least
an $N$-bit program.

The algorithm that I've just described is very slow but it is very
simple.  It's basically just the proof-checking algorithm, which is
$c_{\it FAS}$ bits long, and the number $N$, which is $\log_2 N$ bits
long.  So the total number of bits is, as was claimed, just
\[
    \log_2 N + c_{\it FAS} .
\]
That concludes the proof of my incompleteness result that I wanted to
discuss with G\"odel.

Over the years I've continued to do research on my
information-theoretic approach to incompleteness.  Here are the three
most dramatic results that I've obtained:

\begin{itemize}

\item[1)] Call a program ``elegant'' if no smaller program produces
the same output.  You can't prove that a program is elegant.  More
precisely, $N$ bits of axioms are needed to prove that an $N$-bit
program is elegant.

\item[2)] Consider the binary representation of the halting
probability $\Omega$.  $\Omega$ is the probability that a program
chosen at random halts.  You can't prove what the bits of $\Omega$
are.  More precisely, $N$ bits of axioms are needed to determine $N$
bits of $\Omega$.

\item[3)] I have constructed a perverse algebraic equation
\[
   P(K,X,Y,Z,\ldots) = 0.
\]
Vary the parameter $K$ and ask whether this equation has finitely or
infinitely many whole-number solutions.  In each case this turns out
to be equivalent to determining individual bits of $\Omega$.
Therefore $N$ bits of axioms are needed to be able to settle $N$
cases.

\end{itemize}

These striking examples show that sometimes you have to put more into
a set of axioms in order to get more out.  (2) and (3) are extreme
cases.  They are accidental mathematical assertions that are true for
no reason.  In other words, the questions considered in (2) and (3)
are irreducible; essentially the only way to prove them is to add them
as new axioms.  Thus in this extreme case you get out of a set of
axioms only what you put in.

How do I prove these incompleteness results (1), (2) and (3)?  As
before, the basic idea is the paradox of ``the first positive integer
that cannot be specified in less than a billion words.''  For (1) the
connection with the Berry paradox is obvious.  For (2) and (3) it was
obvious to me only in the case where one is talking about determining
the {\bf first} $N$ bits of $\Omega$.  In the case where the $N$ bits
of $\Omega$ are scattered about, my original proof of (2) and (3) (the
one given in my Cambridge University Press monograph) is decidedly not
along the lines of the Berry paradox.  But a few years later I was
happy to discover a new and more straight-forward proof of (2) and (3)
that is along the lines of the Berry paradox!

In addition to working on incompleteness, I have also devoted a great
deal of thought to the central idea that can be extracted from my
version of the Berry paradox, which is to define the program-size
complexity of something to be the size in bits of the smallest program
that calculates it.  I have developed a general theory dealing with
program-size complexity that I call algorithmic information theory
({\sl AIT\/}).

{\sl AIT\/} is an elegant theory of complexity, perhaps the most
developed of all such theories, but as von Neumann said, pure
mathematics is easy compared to the real world!  {\sl AIT\/} provides
the correct complexity concept for metamathematics, but not necessarily
for other more practical fields.

Program-size complexity in {\sl AIT\/} is analogous to entropy in
statistical mechanics.  Just as thermodynamics gives limits on heat
engines, {\sl AIT\/} gives limits on formal axiomatic systems.

I have recently reformulated {\sl AIT.}

Up to now, the best version of {\sl AIT\/} studied the size of
programs in a computer programming language that was not actually
usable.  Now I obtain the correct program-size complexity measure from
a powerful and easy to use programming language.  This language is a
version of {\sl LISP,} and I have written an interpreter for it in
{\sl C.} A summary of this new work is available as IBM Research
Report RC 19553 ``The limits of mathematics,'' which I am expanding
into a book.

So this is what I would have liked to discuss with G\"odel, if I could
speak with him now.  Of course this is impossible!  But thank you very
much for giving me the opportunity to tell you about these ideas!

\section*{Questions for Future Research}

\begin{itemize}

\item Find questions in algebra, topology and geometry that are
equivalent to determining bits of $\Omega$.

\item What is an interesting or natural mathematical question?

\item How often is such a question independent of the usual axioms?
(I suspect the answer is ``Quite often!'')

\item Show that a classical open question in number theory such as
the Riemann hypothesis is independent of the usual axioms.  (I suspect
that this is often the case, but that it cannot be proven.)

\item Should we take incompleteness seriously or is it a red
herring?  (I believe that we should take incompleteness very seriously
indeed.)

\item Is mathematics quasi-empirical?  In other words, should
mathematics be done more like physics is done?  (I believe the answer
to both questions is ``Yes.'')

\end{itemize}

\section*{Bibliography}

{\bf Books:}

\begin{itemize}

\item
G. J. Chaitin,
{\it Information, Randomness \& Incompleteness,}
second edition, World Scientific, 1990.
Errata:
on page 26, line 25, ``quickly that''
should read ``quickly than'';
on page 31, line 19, ``Here one''
should read ``Here once'';
on page 55, line 17, ``RI, p.\ 35''
should read ``RI, 1962, p.\ 35'';
on page 85, line 14, ``1.\ The problem''
should read ``1.\ The Problem'';
on page 88, line 13, ``4.\ What is life?''
should read ``4.\ What is Life?'';
on page 108, line 13, ``the table in''
should read ``in the table in'';
on page 117, Theorem 2.3(q),
``$H_{C}(s,t)$''
should read ``$H_{C}(s/t)$'';
on page 134, line 7,
``$\# \{ n | H(n) \leq n \} \leq 2^{n}$''
should read
``$\# \{ k | H(k) \leq n \} \leq 2^{n}$'';
on page 274, bottom line, ``$n_{4p+4}$''
should read ``$n_{4p'+4}$''.

\item
G. J. Chaitin,
{\it Algorithmic Information Theory,}
fourth printing, Cambridge University Press, 1992.
Erratum:
on page 111, Theorem I0(q),
``$H_{C}(s,t)$''
should read ``$H_{C}(s/t)$''.

\item
G. J. Chaitin,
{\it Information-Theoretic Incompleteness,}
World Scientific, 1992.
Errata:
on page 67, line 25, ``are there are''
should read ``are there'';
on page 71, line 17, ``that case that''
should read ``the case that'';
on page 75, line 25, ``the the''
should read ``the'';
on page 75, line 31, ``$-\log_2p-\log_2q$''
should read ``$-p\log_2p-q\log_2q$'';
on page 95, line 22, ``This value of''
should read ``The value of'';
on page 98, line 34, ``they way they''
should read ``the way they'';
on page 99, line 16, ``exactly same''
should read ``exactly the same'';
on page 124, line 10, ``are there are''
should read ``are there''.

\end{itemize}
{\bf Recent Papers:}
\begin{itemize}

\item
G. J. Chaitin,
``On the number of $n$-bit strings with maximum complexity,''
{\it Applied Mathematics and Computation\/} {\bf 59} (1993), pp.\ 97--100.

\item
G. J. Chaitin,
``Randomness in arithmetic and the decline and fall of reductionism in
pure mathematics,''
{\it Bulletin of the European Association for Theoretical Computer Science,}
No.\ 50 (June 1993), pp.\ 314--328.

\item
G. J. Chaitin,
``Exhibiting randomness in arithmetic using Mathematica and C,''
{\it IBM Research Report RC-18946,} 94 pp., June 1993.

\item
G. J. Chaitin,
``The limits of mathematics---Course outline \& software,''
{\it IBM Research Report RC-19324,} 127 pp., December 1993.

\item
G. J. Chaitin,
``Randomness and complexity in pure mathematics,''
{\it International Journal of Bifurcation and Chaos\/}
{\bf 4} (1994), pp.\ 3--15.

\item
G. J. Chaitin,
``Responses to `Theoretical Mathematics\ldots',''
{\it Bulletin of the American Mathematical Society\/}
{\bf 30} (1994), pp.\ 181--182.

\item
G. J. Chaitin,
``The limits of mathematics (in C),''
{\it IBM Research Report RC-19553,} 68 pp., May 1994.

\end{itemize}
{\bf See Also:}
\begin{itemize}

\item
M. Davis,
``What is a computation?,'' in
L.A. Steen,
{\it Mathematics Today,}
Springer-Verlag, 1978.

\item
R. Rucker,
{\it Infinity and the Mind,}
Birkh\"auser, 1982.

\item
T. Tymoczko,
{\it New Directions in the Philosophy of Mathematics,}
Birkh\"auser, 1986.

\item
R. Rucker,
{\it Mind Tools,}
Houghton Mifflin, 1987.

\item
H.R. Pagels,
{\it The Dreams of Reason,}
Simon \& Schuster, 1988.

\item
D. Berlinski,
{\it Black Mischief,}
Harcourt Brace Jovanovich, 1988.

\item
R. Herken,
{\it The Universal Turing Machine,}
Oxford University Press, 1988.

\item
I. Stewart,
{\it Game, Set \& Math,}
Blackwell, 1989.

\item
G.S. Boolos and R.C. Jeffrey,
{\it Computability and Logic,}
third edition,
Cambridge University Press, 1989.

\item
J. Ford, ``What is chaos?,'' in
P. Davies,
{\it The New Physics,}
Cambridge University Press, 1989.

\item
J.L. Casti,
{\it Paradigms Lost,}
Morrow, 1989.

\item
G. Nicolis and I. Prigogine,
{\it Exploring Complexity,}
Freeman, 1989.

\item
J.L. Casti,
{\it Searching for Certainty,}
Morrow, 1990.

\item
B.-O. K\"uppers,
{\it Information and the Origin of Life,}
MIT Press, 1990.

\item
J.A. Paulos,
{\it Beyond Numeracy,}
Knopf, 1991.

\item
L. Brisson and
F.W. Meyerstein,
{\it Inventer L'Univers,}
Les Belles Lettres, 1991.
(English edition in press)

\item
J.D. Barrow,
{\it Theories of Everything,}
Oxford University Press, 1991.

\item
D. Ruelle,
{\it Chance and Chaos,}
Princeton University Press, 1991.

\item
T. N{\o}rretranders,
{\it M{\ae}rk Verden,}
Gyldendal, 1991.

\item
M. Gardner,
{\it Fractal Music, Hypercards and More,}
Freeman, 1992.

\item
P. Davies,
{\it The Mind of God,}
Simon \& Schuster, 1992.

\item
J.D. Barrow,
{\it Pi in the Sky,}
Oxford University Press, 1992.

\item
N. Hall,
{\it The New Scientist Guide to Chaos,}
Penguin, 1992.

\item
H.-C. Reichel and E. Prat de la Riba, 
{\it Naturwissenschaft und Weltbild,}
H\"older-Pichler-Tempsky, 1992.

\item
I. Stewart,
{\it The Problems of Mathematics,}
Oxford University Press, 1992.

\item
A.K. Dewdney,
{\it The New Turing Omnibus,}
Freeman, 1993.

\item
A.B. \c{C}ambel,
{\it Applied Chaos Theory,}
Academic Press, 1993.

\item
K. Svozil,
{\it Randomness \& Undecidability in Physics,}
World Scientific, 1993.

\item
J.L. Casti,
{\it Complexification,}
HarperCollins, 1994.

\item
M. Gell-Mann,
{\it The Quark and the Jaguar,}
Freeman, 1994.

\item
T. N{\o}rretranders,
{\it Verden Vokser,}
Aschehdoug, 1994.

\item
S. Wolfram,
{\it Cellular Automata and Complexity,}
Addison-Wesley, 1994.

\item
C. Calude,
{\it Information and Randomness,}
Springer-Verlag, 1994.

\end{itemize}

\chap{The decline and fall of reductionism (excerpt)}

\section*{Bulletin of the EATCS, No.\ 50 \\ (June 1993), pp.\ 314--328}

\section*{}

\section*{Experimental mathematics}

Okay, let me say a little bit in the minutes I have left about what this all
means.

First of all, the connection with physics.
There was a big controversy when quantum mechanics was developed,
because quantum theory is nondeterministic.  Einstein didn't like that.  He
said,
``God doesn't play dice!''  But as I'm sure you all know, with chaos and
nonlinear dynamics we've now realized that even in classical physics we
get randomness and unpredictability.  My work is in the
same spirit.  It shows that pure mathematics, in fact even
elementary number theory, the arithmetic of the natural numbers, 1, 2, 3,
4, 5, is in the same boat.  We get randomness there too.
So, as a newspaper headline would put it,
God not only plays dice in quantum mechanics
and in classical physics, but even in pure mathematics,
even in elementary number theory.  So if a new
paradigm is emerging, randomness is at the heart of it.
By the way, randomness is also at the heart of quantum field theory,
as virtual particles and Feynman path integrals (sums over all histories) show
very clearly.
So my work fits in with a lot of work in physics, which is why I often get
invited
to talk at physics meetings.

However the really important question isn't physics, it's mathematics.
I've heard that G\"odel wrote a letter to his mother who stayed in Europe.
You know, G\"odel and Einstein were friends at the Institute for Advanced
Study.
You'd see them walking down the street together.
Apparently G\"odel
wrote a letter to his mother saying that even though Einstein's work on
physics had really had a tremendous impact on how people did physics,
he was disappointed that his work had not had the same effect on
mathematicians.
It hadn't made a difference
in how mathematicians actually carried on their everyday work.  So I
think that's the key question:  How should you really do mathematics?

I'm claiming I have a much stronger incompleteness result.  If so maybe it'll
be clearer whether mathematics should be done the ordinary way.  What is the
ordinary
way of doing mathematics?  In spite of the fact that everyone knows
that any finite set of axioms is incomplete,
how do mathematicians actually work?    Well suppose you have a conjecture that
you've been
thinking about for a few weeks, and you believe it because you've
tested a large number of cases on a computer.  Maybe it's a conjecture about
the primes and
for two weeks you've tried to prove it.  At the end of two weeks you don't say,
well obviously the reason I haven't been able to show this is because of
G\"odel's incompleteness theorem!  Let us therefore add it
as a new axiom!  But if you took G\"odel's incompleteness theorem
very seriously this might in fact be the way to proceed.  Mathematicians will
laugh
but physicists actually behave this way.

Look at the history of physics.
You start with Newtonian physics.  You cannot get Maxwell's equations from
Newtonian physics.  It's a new domain of experience---you need new postulates
to deal with it.  As for special relativity, well, special relativity is almost
in Maxwell's equations.  But Schr\"odinger's equation does not come from
Newtonian physics and Maxwell's equations.  It's a new domain of experience
and again you need new axioms.  So physicists are used to the idea that when
you start experimenting at a smaller scale, or with new phenomena,
you may need new principles to understand and explain what's going on.

Now in spite of incompleteness mathematicians don't behave at all like
physicists do.
At a subconscious level they still assume that the small
number of principles, of postulates and methods of inference,
that they learned early as mathematics students, are enough.
In their hearts they believe
that if you can't prove a result it's your own fault.  That's
probably a good attitude to take rather than to blame someone else, but let's
look at a question like the Riemann hypothesis.
A physicist would say that there is ample experimental evidence for the Riemann
hypothesis
and would go ahead and take it as a working assumption.

What is the Riemann hypothesis?  There are many unsolved questions involving
the distribution
of the prime numbers that can be settled if you assume the Riemann hypothesis.
Using computers people check these conjectures and they work
beautifully.  They're neat formulas but nobody can prove them.
A lot of them follow from the Riemann hypothesis.  To a physicist this would
be enough:  It's useful, it explains a lot of data.  Of course a physicist then
has
to be prepared to say ``Oh oh, I goofed!''\ because an experiment can
subsequently
contradict a theory.  This happens very often.

In particle physics you throw up theories all the time and most of them
quickly die.  But mathematicians don't like to have to backpedal.
But if you play it safe,
the problem is that you may be losing out, and I believe you are.

I think it should be obvious where I'm leading.
I believe that elementary number theory and the rest of mathematics
should be pursued more in the spirit of experimental science, and that you
should be willing to adopt new principles.  I believe that Euclid's statement
that an axiom is a self-evident truth is a big mistake.
The Schr\"odinger equation certainly isn't a self-evident truth!  And the
Riemann hypothesis isn't self-evident either, but it's very useful.

So I believe that we mathematicians shouldn't ignore incompleteness.  It's a
safe thing to
do but we're losing out on results that we could get.  It would be as if
physicists said, okay no Schr\"odinger equation, no Maxwell's equations, we
stick with Newton, everything must be deduced from Newton's laws.
(Maxwell even tried it.  He had a mechanical model of an electromagnetic
field.  Fortunately they don't teach that in college!)

I proposed all this twenty years ago
when I started getting these information-theoretic incompleteness results.
But independently a new school on the philosophy of
mathematics is emerging called the ``quasi-empirical'' school of thought
regarding the foundations
of mathematics.  There's a book of Tymoczko's called
{\it New Directions in the Philosophy of Mathematics\/}
(Birkh\"auser, Boston, 1986).  It's a good collection of articles.
Another place to look is {\it Searching for Certainty\/} by John Casti
(Morrow, New York, 1990) which has a good chapter on mathematics.  The last
half of the chapter talks about this quasi-empirical view.

By the way, Lakatos, who was one of the people involved in this new movement,
happened to be at Cambridge at that time.  He'd left Hungary.

The main schools of mathematical philosophy
at the beginning of this century were Russell and Whitehead's
view that logic was the basis for everything, the formalist school of Hilbert,
and an ``intuitionist'' constructivist school of Brouwer.
Some people think that Hilbert believed that
mathematics is a meaningless game played with marks of ink on paper.
Not so!
He just said that to be absolutely clear and precise what mathematics is all
about, we have to specify the
rules determining whether a proof is correct so precisely that they become
mechanical.
Nobody who thought that mathematics
is meaningless would have been so energetic and done such important
work and been such an inspiring leader.

Originally
most mathematicians backed Hilbert. Even after G\"odel
and even more emphatically Turing showed that
Hilbert's dream
didn't work, in practice mathematicians carried on as before, in Hilbert's
spirit.
Brouwer's constructivist attitude was mostly considered a nuisance.
As for Russell and Whitehead, they had a lot of
problems getting all of mathematics from logic.  If you
get all of mathematics from set theory you discover that it's nice to define
the whole numbers in terms of sets (von Neumann worked on this).
But then it turns out that there's all kinds of problems with sets.
You're not making the natural numbers more solid by basing them on something
which is
more problematical.

Now everything has gone
topsy-turvy.  It's gone topsy-turvy, not because of any philosophical
argument, not because of G\"odel's
results or Turing's results or my own incompleteness results.
It's gone topsy-turvy for a very simple reason---the computer!

The computer as you all know has changed the way we do everything.
The computer has
enormously and vastly increased mathematical experience.  It's so
easy to do calculations,
to test many cases, to run experiments on the computer.
The computer has
so vastly increased mathematical experience, that in order to cope,
people are forced to proceed in a more pragmatic
fashion.
Mathematicians are proceeding more pragmatically, more
like experimental scientists do.
This new tendency is often called ``experimental mathematics.''
This phrase comes up a lot in the field of chaos, fractals and nonlinear
dynamics.

It's often the case that when doing experiments on the computer,
numerical experiments with equations,
you see that something happens, and you conjecture a result.
Of course it's nice if you can prove it.  Especially if the proof is short.
I'm not sure that a thousand
page proof helps too much.  But if it's a short proof it's
certainly better than not having a proof.  And if you have several proofs from
different viewpoints, that's very good.

But sometimes
you can't find a proof and you can't wait for
someone else to find a proof, and you've got to
carry on as best you can.  So now mathematicians sometimes
go ahead with working hypotheses on the basis of the results
of computer experiments.  Of course if it's physicists doing these computer
experiments, then it's certainly okay; they've always relied heavily on
experiments.
But now even mathematicians sometimes operate in this manner.
I believe that there's a new journal called the {\it Journal of Experimental
Mathematics.}  They should've put me on their editorial board, because
I've been proposing this for twenty years based on my
information-theoretic ideas.

So in the end it wasn't G\"odel,
it wasn't Turing, and it wasn't my results that are making mathematics go in an
experimental mathematics direction, in a quasi-empirical direction.  The reason
that mathematicians are changing their working habits is the computer.
I think it's an excellent joke!
(It's also funny that of the three old schools of mathematical philosophy,
logicist, formalist, and intuitionist, the most neglected was Brouwer,
who had a constructivist attitude
years before the computer gave a tremendous impulse to constructivism.)

Of course, the mere fact that everybody's doing something doesn't mean that
they ought to be.
The change in how people are behaving isn't because of
G\"odel's theorem or Turing's theorems or my theorems, it's because of the
computer.  But I think that the sequence of work that I've outlined does
provide some theoretical justification for what everybody's
doing anyway without worrying about the theoretical justification.
And I think that the question
of how we should actually do mathematics requires {\bf at least} another
generation of work.
That's basically what I wanted to say---thank you very much!

\chap{A Version of Pure LISP}

\section*{Introduction}

In this chapter we present a ``permissive'' simplified version of pure
LISP designed especially for metamathematical applications.  Aside
from the rule that an S-expression must have balanced ()'s, the only
way that an expression can fail to have a value is by looping forever.
This is important because algorithms that simulate other algorithms
chosen at random, must be able to run garbage safely.

This version of LISP developed from one that I originally designed to
use to teach LISP in the early 1970s.  The language was reduced to its
essence and made as easy to learn as possible, and was actually used
in several university courses that I gave in Buenos Aires, Argentina.
Like APL, this version of LISP is so concise that one can write it as
fast as one thinks.  This LISP is so simple that an interpreter for it
can be coded in just six hundred lines of C code.  The C code for
this is given at the end of this book.

{\it How to read this chapter:} This chapter can be quite difficult to
understand, especially if one has never programmed in LISP before.
The correct approach is to read it several times, and to try to work
through all the examples in detail.  Initially the material will seem
completely incomprehensible, but all of a sudden the pieces will snap
together into a coherent whole.  On the other hand, if one has never
experienced LISP before and wishes to master it thoroughly, one should
code a LISP interpreter oneself instead of looking at the
interpreter given at the end of this book; that is how the author
learned LISP.

\section*{Definition of LISP}

LISP is an unusual programming language created around 1960 by John
McCarthy.  It and its descendants are frequently used in research on
artificial intelligence.  And it stands out for its simple design and
for its precisely defined syntax and semantics.

However LISP more closely resembles such fundamental subjects as set
theory and logic than it does a programming language.  As a result
LISP is easy to learn with little previous knowledge.  Contrariwise,
those who know other programming languages may have difficulty
learning to think in the completely different fashion required by
LISP.

LISP is a functional programming language, not an imperative language
like FORTRAN.  In FORTRAN the question is ``In order to do something
what operations must be carried out, and in what order?''  In LISP
the question is ``How can this function be defined?''  The LISP
formalism consists of a handful of primitive functions and certain
rules for defining more complex functions from the initially given
ones.  In a LISP run, after defining functions one requests their
values for specific arguments of interest.  It is the LISP
interpreter's task to deduce these values using the function's
definitions.

LISP functions are technically known as partial recursive functions.
``Partial'' because in some cases they may not have a value (this
situation is analogous to division by zero or an infinite loop).
``Recursive'' because functions re-occur in their own definitions.
The following definition of factorial $n$ is the most familiar example
of a recursive function: if $n$ = 0, then its value is 1, else its
value is $n$ by factorial $n-1$.  From this definition one deduces
that factorial 3 = (3 by factorial 2) = (3 by 2 by factorial 1) = (3
by 2 by 1 by factorial 0) = (3 by 2 by 1 by 1) = 6.

A LISP function whose value is always true or false is called a
predicate.  By means of predicates the LISP formalism encompasses
relations such as ``$x$ is less than $y$.''

Data and function definitions in LISP consist of S-expressions (S
stands for ``symbolic'').  S-expressions are made up of characters
called atoms that are grouped into lists by means of pairs of
parentheses.  The atoms are the printable characters in the ASCII
character set, except for the blank, left parenthesis, right
parenthesis, left bracket, and right bracket.  The simplest kind of
S-expression is an atom all by itself.  All other S-expressions are
lists.  A list consists of a left parenthesis followed by zero or more
elements (which may be atoms or sublists) followed by a right
parenthesis.  Also, the empty list {\tt ()} is considered to be an
atom.

Here are two examples of S-expressions.
{\tt C} is an atom.
\begin{center}
{\tt (d(ef)d((a)))}
\end{center}
is a list with four elements.  The first and third elements are the
atom {\tt d}.  The second element is a list whose elements are the
atoms {\tt e} and {\tt f}, in that order.  The fourth element is a
list with a single element, which is a list with a single element,
which is the atom {\tt a}.

The formal definition is as follows.  The class of S-expressions is
the union of the class of atoms and the class of lists.  A list
consists of a left parenthesis followed by zero or more S-expressions
followed by a right parenthesis.  There is one list that is also an
atom, the empty list {\tt ()}.  All other atoms are individual ASCII
characters.

In LISP the atom {\tt 1} stands for ``true'' and the atom {\tt 0}
stands for ``false.''  Thus a LISP predicate is a function whose value
is always {\tt 0} or {\tt 1}.

It is important to note that we do not identify {\tt 0} and {\tt ()}.
It is usual in LISP to identify falsehood and the empty list; both are
usually called NIL.  Here there is no single-character synonym
for the empty list {\tt ()}; 2 characters are required.

The fundamental semantical concept in LISP is that of the value of an
S-expression in a given environment.  An environment consists of a
so-called ``association list'' in which variables (atoms) and their
values (S-expressions) alternate.  If a variable appears several
times, only its first value is significant.  If a variable does not
appear in the environment, then it itself is its value, so that it is
in effect a literal constant.
\verb|(xa x(a) x((a)) F(&(x)(/(.x)x(F(+x)))))|
is a typical environment.
In this environment the value of {\tt x} is {\tt a}, the
value of {\tt F} is
\verb|(&(x)(/(.x)x(F(+x))))|,
and any other atom, for example {\tt Q}, has itself as value.

Thus the value of an atomic S-expression is obtained by searching odd
elements of the environment for that atom.  What is the value of a
non-atomic S-expression, that is, of a non-empty list?  In this case
the value is defined recursively, in terms of the values of the
elements of the S-expression in the same environment.  The value of
the first element of the S-expression is the function, and the
function's arguments are the values of the remaining elements of the
expression.  Thus in LISP the notation {\tt (fxyz)} is used for what
in FORTRAN would be written {\tt f(x,y,z)}.  Both denote the function
{\tt f} applied to the arguments {\tt xyz}.

There are two kinds of functions: primitive functions and defined
functions.  The ten primitive functions are the atoms {\tt . = + - * ,
' / !} and {\tt ?}.  A defined function is a three-element list
(traditionally called a LAMBDA expression) of the form
\verb|(&vb)|, where {\tt v} is a list of variables.
By definition the result of applying a defined function to arguments
is the value of the body of the function {\tt b} in the environment
resulting from appending a list of the form (variable1 argument1
variable2 argument2\ldots\ ) and the environment of the original
S-expression, in that order.  Appending an $n$-element list
and an $m$-element list yields the $(n+m)$-element list
whose elements are those of the first list followed by those of the
second list.

The primitive functions are now presented. In the examples of their
use the empty environment is assumed.
\begin{itemize}
\item
\begin{description}
\item[Name] Quote
\item[Symbol] {\tt '}
\item[Arguments] 1
\item[Explanation] The result of applying this function is the
unevaluated argument expression.
\item[Examples] {\tt ('(abc))} has value {\tt (abc)}

{\tt ('(*xy))} has value {\tt (*xy)}
\end{description}
\item
\begin{description}
\item[Name] Atom
\item[Symbol] {\tt .}
\item[Arguments] 1
\item[Explanation] The result of applying this function to an
argument is true or false depending on whether or not the argument is
an atom.
\item[Examples] {\tt (.a)} has value {\tt 1}

{\tt (.('(abc)))} has value {\tt 0}
\end{description}
\item
\begin{description}
\item[Name] Equal
\item[Symbol] {\tt =}
\item[Arguments] 2
\item[Explanation] The result of applying this function to two
arguments is true or false depending on whether or not they are the
same S-expression.
\item[Examples] {\tt (=('(abc))('(abc)))} has value {\tt 1}

{\tt (=('(abc))('(abx)))} has value {\tt 0}
\end{description}
\item
\begin{description}
\item[Name] Head/First/CAR
\item[Symbol] {\tt +}
\item[Arguments] 1
\item[Explanation] The result of applying this function to an atom
is the atom itself.
The result of applying this function to a non-empty list
is the first element of the list.
\item[Examples] {\tt (+a)} has value {\tt a}

{\tt (+('(abc)))} has value {\tt a}

{\tt (+('((ab)(cd))))} has value {\tt (ab)}

{\tt (+('(((a)))))} has value {\tt ((a))}
\end{description}
\item
\begin{description}
\item[Name] Tail/Rest/CDR
\item[Symbol] {\tt -}
\item[Arguments] 1
\item[Explanation] The result of applying this function to an atom
is the atom itself.
The result of applying this function to a non-empty list is
what remains if its first element is erased.
Thus the tail of an $(n+1)$-element list is
an $n$-element list.
\item[Examples] {\tt (-a)} has value {\tt a}

{\tt (-('(abc)))} has value {\tt (bc)}

{\tt (-('((ab)(cd))))} has value {\tt ((cd))}

{\tt (-('(((a)))))} has value {\tt ()}
\end{description}
\item
\begin{description}
\item[Name] Join/CONS
\item[Symbol] {\tt *}
\item[Arguments] 2
\item[Explanation] If the second argument is not a list,
then the result of applying this function is the first argument.
If the second argument is an $n$-element list,
then the result of applying this function is
the $(n+1)$-element list whose head is the first argument
and whose tail is the second argument.
\item[Examples] {\tt (*ab)} has value {\tt a}

{\tt (*a())} has value {\tt (a)}

{\tt (*a('(bcd)))} has value {\tt (abcd)}

{\tt (*('(ab))('(cd)))} has value {\tt ((ab)cd)}

{\tt (*('(ab))('((cd))))} has value {\tt ((ab)(cd))}
\end{description}
\item
\begin{description}
\item[Name] Display
\item[Symbol] {\tt ,}
\item[Arguments] 1
\item[Explanation] The result of applying this function is its
argument, in other words, this is an identity function.
The side-effect is to display the argument.
This function is used to display intermediate results.
It is the only primitive function that has a side-effect.
\item[Examples] Evaluation of {\tt (-(,(-(,(-('(abc))))))} displays
{\tt (bc)} and {\tt (c)} and yields value {\tt ()}
\end{description}
\item
\begin{description}
\item[Name] If-then-else
\item[Symbol] {\tt /}
\item[Arguments] 3
\item[Explanation] If the first argument is not false,
then the result is the second argument.
If the first argument is false,
then the result is the third argument.
The argument that is not selected is not evaluated.
\item[Examples] {\tt (/1xy)} has value {\tt x}

{\tt (/0xy)} has value {\tt y}

{\tt (/Xxy)} has value {\tt x}

Evaluation of {\tt (/1x(,y))} does not
have the side-effect of displaying {\tt y}
\end{description}
\item
\begin{description}
\item[Name] Eval
\item[Symbol] {\tt !}
\item[Arguments] 1
\item[Explanation] The expression that is the value of the argument
is evaluated in an empty environment.
This is the only primitive function that is a partial rather than
a total function.
\item[Examples]
{\tt (!(,('(.x))))} diplays {\tt (.x)} and has value {\tt 1}

\verb|(!('(('(&(f)(f)))('(&()(f))))))| has no value.
\end{description}
\item
\begin{description}
\item[Name] Try
\item[Symbol] {\tt ?}
\item[Arguments] 2
\item[Explanation] The expression that is the value of the second
argument is evaluated in an empty environment.
If the evaluation is completed within ``time'' given by
the first argument, the value returned is a list whose sole
element is the value of the value of the second argument.
If the evaluation is not completed within ``time'' given by
the first argument, the value returned is the atom {\tt ?}.
More precisely, the ``time limit'' is
given by the number of
elements of the first argument, and is zero if the first argument
is not a list.
The ``time limit'' actually
limits the depth of the call stack, more precisely,
the maximum number of re-evaluations due to defined functions
or {\tt !} or {\tt ?} which have been
started but have not yet been completed.
The key property of {\tt ?} is that it is a total function,
i.e., is defined for all values of its arguments, and that
{\tt (!x)} is defined if and only if
{\tt (?tx)} is not equal to {\tt ?}
for all sufficiently large values of
{\tt t}.
\item[Examples] {\tt (?0('x))} has value {\tt (x)}

\verb|(?0('(('(&(x)x))a)))| has value {\tt ?}

\verb|(?('(1))('(('(&(x)x))a)))| has value {\tt (a)}
\end{description}
\end{itemize}

The argument of {\tt '} and the unselected argument of
{\tt /} are exceptions to the rule that
the evaluation of an
S-expression that is a non-empty list requires the previous
evaluation of all its elements.
When evaluation of the elements of a list is required, this
is always done one element at a time, from left to right.

\begin{figure}
\begin{verbatim}
Atom        . = + - * , ' / ! ? & :
Arguments   1 2 1 1 2 1 1 3 1 2 2 3
\end{verbatim}
{\bf Figure 1: Atoms with Implicit Parentheses.}
\end{figure}

M-expressions (M stands for ``meta'') are S-expressions in which the
parentheses grouping together primitive functions and their arguments
are omitted as a convenience for the LISP programmer.  See Figure 1.
For these purposes,
\verb|&| (``function/LAMBDA/define'')
is treated as if it were
a primitive function with two arguments, and
``{\tt :}'' (``LET/is'') is treated as if it were
a primitive function with three arguments.
``{\tt :}'' is another meta-notational abbreviation,
but may be thought of as an additional primitive function.

{\tt :vde} denotes the value of {\tt e} in an
environment in which
{\tt v} evaluates to the current value of {\tt d},
and {\tt :(fxyz)de} denotes the value of {\tt e} in an
environment in which {\tt f} evaluates to \verb|(&(xyz)d)|.
More precisely, the M-expression
{\tt :vde} denotes the S-expression \verb|(('(&(v)e))d)|,
and the M-expression
{\tt :(fxyz)de} denotes the S-expression
\verb|(('(&(f)e))('(&(xyz)d)))|, and similarly for functions
with a different number of arguments.
Within the scope of a function definition done via ``{\tt :}'',
i.e., within the
{\tt de} portion
of {\tt :(fxyz)de}, the parentheses that associate the function
{\tt f} with its three arguments are omitted, just as if {\tt f}
were a primitive function, and similarly for functions {\tt f} with a
different number of arguments.

Also, we shall often use unary arithmetic.  To make this easier,
the M-expression {\tt \{ddd\}} denotes the list of {\tt ddd} 1's; the
number of 1's is given in decimal digits.  E.g., \{99\} is a list
of ninety-nine 1's.

A {\tt "} (``literally'') is written before a self-contained
portion of an M-expression to indicate that the convention regarding
invisible parentheses and the meaning of ``{\tt :}'' and {\tt \{\}}
does not apply within it, i.e., that
there follows an S-expression ``as is''.

Input to the LISP interpreter consists of a list of M-expressions.
All blanks are ignored, and comments may be inserted anywhere by
placing them between balanced {\tt [}'s and {\tt ]}'s,
so that comments may include other comments.
Anything remaining on the last line of an M-expression that is
excess is ignored.  Thus M-expressions can end with excess
right parentheses to make sure that all lists are closed, but on the
other hand two M-expressions cannot be given on the same line.
Two kinds of M-expressions are read by the interpreter: expressions to
be evaluated, and others that indicate the environment to be used for
these evaluations.
The initial environment is the empty list {\tt ()}.

Each M-expression is transformed into the corresponding
S-e\-x\-p\-r\-e\-s\-s\-i\-o\-n
and displayed:
\begin{itemize}
\item[(1)]
If the S-expression is of the form
\verb|(&xe)|
where {\tt x} is an atom and {\tt e} is an S-expression, then
{\tt (xv)}
is concatenated with the current environment to obtain a new environment,
where {\tt v} is the value of {\tt e}.
Thus
\verb|(&xe)|
is used to define the value of a variable
{\tt x} to be equal to the value of an S-expression {\tt e}.
\item[(2)]
If the S-expression is of the form
\verb|(&(fxyz)d)|
where {\tt fxyz} is one or more atoms
and {\tt d} is an S-expression, then
\verb|(f(&(xyz)d))| is
concatenated with the current environment to obtain a new environment.
Thus
\verb|(&(fxyz)d)|
is used to establish function definitions, in this case
the function {\tt f} of the variables {\tt xyz}.
Within {\tt d} and all the remaining interpreter input,
the parentheses that associate the function
{\tt f} with its three arguments are omitted, just as if {\tt f}
were a primitive function, and similarly for functions {\tt f} with a
different number of arguments.
\item[(3)]
If the S-expression is not of the form
\verb|(&...)| then it is
evaluated in the current environment and its value is displayed.
The primitive function ``{\tt ,}'' may cause the interpreter to
display additional S-expressions before this value.
\end{itemize}

\section*{Examples}

Here are five elementary examples of expressions and their values.
\begin{itemize}
\item
The M-expression
{\tt *a*b*c()}
denotes the S-expression
\begin{center}
{\tt (*a(*b(*c())))}
\end{center}
whose value is the S-expression
{\tt (abc)}.
\item
The M-expression
{\tt +---'(abcde)}
denotes the S-expression
\begin{center}
{\tt (+(-(-(-('(abcde))))))}
\end{center}
whose value is the S-expression
{\tt d}.
\item
The M-expression
{\tt *"+*"=*"-()}
denotes the S-expression
\begin{center}
{\tt (*+(*=(*-())))}
\end{center}
whose value is the S-expression
{\tt (+=-)}.
\item
The M-expression
\begin{center}
\verb|('&(xyz)*z*y*x()abc)|
\end{center}
denotes the S-expression
\begin{center}
\verb|(('(&(xyz)(*z(*y(*x())))))abc)|
\end{center}
whose value is the S-expression
{\tt (cba)}.
\item
The M-expression
\begin{center}
{\tt :(Cxy)/.xy*+xC-xy C'(abcdef)'(ghijkl)}
\end{center}
denotes the S-expression
\begin{center}
\verb|(| \\
\verb| ('(&(C)(C('(abcdef))('(ghijkl)))))| \\
\verb| ('(&(xy)(/(.x)y(*(+x)(C(-x)y)))))| \\
\verb|)|
\end{center}
whose value is the S-expression
{\tt (abcdefghijkl)}.
In this example {\tt C} is the concatenation function.
It is instructive to state the definition of
concatenation, usually called APPEND, in words:
``Let concatenation
be a function of two variables $x$ and $y$ defined as follows:
if $x$ is an atom, then the value is $y$;
otherwise join the head of $x$ to
the concatenation of the tail of $x$ with $y$.''
\end{itemize}

Here are a number of important details about the way our LISP achieves
its ``permissiveness.''  Most important, extra arguments to functions
are evaluated but ignored, and empty lists are supplied for missing
arguments.  E.g., parameters in a function definition which are not
supplied with an argument expression when the function is applied will
be bound to the empty list {\tt ()}.  This works this way because when
the interpreter runs off the end of a list of arguments, the list of
arguments has been reduced to the empty argument list, and head and
tail applied to this empty list will continue to give the empty list.
Also if an atom is repeated in the parameter list of a function
definition, the binding corresponding to the first occurrence will
shadow the later occurrences of the same variable.  In our LISP there
are no erroneous expressions, only expressions that fail to have a
value because the interpreter never finishes evaluating them: it goes
into an infinite loop and never returns a value.

\newpage
\begin{center}
\begin{tabular}{||c|l|l|l||}   \hline\hline
' & quote     & 1 arg  & '(abc) $\longrightarrow$ (abc)             \\ \hline
+ & head      & 1 arg  & +'(abc) $\longrightarrow$ a                \\
  &           &        & +a $\longrightarrow$ a                     \\ \hline
--& tail      & 1 arg  & --'(abc) $\longrightarrow$ (bc)            \\
  &           &        & --a $\longrightarrow$ a                    \\ \hline
* & join      & 2 args & *a'(bc) $\longrightarrow$ (abc)            \\
  &           &        & *ab $\longrightarrow$ a                    \\ \hline
. & atom      & 1 arg  & .a $\longrightarrow$ 1                     \\
  &           &        & .'(a) $\longrightarrow$ 0                  \\ \hline
= & equal     & 2 args & =aa $\longrightarrow$ 1                    \\
  &           &        & =ab $\longrightarrow$ 0                    \\ \hline
/ & if        & 3 args & /0ab $\longrightarrow$ b                   \\
  &           &        & /xab $\longrightarrow$ a                   \\ \hline
\& & function & 2 args & ('\&(xy)y ab) $\longrightarrow$ b          \\ \hline
, & display   & 1 arg  & ,x $\longrightarrow$ x and displays x      \\ \hline
! & eval      & 1 arg  & !e $\longrightarrow$ evaluate e            \\ \hline
? & try       & 3 args & ?teb $\longrightarrow$ evaluate e time t with bits b
\\
  &           &        & ?teb $\longrightarrow$
$
  \left(
     \begin{array}{c}
        \mbox{!} \\
        \mbox{?} \\
        \mbox{(value)}
     \end{array}
     \begin{array}{l}
        \mbox{captured} \\
        \mbox{displays\ldots}
     \end{array}
  \right)
$
                                                                    \\ \hline
@ & read bit  & 0 args & @ $\longrightarrow$ 0 or 1                 \\ \hline
\% & read exp & 0 args & \% $\longrightarrow$ any m-expression      \\ \hline
\# & bits for & 1 arg  & \#x $\longrightarrow$ bit string for x     \\ \hline
\verb|^| & append    & 2 args & \verb|^|'(ab)'(cd) $\longrightarrow$ (abcd)
  \\ \hline
\verb|~| & show      & 1 arg  & \verb|~|x $\longrightarrow$ x and may show x
  \\ \hline
: & let       & 3 args & :xv e    $\longrightarrow$ ('\&(x)e v)     \\
  &           &        & :(fx)d e $\longrightarrow$ ('\&(f)e '\&(x)d) \\ \hline
\& & define   & 2 args & \&xv    $\longrightarrow$ x is v           \\
  &           &        & \&(fx)d $\longrightarrow$ f is \&(x)d      \\ \hline
" & literally & 1 arg  & "+ $\longrightarrow$ +                     \\ \hline
\{\} & unary  &        & \{3\} $\longrightarrow$ (111)              \\ \hline
[ ]& comment  &        & [ignored]                   \\ \hline
( )& empty    &        &                             \\ \hline
0  & false    &        &                             \\ \hline
1  & true     &        &                             \\ \hline\hline
\end{tabular}
\end{center}

\chap{How to Give Binary Data to LISP Expressions}

Here is a quick summary of the toy LISP presented in the previous
chapter.  Each LISP primitive function and variable is a single
ASCII character.  These primitive functions, all of which have a fixed
number of arguments, are now {\tt '} for QUOTE (1 argument), {\tt .}
for ATOM (1 argument), {\tt =} for EQ (2 arguments), {\tt +} for HEAD
(1 argument), {\tt -} for TAIL (1 argument), {\tt *} for JOIN (2
arguments), {\tt \&} for FUNCTION and DEFINE (2 arguments), {\tt :} for
LET-BE-IN (3 arguments), {\tt /} for IF-THEN-ELSE (3 arguments), {\tt
,} for DISPLAY (1 argument), {\tt !} for EVAL (1 argument), and {\tt ?}
for TRY (which had 2 arguments, but will soon have 3).  The
meta-notation {\tt "} (LITERALLY)
indicates that an S-expression with explicit
parentheses follows, not what is usually the case in my toy LISP, an
M-expression, in which parentheses are often
implicit.  Finally comments are in {\tt []}'s,
the empty list is written {\tt ()}, TRUE
and FALSE are {\tt 1} and {\tt 0}, and in M-expressions
the list of {\tt ddd} 1's is
written {\tt \{ddd\}}.

The new idea is this.  We define our standard self-delimiting
universal Turing machine as follows.  Its program is in binary, and
appears on a tape in the following form.  First comes a toy LISP
M-expression, written in ASCII with 7 bits per character.  Note that
this expression is self-delimiting.
The TM reads in this LISP expression, and then evaluates it.  As it
does this, two new primitive functions {\tt @} and {\tt \%} with no
arguments may be used to read more from the TM tape.  Both of these
functions explode if the tape is exhausted, killing the computation.
{\tt @} reads a single bit from the tape, and {\tt \%} reads in an
entire LISP M-expression, in 7-bit character chunks.

This is the only way that information on the TM tape may be accessed,
which forces it to be used in a self-delimiting fashion.  This is
because no algorithm can search for the end of the tape and then use
the length of the tape as data in the computation.  If an algorithm
attempts to read a bit that is not on the tape, the algorithm aborts.

How is information placed on the TM tape in the first place?  Well, in
the starting environment, the tape is empty and any attempt to read it
will give an error message.  To place information on the tape, a new
argument is added to the primitive function {\tt ?} for depth-limited
evaluation.

Consider the three arguments $\alpha$, $\beta$ and $\gamma$ of {\tt
?}.  The meaning of the first argument, the depth limit $\alpha$, has
been changed slightly.  If $\alpha$ is a non-null atom, then there is
no depth limit.  If $\alpha$ is the empty list, this means zero depth
limit (no function calls or re-evaluations).  And an $N$-element list
$\alpha$ means depth limit $N$.  The second argument $\beta$ of {\tt
?} is as before the expression to be evaluated as long as the depth
limit $\alpha$ is not exceeded.  The new third argument $\gamma$ of
{\tt ?} is a list of bits to be used as the TM tape.

The value $\nu$ returned by the primitive function {\tt ?} is also
changed.  $\nu$ is a list.  The first element of $\nu$ is {\tt !} if
the evaluation of $\beta$ aborted because an attempt was made to read
a non-existent bit from the TM tape.  The first element of $\nu$ is
{\tt ?} if evaluation of $\beta$ aborted because the depth limit
$\alpha$ was exceeded.  {\tt !} and {\tt ?} are the only possible
error flags, because my toy LISP is designed with maximally permissive
semantics.  If the computation $\beta$ terminated normally instead of
aborting, the first element of $\nu$ will be a list with only one
element, which is the result produced by the computation $\beta$.
That's the first element of the list $\nu$ produced by the {\tt ?}
primitive function.

The rest of the value $\nu$ is a stack of all the arguments to the
primitive function ``{\tt ,}'' that were encountered during the evaluation
of $\beta$.  More precisely, if ``{\tt ,}'' was called $N$ times during
the evaluation of $\beta$, then $\nu$ will be a list of $N+1$
elements.  The $N$ arguments of ``{\tt ,}'' appear in $\nu$ in inverse
chronological order.  Thus {\tt ?} can not only be used to determine
if a computation $\beta$ reads too much tape or goes on too long
(i.e., to greater depth than $\alpha$), but {\tt ?} can also be used
to capture all the output that $\beta$ displayed as it went along,
whether the computation $\beta$ aborted or not.

In summary, all that one has to do to simulate a self-delimiting
universal Turing machine $U(p)$ running on the binary program $p$ is
to write
\begin{verbatim}
                         ?0'!%p
\end{verbatim}
This is an M-expression with parentheses omitted from primitive
functions (which can be done because
all primitive functions have a fixed number of
arguments).  With all parentheses supplied, it becomes the S-expression
\begin{verbatim}
                     (?0('(!(%)))p)
\end{verbatim}
This says that one is to read a complete LISP M-expression from the TM
tape $p$ and then evaluate it without any time limit and using
whatever is left on the tape $p$.

Two more primitive functions have also been added, the 2-argument
function \verb|^| for APPEND, i.e., list concatenation, and the
1-argument function {\tt \#} for converting an M-expression into the
list of the bits in its ASCII character string representation.  These
are used for constructing the bit strings that are then put on the TM
tape using {\tt ?}'s third argument $\gamma$.  Note that the functions
\verb|^|, {\tt \#} and {\tt \%} could be programmed rather than
included as built-in primitive functions, but it is extremely
convenient and much much faster to provide them built in.

Finally a new 1-argument identity function \verb|~| for SHOW
is provided for debugging;  the {\tt show} version of the interpreter prints
this argument.
Output produced by \verb|~| is invisible to the ``official'' {\tt ?}
and ``{\tt ,}'' output mechanism.  \verb|~| is needed because {\tt
?}$\alpha\beta\gamma$ suppresses all output $\theta$ produced within
its depth-controlled evaluation of $\beta$.  Instead {\tt ?} stacks
all output $\theta$ from within $\beta$ for inclusion in the final
value $\nu$ that {\tt ?} returns, namely $\nu = $ (atomic error flag
or (value of $\beta$) followed by the output $\theta$).

Each time the interpreter {\tt show}'s an expression, it also
indicates its size in characters and bits, in decimal and octal,
like this:
``decimal(octal)/decimal(octal)''.  The size in bits is seven times
the size in characters.  This is intended for use in sizing M-expressions,
which in order to be written as S-expressions must be encased in
an extra pair of parentheses.  Hence
the size in characters does not include
the outermost containing parentheses.

To summarize, we've added the following five primitive functions to
our LISP: {\tt \#} for BITS FOR (1 argument), {\tt
@} for READ BIT (0 arguments), {\tt \%} for READ M-EXP (0 arguments),
\verb|^| for APPEND (2 arguments), and \verb|~| for SHOW (1
argument).  And now {\tt ?} has three arguments instead of two and
returns a more complicated value.

\chap{Course Outline}

The course begins by explaining with examples my toy LISP.  See {\tt
test.l} and {\tt example.l}.

Then the theory of LISP program-size complexity is developed a little.
The LISP program-size complexity $H_L(x)$ of an S-expression $x$ is
defined to be the size in characters of the smallest self-contained
LISP M-expression whose value is $x$.  ``Self-contained'' means that
we assume the initial empty environment with no function definitions,
and that there is no binary data available.  The LISP program-size
complexity of a formal axiomatic system is the size in characters of
the smallest self-contained LISP M-expression that displays the
theorems of the formal axiomatic system (in any order, and perhaps
with duplications) using the DISPLAY primitive function ``{\tt ,}''.

LISP program-size complexity is extremely simple and concrete.

First of all, LISP program-size complexity is subadditive, because
expressions are self-delimiting and can be concatenated, and also
because we are dealing with pure LISP and no side-effects get in the
way.  More precisely, consider the LISP M-expression {\tt *p*q()},
where $p$ is a minimum-size LISP M-expression for $x$, and $q$ is a
minimum-size LISP M-expression for $y$.  The value of this M-expression
is the pair $(xy)$ consisting of $x$ and $y$.  This shows that
\[
 H_L ((xy)) \le H_L(x) + H_L(y) + 4 .
\]
And the probability $\Omega_L$ that a LISP M-expression ``halts'' or has
a value is well-defined, also because programs are self-delimiting.

Secondly, LISP complexity easily yields elegant incompleteness
theorems.  It is easy to show that it is impossible to prove that a
self-contained LISP M-expression is elegant, i.e., that no smaller
M-expression has the same output.  More precisely, a formal axiomatic
system of LISP complexity $N$ cannot enable us to prove that a LISP
M-expression that has more than $N+163$ characters is elegant.  See
{\tt lgodel.l} and {\tt lgodel2.l}.  And it is easy to show that
it is impossible to establish lower bounds on LISP complexity.  More
precisely, a formal axiomatic system of LISP complexity $N$ cannot
enable us to exhibit an S-expression with LISP complexity greater than
$N+157$.  See {\tt lgodel3.l} and {\tt lgodel4.l}.

But LISP programs have severe information-theoretic limitations
because they do not encode information very efficiently in 7-bit ASCII
characters subject to LISP syntax constraints.  So next we give binary
data to LISP programs and measure the size of these programs as (7
times the number of characters in the LISP program) plus (the number
of bits in the binary data).  More precisely, we define a new
complexity $H \equiv H_U$ measured in bits via minimum-size binary
programs for a self-delimiting universal Turing machine $U$.  I.e.,
$H(\cdots)$ denotes the size in bits of the smallest program that
makes our standard universal Turing machine $U$ compute $\cdots$.
Binary programs pack information more densely than LISP M-expressions,
but they must be kept self-delimiting.  We define our standard
self-delimiting universal Turing machine $U(p)$ with binary program
(``tape'') $p$ as
\begin{verbatim}
                         ?0'!%p
\end{verbatim}
as was explained in the previous chapter.

Next we show that
\[
   H(x,y) \le H(x) + H(y) + 56.
\]
This inequality states that the information needed to compute the pair
$(x,y)$ is bounded by the constant 56 plus the sum of the information
needed to compute $x$ and the information needed to compute $y$.
Consider
\begin{verbatim}
                        *!%*!%()
\end{verbatim}
This is an M-expression with parentheses omitted from primitive
functions.
The constant 56 is just the size of this LISP M-expression, which is
exactly 8 characters = 56 bits.  See {\tt univ.l}.  Note that in
standard LISP this would be something like
\begin{verbatim}
              (CONS (EVAL (READ-EXPRESSION))
              (CONS (EVAL (READ-EXPRESSION))
                    NIL))
\end{verbatim}
which is much more than 8 characters long.

Next let's look at a binary string $x$ of $|x|$ bits.  We show that
\[
   H(x) \le 2|x| + 140
\]
and
\[
   H(x) \le |x| + H(|x|) + 441.
\]
As before, the programs for doing this are exhibited and run.
See {\tt univ.l}.

Next we show how to calculate the halting probability $\Omega$ of our
standard self-delimiting universal Turing machine in the limit from
below.  The LISP program for doing this, {\tt omega.l}, is now
remarkably clear and fast, and much better than the one given in
my Cambridge University Press monograph.  The $N$th lower bound on
$\Omega$ is (the number of $N$-bit programs that halt on $U$ within time
$N$) divided by $2^N$.  Using a big (512 megabyte) version of our LISP
interpreter, we calculate these lower bounds for $N = 0$ to $N =22$.
See {\tt omega.r}.  Using this algorithm for computing $\Omega$ in
the limit from below, we show that if $\Omega_N$ denotes the first $N$
bits of the fractional part of the base-two real number $\Omega$, then
\[
   H(\Omega_N) > N - 1883.
\]
Again this is done with a program that can actually be run and whose
size in bits is the constant 1883.  See {\tt omega2.l}.

Next we turn to the self-delimiting program-size complexity $H(X)$ for
infinite r.e.\ sets $X$, which is not considered at all in my
Cambridge University Press monograph.  This is now defined to be the
size in bits of the smallest LISP M-expression $\xi$ that displays the
members of the r.e.\ set $X$ using the LISP primitive ``{\tt ,}''.  ``{\tt
,}'' is an identity function with the side-effect of displaying the
value of its argument.  Note that this LISP M-expression $\xi$ is
allowed to read additional bits or M-expressions from the UTM $U$'s tape
using the primitive functions {\tt @} and {\tt \%} if $\xi$ so
desires.  But of course $\xi$ is charged for this; this adds to
$\xi$'s program size.  In this context it doesn't matter whether $\xi$
halts or not or what its value, if any, is.

It was in order to deal with such unending expressions $\xi$ that the
LISP primitive function TRY for time-limited evaluation {\tt
?}$\alpha\beta\gamma$ now captures all output from ``{\tt ,}'' within its
second argument $\beta$.

To illustrate these new concepts, we show that
\[
   H(X \cap Y) \le H(X) + H(Y) + 1799
\]
and
\[
   H(X \cup Y) \le H(X) + H(Y) + 1799.
\]
for infinite r.e.\ sets $X$ and $Y$.  As before, we run examples.  See
{\tt sets0.l} through {\tt sets4.l}.

Now consider a formal axiomatic system $A$ of complexity $N$, i.e.,
with a set of theorems $T_A$ that considered as an r.e.\ set as above
has self-delimiting program-size complexity $H(T_A) = N$.  We show that
$A$ has the following limitations.  First of all, we show directly
that $A$ cannot enable us to exhibit a specific S-expression $s$ with
self-delimiting complexity $H(s)$ greater than $N+994$.  See
{\tt godel.l} and {\tt godel2.l}.

Secondly, using the $H(\Omega_N) > N-1883$ inequality established in {\tt
omega2.l}, we show that $A$ cannot enable us to determine more than
$N+3192$ bits of the binary representation of the halting
probability $\Omega$, even if these bits are scattered and we leave
gaps.  (See {\tt godel3.l} through {\tt godel5.l}.)  In my
Cambridge University Press monograph, this took a hundred pages to
show, and involved the systematic development of a general theory
using measure-theoretic arguments.  Following ``Information-theoretic
incompleteness,'' {\it Applied Mathematics and Computation\/} {\bf 52}
(1992), pp.\ 83--101, the proof is now a straight-forward
Berry-paradox-like program-size argument.  Moreover we are using a
deeper definition of $H(A) \equiv H(T_A)$ via infinite self-delimiting
computations.

And last but not least, the philosophical implications of all this
should be discussed, especially the extent to which it tends to
justify experimental mathematics.  This would be along the lines of
the discussion in my talk transcript ``Randomness in arithmetic and
the decline and fall of reductionism in pure mathematics,'' {\it
Bulletin of the European Association for Theoretical Computer
Science,} No.\ 50 (June 1993), pp.\ 314--328, later published as
``Randomness and complexity in pure mathematics,'' {\it International
Journal of Bifurcation and Chaos\/} {\bf 4} (1994), pp.\ 3--15.

This concludes our ``hand-on'' course on the information-theoretic
limits of mathematics.

\chap{Software User Guide}

All the software for this course is written in a toy version of {\sl
LISP}.  {\tt *.l} files are toy {\sl LISP} code, and {\tt *.r}
files are interpreter output.  Three {\sl C} programs, {\tt lisp.c,}
{\tt show.c,} and {\tt big.c} are provided; they are slightly
different versions of the {\sl LISP} interpreter.  In these
instructions we assume that this software is being run in the {\sl AIX}
environment.

To run the standard version {\tt lisp} of the interpreter, first
compile {\tt lisp.c}.  This is done using the command {\tt cc -O
-olisp lisp.c}.  The resulting interpreter is about 128 megabytes big.
If this is too large, reduce the \verb|#define SIZE| before compiling
it.

There are three different ways to use this interpreter: To use the
interpreter {\tt lisp }interactively, that is, with input and output on the
screen, enter
\[
\mbox{\tt lisp}
\]
To run a {\sl LISP} program {\tt X.l} with output on the screen, enter
\[
\mbox{\tt lisp <X.l}
\]
To run a {\sl LISP} program {\tt X.l} with output in a file
rather than on the screen, enter
\[
\mbox{\tt lisp <X.l >X.r}
\]

Similarly, the {\tt show} and {\tt big} (512 megabyte) versions of the
interpreter are obtained by compiling {\tt show.c} and {\tt big.c,}
respectively, and they are used in the same way as the standard
version {\tt lisp} is used.  The {\tt show} version of the interpreter
is the only one that shows and sizes the arguments of
SHOW \verb|~|.

\chap{test.l}{\Size\begin{verbatim}
[ LISP test run ]
'(abc)
+'(abc)
-'(abc)
*'(ab)'(cd)
.'a
.'(abc)
='(ab)'(ab)
='(ab)'(ac)
-,-,-,-,-,-,'(abcdef)
/0'x'y
/1'x'y
!,'/1'x'y
(*"&*()*,'/1'x'y())
('&(xy)y 'a 'b)
: x 'a : y 'b *x*y()
[ first atom ]
: (Fx)/.,xx F+x F'((((a)b)c)d)
[ concatenation ]
:(Cxy) /.,xy *+xC-xy C'(ab)'(cd)
?'()'
:(Cxy) /.,xy *+xC-xy C'(ab)'(cd)
'()
?'(1)'
:(Cxy) /.,xy *+xC-xy C'(ab)'(cd)
'()
?'(11)'
:(Cxy) /.,xy *+xC-xy C'(ab)'(cd)
'()
?'(111)'
:(Cxy) /.,xy *+xC-xy C'(ab)'(cd)
'()
?'(1111)'
:(Cxy) /.,xy *+xC-xy C'(ab)'(cd)
'()

[ d: x goes to (xx) ]
& (dx) *,x*x()
[ e really doubles length of string each time ]
& (ex) *,xx
dddddddd()
eeeeeeee()
\end{verbatim}
}\chap{test.r}{\Size\begin{verbatim}
lisp.c

LISP Interpreter Run

[ LISP test run ]
'(abc)

expression  ('(abc))
value       (abc)

+'(abc)

expression  (+('(abc)))
value       a

-'(abc)

expression  (-('(abc)))
value       (bc)

*'(ab)'(cd)

expression  (*('(ab))('(cd)))
value       ((ab)cd)

.'a

expression  (.('a))
value       1

.'(abc)

expression  (.('(abc)))
value       0

='(ab)'(ab)

expression  (=('(ab))('(ab)))
value       1

='(ab)'(ac)

expression  (=('(ab))('(ac)))
value       0

-,-,-,-,-,-,'(abcdef)

expression  (-(,(-(,(-(,(-(,(-(,(-(,('(abcdef))))))))))))))
display     (abcdef)
display     (bcdef)
display     (cdef)
display     (def)
display     (ef)
display     (f)
value       ()

/0'x'y

expression  (/0('x)('y))
value       y

/1'x'y

expression  (/1('x)('y))
value       x

!,'/1'x'y

expression  (!(,('(/1('x)('y)))))
display     (/1('x)('y))
value       x

(*"&*()*,'/1'x'y())

expression  ((*&(*()(*(,('(/1('x)('y))))()))))
display     (/1('x)('y))
value       x

('&(xy)y 'a 'b)

expression  (('(&(xy)y))('a)('b))
value       b

: x 'a : y 'b *x*y()

expression  (('(&(x)(('(&(y)(*x(*y()))))('b))))('a))
value       (ab)

[ first atom ]
: (Fx)/.,xx F+x F'((((a)b)c)d)

expression  (('(&(F)(F('((((a)b)c)d)))))('(&(x)(/(.(,x))x(F(+x
            ))))))
display     ((((a)b)c)d)
display     (((a)b)c)
display     ((a)b)
display     (a)
display     a
value       a

[ concatenation ]
:(Cxy) /.,xy *+xC-xy C'(ab)'(cd)

expression  (('(&(C)(C('(ab))('(cd)))))('(&(xy)(/(.(,x))y(*(+x
            )(C(-x)y))))))
display     (ab)
display     (b)
display     ()
value       (abcd)

?'()'
:(Cxy) /.,xy *+xC-xy C'(ab)'(cd)
'()

expression  (?('())('(('(&(C)(C('(ab))('(cd)))))('(&(xy)(/(.(,
            x))y(*(+x)(C(-x)y)))))))('()))
value       (?)

?'(1)'
:(Cxy) /.,xy *+xC-xy C'(ab)'(cd)
'()

expression  (?('(1))('(('(&(C)(C('(ab))('(cd)))))('(&(xy)(/(.(
            ,x))y(*(+x)(C(-x)y)))))))('()))
value       (?)

?'(11)'
:(Cxy) /.,xy *+xC-xy C'(ab)'(cd)
'()

expression  (?('(11))('(('(&(C)(C('(ab))('(cd)))))('(&(xy)(/(.
            (,x))y(*(+x)(C(-x)y)))))))('()))
value       (?(ab))

?'(111)'
:(Cxy) /.,xy *+xC-xy C'(ab)'(cd)
'()

expression  (?('(111))('(('(&(C)(C('(ab))('(cd)))))('(&(xy)(/(
            .(,x))y(*(+x)(C(-x)y)))))))('()))
value       (?(b)(ab))

?'(1111)'
:(Cxy) /.,xy *+xC-xy C'(ab)'(cd)
'()

expression  (?('(1111))('(('(&(C)(C('(ab))('(cd)))))('(&(xy)(/
            (.(,x))y(*(+x)(C(-x)y)))))))('()))
value       (((abcd))()(b)(ab))


[ d: x goes to (xx) ]
& (dx) *,x*x()

d:          (&(x)(*(,x)(*x())))

[ e really doubles length of string each time ]
& (ex) *,xx

e:          (&(x)(*(,x)x))

dddddddd()

expression  (d(d(d(d(d(d(d(d()))))))))
display     ()
display     (()())
display     ((()())(()()))
display     (((()())(()()))((()())(()())))
display     ((((()())(()()))((()())(()())))(((()())(()()))((()
            ())(()()))))
display     (((((()())(()()))((()())(()())))(((()())(()()))(((
            )())(()()))))((((()())(()()))((()())(()())))(((()(
            ))(()()))((()())(()())))))
display     ((((((()())(()()))((()())(()())))(((()())(()()))((
            ()())(()()))))((((()())(()()))((()())(()())))(((()
            ())(()()))((()())(()())))))(((((()())(()()))((()()
            )(()())))(((()())(()()))((()())(()()))))((((()())(
            ()()))((()())(()())))(((()())(()()))((()())(()()))
            ))))
display     (((((((()())(()()))((()())(()())))(((()())(()()))(
            (()())(()()))))((((()())(()()))((()())(()())))((((
            )())(()()))((()())(()())))))(((((()())(()()))((()(
            ))(()())))(((()())(()()))((()())(()()))))((((()())
            (()()))((()())(()())))(((()())(()()))((()())(()())
            )))))((((((()())(()()))((()())(()())))(((()())(()(
            )))((()())(()()))))((((()())(()()))((()())(()())))
            (((()())(()()))((()())(()())))))(((((()())(()()))(
            (()())(()())))(((()())(()()))((()())(()()))))(((((
            )())(()()))((()())(()())))(((()())(()()))((()())((
            )())))))))
value       ((((((((()())(()()))((()())(()())))(((()())(()()))
            ((()())(()()))))((((()())(()()))((()())(()())))(((
            ()())(()()))((()())(()())))))(((((()())(()()))((()
            ())(()())))(((()())(()()))((()())(()()))))((((()()
            )(()()))((()())(()())))(((()())(()()))((()())(()()
            ))))))((((((()())(()()))((()())(()())))(((()())(()
            ()))((()())(()()))))((((()())(()()))((()())(()()))
            )(((()())(()()))((()())(()())))))(((((()())(()()))
            ((()())(()())))(((()())(()()))((()())(()()))))((((
            ()())(()()))((()())(()())))(((()())(()()))((()())(
            ()())))))))(((((((()())(()()))((()())(()())))(((()
            ())(()()))((()())(()()))))((((()())(()()))((()())(
            ()())))(((()())(()()))((()())(()())))))(((((()())(
            ()()))((()())(()())))(((()())(()()))((()())(()()))
            ))((((()())(()()))((()())(()())))(((()())(()()))((
            ()())(()()))))))((((((()())(()()))((()())(()())))(
            ((()())(()()))((()())(()()))))((((()())(()()))((()
            ())(()())))(((()())(()()))((()())(()())))))(((((()
            ())(()()))((()())(()())))(((()())(()()))((()())(()
            ()))))((((()())(()()))((()())(()())))(((()())(()()
            ))((()())(()()))))))))

eeeeeeee()

expression  (e(e(e(e(e(e(e(e()))))))))
display     ()
display     (())
display     ((())())
display     (((())())(())())
display     ((((())())(())())((())())(())())
display     (((((())())(())())((())())(())())(((())())(())())(
            (())())(())())
display     ((((((())())(())())((())())(())())(((())())(())())
            ((())())(())())((((())())(())())((())())(())())(((
            ())())(())())((())())(())())
display     (((((((())())(())())((())())(())())(((())())(())()
            )((())())(())())((((())())(())())((())())(())())((
            (())())(())())((())())(())())(((((())())(())())(((
            ))())(())())(((())())(())())((())())(())())((((())
            ())(())())((())())(())())(((())())(())())((())())(
            ())())
value       ((((((((())())(())())((())())(())())(((())())(())(
            ))((())())(())())((((())())(())())((())())(())())(
            ((())())(())())((())())(())())(((((())())(())())((
            ())())(())())(((())())(())())((())())(())())((((()
            )())(())())((())())(())())(((())())(())())((())())
            (())())((((((())())(())())((())())(())())(((())())
            (())())((())())(())())((((())())(())())((())())(()
            )())(((())())(())())((())())(())())(((((())())(())
            ())((())())(())())(((())())(())())((())())(())())(
            (((())())(())())((())())(())())(((())())(())())(((
            ))())(())())

End of LISP Run

Elapsed time is 0 seconds.
\end{verbatim}
}\chap{example.l}{\Size\begin{verbatim}
[[[(Fx) = flatten x by removing all interior parentheses]]]
[Define F of x as follows: if x is empty then return empty, if
 x is an atom then join x to the empty list, otherwise split
 x into its head and tail, flatten each, and append the results.]
& (Fx) /=x()() /.x*x() ^F+xF-x
F,"F [use F to flatten itself]
[[[(Gx) = size of x in unary]]]
[Let G of x be [if x is empty, then unary two, if x is an atom,
 then unary one, otherwise split x into its head and tail,
 size each, and add the results] in ...]
: (Gx) /=x()'{2} /.x'{1} ^G+xG-x
[Let G of x be [...] in:]
G,"G [apply G to itself]
\end{verbatim}
}\chap{example.r}{\Size\begin{verbatim}
lisp.c

LISP Interpreter Run

[[[(Fx) = flatten x by removing all interior parentheses]]]
[Define F of x as follows: if x is empty then return empty, if
 x is an atom then join x to the empty list, otherwise split
 x into its head and tail, flatten each, and append the results.]
& (Fx) /=x()() /.x*x() ^F+xF-x

F:          (&(x)(/(=x())()(/(.x)(*x())(^(F(+x))(F(-x))))))

F,"F [use F to flatten itself]

expression  (F(,F))
display     (&(x)(/(=x())()(/(.x)(*x())(^(F(+x))(F(-x))))))
value       (&x/=x/.x*x^F+xF-x)

[[[(Gx) = size of x in unary]]]
[Let G of x be [if x is empty, then unary two, if x is an atom,
 then unary one, otherwise split x into its head and tail,
 size each, and add the results] in ...]
: (Gx) /=x()'{2} /.x'{1} ^G+xG-x
[Let G of x be [...] in:]
G,"G [apply G to itself]

expression  (('(&(G)(G(,G))))('(&(x)(/(=x())('(11))(/(.x)('(1)
            )(^(G(+x))(G(-x))))))))
display     (&(x)(/(=x())('(11))(/(.x)('(1))(^(G(+x))(G(-x))))
            ))
value       (1111111111111111111111111111111111111111111111111
            111)

End of LISP Run

Elapsed time is 0 seconds.
\end{verbatim}
}\chap{lgodel.l}{\Size\begin{verbatim}
[[[ Show that a formal system of lisp complexity H_L (FAS) = N
    cannot enable us to exhibit an elegant M-expression of
    size greater than N + 163.
    An elegant lisp M-expression is one with the property that no
    smaller M-expression has the same output.  One may consider
    the output of an M-expression to be either its value or what
    it displays.  The proof below works in either case.
    Setting: formal axiomatic system is never-ending lisp M-expression
    that displays elegant M-expressions.
]]]

[ Idea is to have a program P search for something X that can be proved
  to be more complex than P is, and therefore P can never find X.
  I.e., idea is to show that if this program halts we get a contradiction,
  and therefore the program doesn't halt. ]

[m-expression:]
'"(*a*b*c())
[convert m-exp to s-exp]
++?0'%#'"(*a*b*c())
[convert m-exp to s-exp and run it]
!,++?0'%#'"(*a*b*c())
[display m-exp, convert to s-exp and run it]
!,++?0'%#,'"(*a*b*c())

!++?0'%#~'"( [begin literally]

[ | = |x| = size of s-expression x in characters ]
:(|x) /=x()'{2} /.x'{1} ^|+x|-x

[ < = unary number x is less than unary number y ]
[ x and y are lists of 1's ]
:(<xy) /.y0 /.x1 <-x-y

[ E = examine list x for element that is more than n characters in size. ]
[ If not found returns false/0. ]
:(Exn) /.x0 /<n|+x +x E-xn

[ (2n) = convert reversed binary to unary ]
:(2n) /.n() :k2-n /+n *1^kk ^kk

[ Here we are given the formal axiomatic system FAS. ]
:f '"("?)  [insert FAS m-expression here inside ()]

[ n = the number of characters in program including the FAS. ]
:n ~^|f 2'(110 001 01) [show that this m-exp knows its own size]
[ n = 163 base 10 + |FAS| = 243 base 8 + |FAS| ]

[ L = loop running the formal axiomatic system ]
:(Lt)
  :v ?t'!%#f    [Run the formal system for t time steps.]
  :s E-vn       [Did it output an elegant m-exp larger than this program?]
  /s !++?0'%#s  [If found elegant m-exp bigger than this program,
                 run it so that its output is our output (contradiction!)]
  /.+v L*1t     [If not, keep looping]
  "?            [or halt if formal system halted.]
L()             [Start loop running with t = 0.]

)               [end literally]
\end{verbatim}
}\chap{lgodel.r}{\Size\begin{verbatim}
show.c

LISP Interpreter Run

[[[ Show that a formal system of lisp complexity H_L (FAS) = N
    cannot enable us to exhibit an elegant M-expression of
    size greater than N + 163.
    An elegant lisp M-expression is one with the property that no
    smaller M-expression has the same output.  One may consider
    the output of an M-expression to be either its value or what
    it displays.  The proof below works in either case.
    Setting: formal axiomatic system is never-ending lisp M-expression
    that displays elegant M-expressions.
]]]

[ Idea is to have a program P search for something X that can be proved
  to be more complex than P is, and therefore P can never find X.
  I.e., idea is to show that if this program halts we get a contradiction,
  and therefore the program doesn't halt. ]

[m-expression:]
'"(*a*b*c())

expression  ('(*a*b*c()))
value       (*a*b*c())

[convert m-exp to s-exp]
++?0'%#'"(*a*b*c())

expression  (+(+(?0('(%))(#('(*a*b*c()))))))
value       (*a(*b(*c())))

[convert m-exp to s-exp and run it]
!,++?0'%#'"(*a*b*c())

expression  (!(,(+(+(?0('(%))(#('(*a*b*c()))))))))
display     (*a(*b(*c())))
value       (abc)

[display m-exp, convert to s-exp and run it]
!,++?0'%#,'"(*a*b*c())

expression  (!(,(+(+(?0('(%))(#(,('(*a*b*c())))))))))
display     (*a*b*c())
display     (*a(*b(*c())))
value       (abc)


!++?0'%#~'"( [begin literally]

[ | = |x| = size of s-expression x in characters ]
:(|x) /=x()'{2} /.x'{1} ^|+x|-x

[ < = unary number x is less than unary number y ]
[ x and y are lists of 1's ]
:(<xy) /.y0 /.x1 <-x-y

[ E = examine list x for element that is more than n characters in size. ]
[ If not found returns false/0. ]
:(Exn) /.x0 /<n|+x +x E-xn

[ (2n) = convert reversed binary to unary ]
:(2n) /.n() :k2-n /+n *1^kk ^kk

[ Here we are given the formal axiomatic system FAS. ]
:f '"("?)  [insert FAS m-expression here inside ()]

[ n = the number of characters in program including the FAS. ]
:n ~^|f 2'(110 001 01) [show that this m-exp knows its own size]
[ n = 163 base 10 + |FAS| = 243 base 8 + |FAS| ]

[ L = loop running the formal axiomatic system ]
:(Lt)
  :v ?t'!%#f    [Run the formal system for t time steps.]
  :s E-vn       [Did it output an elegant m-exp larger than this program?]
  /s !++?0'%#s  [If found elegant m-exp bigger than this program,
                 run it so that its output is our output (contradiction!)]
  /.+v L*1t     [If not, keep looping]
  "?            [or halt if formal system halted.]
L()             [Start loop running with t = 0.]

)               [end literally]

expression  (!(+(+(?0('(%))(#(~('(:(|x)/=x()'{2}/.x'{1}^|+x|-x
            :(<xy)/.y0/.x1<-x-y:(Exn)/.x0/<n|+x+xE-xn:(2n)/.n(
            ):k2-n/+n*1^kk^kk:f'"("?):n~^|f2'(11000101):(Lt):v
            ?t'!%#f:sE-vn/s!++?0'%#s/.+vL*1t"?L()))))))))
show        (:(|x)/=x()'{2}/.x'{1}^|+x|-x:(<xy)/.y0/.x1<-x-y:(
            Exn)/.x0/<n|+x+xE-xn:(2n)/.n():k2-n/+n*1^kk^kk:f'"
            ("?):n~^|f2'(11000101):(Lt):v?t'!%#f:sE-vn/s!++?0'
            %#s/.+vL*1t"?L())
size        165(245)/1155(2203)
show        (1111111111111111111111111111111111111111111111111
            11111111111111111111111111111111111111111111111111
            11111111111111111111111111111111111111111111111111
            111111111111111111)
size        167(247)/1169(2221)
value       ?

End of LISP Run

Elapsed time is 0 seconds.
\end{verbatim}
}\chap{lgodel2.l}{\Size\begin{verbatim}
[[[ Show that a formal system of lisp complexity H_L (FAS) = N
    cannot enable us to exhibit an elegant M-expression of
    size greater than N + 163.
    An elegant lisp M-expression is one with the property that no
    smaller M-expression has the same output.  One may consider
    the output of an M-expression to be either its value or what
    it displays.  The proof below works in either case.
    Setting: formal axiomatic system is never-ending lisp M-expression
    that displays elegant M-expressions.
]]]

[ Idea is to have a program P search for something X that can be proved
  to be more complex than P is, and therefore P can never find X.
  I.e., idea is to show that if this program halts we get a contradiction,
  and therefore the program doesn't halt. ]

[m-expression:]
'"(*a*b*c())
[convert m-exp to s-exp]
++?0'%#'"(*a*b*c())
[convert m-exp to s-exp and run it]
!,++?0'%#'"(*a*b*c())
[display m-exp, convert to s-exp and run it]
!,++?0'%#,'"(*a*b*c())

!++?0'%#~'"( [begin literally]

[ | = |x| = size of s-expression x in characters ]
:(|x) /=x()'{2} /.x'{1} ^|+x|-x

[ < = unary number x is less than unary number y ]
[ x and y are lists of 1's ]
:(<xy) /.y1 /.x1 <-x-y            [[ MAKE ALWAYS "TRUE" FOR TEST ]]

[ E = examine list x for element that is more than n characters in size. ]
[ If not found returns false/0. ]
:(Exn) /.x0 /<n|+x +x E-xn

[ (2n) = convert reversed binary to unary ]
:(2n) /.n() :k2-n /+n *1^kk ^kk

[ Here we are given the formal axiomatic system FAS. ]
:f '"(,'"(*a*b*c()))    [[ FAS = {The m-exp *a*b*c() is elegant} ]]

[ n = the number of characters in program including the FAS. ]
:n ~^|f 2'(110 001 01) [show that this m-exp knows its own size]
[ n = 163 base 10 + |FAS| = 243 base 8 + |FAS| ]

[ L = loop running the formal axiomatic system ]
:(Lt)
  :v ?t'!%#f    [Run the formal system for t time steps.]
  :s E-vn       [Did it output an elegant m-exp larger than this program?]
  /s !++?0'%#s  [If found elegant m-exp bigger than this program,
                 run it so that its output is our output (contradiction!)]
  /.+v L*1t     [If not, keep looping]
  "?            [or halt if formal system halted.]
L()             [Start loop running with t = 0.]

)               [end literally]
\end{verbatim}
}\chap{lgodel2.r}{\Size\begin{verbatim}
show.c

LISP Interpreter Run

[[[ Show that a formal system of lisp complexity H_L (FAS) = N
    cannot enable us to exhibit an elegant M-expression of
    size greater than N + 163.
    An elegant lisp M-expression is one with the property that no
    smaller M-expression has the same output.  One may consider
    the output of an M-expression to be either its value or what
    it displays.  The proof below works in either case.
    Setting: formal axiomatic system is never-ending lisp M-expression
    that displays elegant M-expressions.
]]]

[ Idea is to have a program P search for something X that can be proved
  to be more complex than P is, and therefore P can never find X.
  I.e., idea is to show that if this program halts we get a contradiction,
  and therefore the program doesn't halt. ]

[m-expression:]
'"(*a*b*c())

expression  ('(*a*b*c()))
value       (*a*b*c())

[convert m-exp to s-exp]
++?0'%#'"(*a*b*c())

expression  (+(+(?0('(%))(#('(*a*b*c()))))))
value       (*a(*b(*c())))

[convert m-exp to s-exp and run it]
!,++?0'%#'"(*a*b*c())

expression  (!(,(+(+(?0('(%))(#('(*a*b*c()))))))))
display     (*a(*b(*c())))
value       (abc)

[display m-exp, convert to s-exp and run it]
!,++?0'%#,'"(*a*b*c())

expression  (!(,(+(+(?0('(%))(#(,('(*a*b*c())))))))))
display     (*a*b*c())
display     (*a(*b(*c())))
value       (abc)


!++?0'%#~'"( [begin literally]

[ | = |x| = size of s-expression x in characters ]
:(|x) /=x()'{2} /.x'{1} ^|+x|-x

[ < = unary number x is less than unary number y ]
[ x and y are lists of 1's ]
:(<xy) /.y1 /.x1 <-x-y            [[ MAKE ALWAYS "TRUE" FOR TEST ]]

[ E = examine list x for element that is more than n characters in size. ]
[ If not found returns false/0. ]
:(Exn) /.x0 /<n|+x +x E-xn

[ (2n) = convert reversed binary to unary ]
:(2n) /.n() :k2-n /+n *1^kk ^kk

[ Here we are given the formal axiomatic system FAS. ]
:f '"(,'"(*a*b*c()))    [[ FAS = {The m-exp *a*b*c() is elegant} ]]

[ n = the number of characters in program including the FAS. ]
:n ~^|f 2'(110 001 01) [show that this m-exp knows its own size]
[ n = 163 base 10 + |FAS| = 243 base 8 + |FAS| ]

[ L = loop running the formal axiomatic system ]
:(Lt)
  :v ?t'!%#f    [Run the formal system for t time steps.]
  :s E-vn       [Did it output an elegant m-exp larger than this program?]
  /s !++?0'%#s  [If found elegant m-exp bigger than this program,
                 run it so that its output is our output (contradiction!)]
  /.+v L*1t     [If not, keep looping]
  "?            [or halt if formal system halted.]
L()             [Start loop running with t = 0.]

)               [end literally]

expression  (!(+(+(?0('(%))(#(~('(:(|x)/=x()'{2}/.x'{1}^|+x|-x
            :(<xy)/.y1/.x1<-x-y:(Exn)/.x0/<n|+x+xE-xn:(2n)/.n(
            ):k2-n/+n*1^kk^kk:f'"(,'"(*a*b*c())):n~^|f2'(11000
            101):(Lt):v?t'!%#f:sE-vn/s!++?0'%#s/.+vL*1t"?L()))
            ))))))
show        (:(|x)/=x()'{2}/.x'{1}^|+x|-x:(<xy)/.y1/.x1<-x-y:(
            Exn)/.x0/<n|+x+xE-xn:(2n)/.n():k2-n/+n*1^kk^kk:f'"
            (,'"(*a*b*c())):n~^|f2'(11000101):(Lt):v?t'!%#f:sE
            -vn/s!++?0'%#s/.+vL*1t"?L())
size        176(260)/1232(2320)
show        (1111111111111111111111111111111111111111111111111
            11111111111111111111111111111111111111111111111111
            11111111111111111111111111111111111111111111111111
            11111111111111111111111111111)
size        178(262)/1246(2336)
value       (abc)

End of LISP Run

Elapsed time is 0 seconds.
\end{verbatim}
}\chap{lgodel3.l}{\Size\begin{verbatim}
[[[ Show that a formal system of lisp complexity H_L (FAS) = N
    cannot enable us to exhibit an S-expression with lisp complexity
    greater than N + 157.
    Setting: formal axiomatic system is never-ending lisp expression
    that displays pairs (s-expression x, unary number n)
    with H_L (x) >= n.
]]]

[ Idea is to have a program P search for something X that can be proved
  to be more complex than P is, and therefore P can never find X.
  I.e., idea is to show that if this program halts we get a contradiction,
  and therefore the program doesn't halt. ]

!++?0'%#~'"( [begin literally]

[ | = |x| = size in characters of s-expression x. ]
:(|x) /=x()'{2} /.x'{1} ^|+x|-x

[ < = unary number x is less than unary number y ]
[ x and y are lists of 1's ]
:(<xy) /.y0 /.x1 <-x-y

[ E = examine list x for first s in pair (s,m) with H_L (s) >= m > n. ]
[ If not found returns false/0. ]
:(Exn) /.x0 /<n+-+x ++x E-xn

[ (2n) = convert reversed binary to unary ]
:(2n) /.n() :k2-n /+n *1^kk ^kk

[ Here we are given the formal axiomatic system FAS. ]
:f '"("?)  [Replace "? here by m-exp for FAS]

[ n = number of characters in program including the FAS. ]
:n ~^|f 2'(110 110 01)
[ n = 155 base 10 + (|FAS| + 2) = 233 base 8 + (|FAS| + 2) ]

[ L = loop running the formal axiomatic system ]
:(Lt)
  :v ?t'!%#f  [Run the formal system for t time steps.]
  :s E-vn     [Have we found an s-exp more complex than this program?]
  /s s        [If found s-exp more complex than this program,
               return it as our value (contradiction!)]
         [This m-exp's value is an s-exp of complexity > 157 + |FAS|.
          But this m-exp's size is 157 + |FAS|!  Impossible!]
  /.+v L*1t   [If not, keep looping]
  "?          [or halt if formal system halted.]
L()           [Start loop running with t = 0.]

) [end literally]
\end{verbatim}
}\chap{lgodel3.r}{\Size\begin{verbatim}
show.c

LISP Interpreter Run

[[[ Show that a formal system of lisp complexity H_L (FAS) = N
    cannot enable us to exhibit an S-expression with lisp complexity
    greater than N + 157.
    Setting: formal axiomatic system is never-ending lisp expression
    that displays pairs (s-expression x, unary number n)
    with H_L (x) >= n.
]]]

[ Idea is to have a program P search for something X that can be proved
  to be more complex than P is, and therefore P can never find X.
  I.e., idea is to show that if this program halts we get a contradiction,
  and therefore the program doesn't halt. ]

!++?0'%#~'"( [begin literally]

[ | = |x| = size in characters of s-expression x. ]
:(|x) /=x()'{2} /.x'{1} ^|+x|-x

[ < = unary number x is less than unary number y ]
[ x and y are lists of 1's ]
:(<xy) /.y0 /.x1 <-x-y

[ E = examine list x for first s in pair (s,m) with H_L (s) >= m > n. ]
[ If not found returns false/0. ]
:(Exn) /.x0 /<n+-+x ++x E-xn

[ (2n) = convert reversed binary to unary ]
:(2n) /.n() :k2-n /+n *1^kk ^kk

[ Here we are given the formal axiomatic system FAS. ]
:f '"("?)  [Replace "? here by m-exp for FAS]

[ n = number of characters in program including the FAS. ]
:n ~^|f 2'(110 110 01)
[ n = 155 base 10 + (|FAS| + 2) = 233 base 8 + (|FAS| + 2) ]

[ L = loop running the formal axiomatic system ]
:(Lt)
  :v ?t'!%#f  [Run the formal system for t time steps.]
  :s E-vn     [Have we found an s-exp more complex than this program?]
  /s s        [If found s-exp more complex than this program,
               return it as our value (contradiction!)]
         [This m-exp's value is an s-exp of complexity > 157 + |FAS|.
          But this m-exp's size is 157 + |FAS|!  Impossible!]
  /.+v L*1t   [If not, keep looping]
  "?          [or halt if formal system halted.]
L()           [Start loop running with t = 0.]

) [end literally]

expression  (!(+(+(?0('(%))(#(~('(:(|x)/=x()'{2}/.x'{1}^|+x|-x
            :(<xy)/.y0/.x1<-x-y:(Exn)/.x0/<n+-+x++xE-xn:(2n)/.
            n():k2-n/+n*1^kk^kk:f'"("?):n~^|f2'(11011001):(Lt)
            :v?t'!%#f:sE-vn/ss/.+vL*1t"?L()))))))))
show        (:(|x)/=x()'{2}/.x'{1}^|+x|-x:(<xy)/.y0/.x1<-x-y:(
            Exn)/.x0/<n+-+x++xE-xn:(2n)/.n():k2-n/+n*1^kk^kk:f
            '"("?):n~^|f2'(11011001):(Lt):v?t'!%#f:sE-vn/ss/.+
            vL*1t"?L())
size        159(237)/1113(2131)
show        (1111111111111111111111111111111111111111111111111
            11111111111111111111111111111111111111111111111111
            11111111111111111111111111111111111111111111111111
            1111111111)
size        159(237)/1113(2131)
value       ?

End of LISP Run

Elapsed time is 0 seconds.
\end{verbatim}
}\chap{lgodel4.l}{\Size\begin{verbatim}
[[[ Show that a formal system of lisp complexity H_L (FAS) = N
    cannot enable us to exhibit an S-expression with lisp complexity
    greater than N + 157.
    Setting: formal axiomatic system is never-ending lisp expression
    that displays pairs (s-expression x, unary number n)
    with H_L (x) >= n.
]]]

[ Idea is to have a program P search for something X that can be proved
  to be more complex than P is, and therefore P can never find X.
  I.e., idea is to show that if this program halts we get a contradiction,
  and therefore the program doesn't halt. ]

!++?0'%#~'"( [begin literally]

[ | = |x| = size in characters of s-expression x. ]
:(|x) /=x()'{2} /.x'{1} ^|+x|-x

[ < = unary number x is less than unary number y ]
[ x and y are lists of 1's ]
:(<xy) /.y1 /.x1 <-x-y               [[FORCE VALUE "TRUE" FOR TEST]]

[ E = examine list x for first s in pair (s,m) with H_L (s) >= m > n. ]
[ If not found returns false/0. ]
:(Exn) /.x0 /<n+-+x ++x E-xn

[ (2n) = convert reversed binary to unary ]
:(2n) /.n() :k2-n /+n *1^kk ^kk

[ Here we are given the formal axiomatic system FAS. ]
:f '"( ,'((xy)(111)) )                    [[FAS = {"H((xy)) >= 3"}]]

[ n = number of characters in program including the FAS. ]
:n ~^|f 2'(110 110 01)
[ n = 155 base 10 + (|FAS| + 2) = 233 base 8 + (|FAS| + 2) ]

[ L = loop running the formal axiomatic system ]
:(Lt)
  :v ?t'!%#f  [Run the formal system for t time steps.]
  :s E-vn     [Have we found an s-exp more complex than this program?]
  /s s        [If found s-exp more complex than this program,
               return it as our value (contradiction!)]
         [This m-exp's value is an s-exp of complexity > 157 + |FAS|.
          But this m-exp's size is 157 + |FAS|!  Impossible!]
  /.+v L*1t   [If not, keep looping]
  "?          [or halt if formal system halted.]
L()           [Start loop running with t = 0.]

) [end literally]
\end{verbatim}
}\chap{lgodel4.r}{\Size\begin{verbatim}
show.c

LISP Interpreter Run

[[[ Show that a formal system of lisp complexity H_L (FAS) = N
    cannot enable us to exhibit an S-expression with lisp complexity
    greater than N + 157.
    Setting: formal axiomatic system is never-ending lisp expression
    that displays pairs (s-expression x, unary number n)
    with H_L (x) >= n.
]]]

[ Idea is to have a program P search for something X that can be proved
  to be more complex than P is, and therefore P can never find X.
  I.e., idea is to show that if this program halts we get a contradiction,
  and therefore the program doesn't halt. ]

!++?0'%#~'"( [begin literally]

[ | = |x| = size in characters of s-expression x. ]
:(|x) /=x()'{2} /.x'{1} ^|+x|-x

[ < = unary number x is less than unary number y ]
[ x and y are lists of 1's ]
:(<xy) /.y1 /.x1 <-x-y               [[FORCE VALUE "TRUE" FOR TEST]]

[ E = examine list x for first s in pair (s,m) with H_L (s) >= m > n. ]
[ If not found returns false/0. ]
:(Exn) /.x0 /<n+-+x ++x E-xn

[ (2n) = convert reversed binary to unary ]
:(2n) /.n() :k2-n /+n *1^kk ^kk

[ Here we are given the formal axiomatic system FAS. ]
:f '"( ,'((xy)(111)) )                    [[FAS = {"H((xy)) >= 3"}]]

[ n = number of characters in program including the FAS. ]
:n ~^|f 2'(110 110 01)
[ n = 155 base 10 + (|FAS| + 2) = 233 base 8 + (|FAS| + 2) ]

[ L = loop running the formal axiomatic system ]
:(Lt)
  :v ?t'!%#f  [Run the formal system for t time steps.]
  :s E-vn     [Have we found an s-exp more complex than this program?]
  /s s        [If found s-exp more complex than this program,
               return it as our value (contradiction!)]
         [This m-exp's value is an s-exp of complexity > 157 + |FAS|.
          But this m-exp's size is 157 + |FAS|!  Impossible!]
  /.+v L*1t   [If not, keep looping]
  "?          [or halt if formal system halted.]
L()           [Start loop running with t = 0.]

) [end literally]

expression  (!(+(+(?0('(%))(#(~('(:(|x)/=x()'{2}/.x'{1}^|+x|-x
            :(<xy)/.y1/.x1<-x-y:(Exn)/.x0/<n+-+x++xE-xn:(2n)/.
            n():k2-n/+n*1^kk^kk:f'"(,'((xy)(111))):n~^|f2'(110
            11001):(Lt):v?t'!%#f:sE-vn/ss/.+vL*1t"?L()))))))))
show        (:(|x)/=x()'{2}/.x'{1}^|+x|-x:(<xy)/.y1/.x1<-x-y:(
            Exn)/.x0/<n+-+x++xE-xn:(2n)/.n():k2-n/+n*1^kk^kk:f
            '"(,'((xy)(111))):n~^|f2'(11011001):(Lt):v?t'!%#f:
            sE-vn/ss/.+vL*1t"?L())
size        170(252)/1190(2246)
show        (1111111111111111111111111111111111111111111111111
            11111111111111111111111111111111111111111111111111
            11111111111111111111111111111111111111111111111111
            111111111111111111111)
size        170(252)/1190(2246)
value       (xy)

End of LISP Run

Elapsed time is 0 seconds.
\end{verbatim}
}\chap{univ.l}{\Size\begin{verbatim}
[[[
 First steps with my new construction for
 a self-delimiting universal Turing machine.
 We show that
    H((xy)) <= H(x) + H(y) + 56.
 Consider a bit string x of length |x|.
 We also show that
    H(x) <= 2|x| + 140
 and that
    H(x) <= |x| + H(|x|) + 441.
]]]

?0
':(f) :x@ :y@ /=xy *xf () f
'(0011001101)
[ This is 140 bits long: ]
 ~'"( :(f) :x@ :y@ /=xy *xf () f )
[ Here are the 140 bits: ]
#'"( :(f) :x@ :y@ /=xy *xf () f )
[ Here is the self-delimiting universal Turing machine! ]
[ (with slightly funny handling of out-of-tape condition) ]
& (Up) ++?0'!%p
[Show that H(x) <= 2|x| + 140]
U
 ^ #'"( :(f) :x@ :y@ /=xy *xf () f )
   '(0011001101)
U
 ^ #'"( :(f) :x@ :y@ /=xy *xf () f )
   '(0011001100)
U
 ^ #~'"( *!%*!%() ) [The length of this bit string is the]
                    [constant c = 56 in H(x,y) <= H(x)+H(y)+c.]
 ^ #'"( :(f) :x@ :y@ /=xy *xf () f )
 ^ '(0011001101)
 ^ #'"( :(f) :x@ :y@ /=xy *xf () f )
   '(1100110001)
& (Dk) /=0+k *1D-k /.-k () *0-k [D = decrement reverse binary]
,D,D,D,D,'(001)
U
 ^ #~'       [Show that H(x) <= |x| + H(|x|) + 441.]
"(           [begin literally]
   : (Re) /.e() ^R-e*+e()            [R = reverse  ]
   : (Dk) /=0+k *1D-k /.-k () *0-k   [D = decrement]
   : (Lk) /.k () *@LDk               [L = loop     ]
   LR!%
 )           [end literally]
 ^ #'"( '(1000) )  [ lisp expression that evaluates to binary 8, ]
   '(0000 0001)    [ so 8 bits of data follows the header ]
\end{verbatim}
}\chap{univ.r}{\Size\begin{verbatim}
show.c

LISP Interpreter Run

[[[
 First steps with my new construction for
 a self-delimiting universal Turing machine.
 We show that
    H((xy)) <= H(x) + H(y) + 56.
 Consider a bit string x of length |x|.
 We also show that
    H(x) <= 2|x| + 140
 and that
    H(x) <= |x| + H(|x|) + 441.
]]]

?0
':(f) :x@ :y@ /=xy *xf () f
'(0011001101)

expression  (?0('(('(&(f)(f)))('(&()(('(&(x)(('(&(y)(/(=xy)(*x
            (f))())))(@))))(@))))))('(0011001101)))
value       (((0101)))

[ This is 140 bits long: ]
 ~'"( :(f) :x@ :y@ /=xy *xf () f )

expression  (~('(:(f):x@:y@/=xy*xf()f)))
show        (:(f):x@:y@/=xy*xf()f)
size        20(24)/140(214)
value       (:(f):x@:y@/=xy*xf()f)

[ Here are the 140 bits: ]
#'"( :(f) :x@ :y@ /=xy *xf () f )

expression  (#('(:(f):x@:y@/=xy*xf()f)))
value       (0111010010100011001100101001011101011110001000000
            01110101111001100000001011110111101111100011110010
            10101011110001100110010100001010011100110)

[ Here is the self-delimiting universal Turing machine! ]
[ (with slightly funny handling of out-of-tape condition) ]
& (Up) ++?0'!%p

U:          (&(p)(+(+(?0('(!(%)))p))))

[Show that H(x) <= 2|x| + 140]
U
 ^ #'"( :(f) :x@ :y@ /=xy *xf () f )
   '(0011001101)

expression  (U(^(#('(:(f):x@:y@/=xy*xf()f)))('(0011001101))))
value       (0101)

U
 ^ #'"( :(f) :x@ :y@ /=xy *xf () f )
   '(0011001100)

expression  (U(^(#('(:(f):x@:y@/=xy*xf()f)))('(0011001100))))
value       !

U
 ^ #~'"( *!%*!%() ) [The length of this bit string is the]
                    [constant c = 56 in H(x,y) <= H(x)+H(y)+c.]
 ^ #'"( :(f) :x@ :y@ /=xy *xf () f )
 ^ '(0011001101)
 ^ #'"( :(f) :x@ :y@ /=xy *xf () f )
   '(1100110001)

expression  (U(^(#(~('(*!%*!%()))))(^(#('(:(f):x@:y@/=xy*xf()f
            )))(^('(0011001101))(^(#('(:(f):x@:y@/=xy*xf()f)))
            ('(1100110001)))))))
show        (*!%*!%())
size        8(10)/56(70)
value       ((0101)(1010))

& (Dk) /=0+k *1D-k /.-k () *0-k [D = decrement reverse binary]

D:          (&(k)(/(=0(+k))(*1(D(-k)))(/(.(-k))()(*0(-k)))))

,D,D,D,D,'(001)

expression  (,(D(,(D(,(D(,(D(,('(001)))))))))))
display     (001)
display     (11)
display     (01)
display     (1)
display     ()
value       ()

U
 ^ #~'       [Show that H(x) <= |x| + H(|x|) + 441.]
"(           [begin literally]
   : (Re) /.e() ^R-e*+e()            [R = reverse  ]
   : (Dk) /=0+k *1D-k /.-k () *0-k   [D = decrement]
   : (Lk) /.k () *@LDk               [L = loop     ]
   LR!%
 )           [end literally]
 ^ #'"( '(1000) )  [ lisp expression that evaluates to binary 8, ]
   '(0000 0001)    [ so 8 bits of data follows the header ]

expression  (U(^(#(~('(:(Re)/.e()^R-e*+e():(Dk)/=0+k*1D-k/.-k(
            )*0-k:(Lk)/.k()*@LDkLR!%))))(^(#('('(1000))))('(00
            000001)))))
show        (:(Re)/.e()^R-e*+e():(Dk)/=0+k*1D-k/.-k()*0-k:(Lk)
            /.k()*@LDkLR!%)
size        63(77)/441(671)
value       (00000001)

End of LISP Run

Elapsed time is 0 seconds.
\end{verbatim}
}\chap{omega.l}{\Size\begin{verbatim}
[[[[ Omega in the limit from below! ]]]]

[[[
 (Wn) = the nth lower bound on the halting probability
 Omega = (the number of n-bit programs that halt when
 run on U for time n) divided by (2 to the power n).
]]]

[S = sum of three bits]
&(Sxyz) =x=yz

[C = carry of three bits]
&(Cxyz) /x/y1z/yz0

[A = addition of reversed binary x and y]
[c = carryin]
&(Axyc) /.x /.y /c'(1)() A'(0)yc
        /.y Ax'(0)c
        * S+x+yc A-x-yC+x+yc

[Pad x to length k on right and reverse]
&(Rxk) /.x/.k() *0Rx-k ^ R-x-k *+x()

&(Vkp) /.k /.+?n'!%p () '(1)
       A V-k*0p V-k*1p 0

&(Wn) *0*". R Vn() n

[These lower bounds are zero until we
get a complete 7-bit character:]
W'{0}
W'{1}
W'{2}
W'{3}
W'{4}
W'{5}
W'{6}
[Programs are now one 7-bit ASCII character.]
W'{7}
W'{8}
W'{9}
W'{10}
W'{11}
W'{12}
W'{13}
[Programs are now two 7-bit ASCII characters.]
W'{14}
W'{15}
W'{16}
W'{17}
W'{18}
W'{19}
W'{20}
[Programs are now three 7-bit ASCII characters.]
W'{21}
W'{22}
[At this point we run out of memory, even with
 512 megabytes of lisp storage.  The solution
 is either more storage, or a much more
 complicated lisp interpreter with garbage
 collection.  We will never reach the first
 program that loops forever.  The smallest
 example I know is  :(f)f f  which is
 6 characters = 42 bits.]
[
W'{23}
]
\end{verbatim}
}\chap{omega.r}{\Size\begin{verbatim}
big.c

LISP Interpreter Run

[[[[ Omega in the limit from below! ]]]]

[[[
 (Wn) = the nth lower bound on the halting probability
 Omega = (the number of n-bit programs that halt when
 run on U for time n) divided by (2 to the power n).
]]]

[S = sum of three bits]
&(Sxyz) =x=yz

S:          (&(xyz)(=x(=yz)))


[C = carry of three bits]
&(Cxyz) /x/y1z/yz0

C:          (&(xyz)(/x(/y1z)(/yz0)))


[A = addition of reversed binary x and y]
[c = carryin]
&(Axyc) /.x /.y /c'(1)() A'(0)yc
        /.y Ax'(0)c
        * S+x+yc A-x-yC+x+yc

A:          (&(xyc)(/(.x)(/(.y)(/c('(1))())(A('(0))yc))(/(.y)(
            Ax('(0))c)(*(S(+x)(+y)c)(A(-x)(-y)(C(+x)(+y)c)))))
            )


[Pad x to length k on right and reverse]
&(Rxk) /.x/.k() *0Rx-k ^ R-x-k *+x()

R:          (&(xk)(/(.x)(/(.k)()(*0(Rx(-k))))(^(R(-x)(-k))(*(+
            x)()))))


&(Vkp) /.k /.+?n'!%p () '(1)
       A V-k*0p V-k*1p 0

V:          (&(kp)(/(.k)(/(.(+(?n('(!(%)))p)))()('(1)))(A(V(-k
            )(*0p))(V(-k)(*1p))0)))


&(Wn) *0*". R Vn() n

W:          (&(n)(*0(*.(R(Vn())n))))


[These lower bounds are zero until we
get a complete 7-bit character:]
W'{0}

expression  (W('()))
value       (0.)

W'{1}

expression  (W('(1)))
value       (0.0)

W'{2}

expression  (W('(11)))
value       (0.00)

W'{3}

expression  (W('(111)))
value       (0.000)

W'{4}

expression  (W('(1111)))
value       (0.0000)

W'{5}

expression  (W('(11111)))
value       (0.00000)

W'{6}

expression  (W('(111111)))
value       (0.000000)

[Programs are now one 7-bit ASCII character.]
W'{7}

expression  (W('(1111111)))
value       (0.1001001)

W'{8}

expression  (W('(11111111)))
value       (0.10010100)

W'{9}

expression  (W('(111111111)))
value       (0.100101000)

W'{10}

expression  (W('(1111111111)))
value       (0.1001010000)

W'{11}

expression  (W('(11111111111)))
value       (0.10010100000)

W'{12}

expression  (W('(111111111111)))
value       (0.100101000000)

W'{13}

expression  (W('(1111111111111)))
value       (0.1001010000000)

[Programs are now two 7-bit ASCII characters.]
W'{14}

expression  (W('(11111111111111)))
value       (0.10100000111100)

W'{15}

expression  (W('(111111111111111)))
value       (0.101000010001000)

W'{16}

expression  (W('(1111111111111111)))
value       (0.1010000100010000)

W'{17}

expression  (W('(11111111111111111)))
value       (0.10100001000100000)

W'{18}

expression  (W('(111111111111111111)))
value       (0.101000010001000000)

W'{19}

expression  (W('(1111111111111111111)))
value       (0.1010000100010000000)

W'{20}

expression  (W('(11111111111111111111)))
value       (0.10100001000100000000)

[Programs are now three 7-bit ASCII characters.]
W'{21}

expression  (W('(111111111111111111111)))
value       (0.101001001011001101110)

W'{22}

expression  (W('(1111111111111111111111)))
value       (0.1010010011000111110100)

[At this point we run out of memory, even with
 512 megabytes of lisp storage.  The solution
 is either more storage, or a much more
 complicated lisp interpreter with garbage
 collection.  We will never reach the first
 program that loops forever.  The smallest
 example I know is  :(f)f f  which is
 6 characters = 42 bits.]
[
W'{23}
]
End of LISP Run

Elapsed time is 5170 seconds.
\end{verbatim}
}\chap{omega2.l}{\Size\begin{verbatim}

[[[
 Show that H(Omega_n) > n - 1883.
 Omega_n is the first n bits of Omega.
]]]

[First test new stuff]

[<= for fractional binary numbers, i.e., is 0.x <= 0.y ?]
& (<xy) /.x 1
        /.y <x'(0)
        /=+x+y <-x-y
        +y
<'(000)'(000)
<'(000)'(001)
<'(001)'(000)
<'(001)'(001)
<'(110)'(110)
<'(110)'(111)
<'(111)'(110)
<'(111)'(111)
<'()'(000)
<'()'(001)
<'(000)'()
<'(001)'()

[Now run it all!]

++?0'!%  [ Universal binary computer U ]
         [ Put together program for U :]

^#~'     [ Show prefix prepended to program for first n bits of Omega ]
         [ so that we can see how many bits long the prepended prefix is.]

"(       [ begin literally ]

[S = sum of three bits]
:(Sxyz) =x=yz
[C = carry of three bits]
:(Cxyz) /x/y1z/yz0
[A = addition of reversed binary x and y]
[c = carryin]
:(Axyc) /.x /.y /c'(1)() A'(0)yc
        /.y Ax'(0)c
        * S+x+yc A-x-yC+x+yc
[Pad x to length k on right and reverse]
:(Rxk) /.x/.k() *0Rx-k ^ R-x-k *+x()
[W = lower bound on Omega obtained by
 running all n-bit programs for time n]
:(Vkp) /.k /.+?n'!%p () '(1)
       A V-k*0p V-k*1p 0
:(Wn) [[[*0*".]]]R Vn() n  [No "0." in front]

[<= for fractional binary numbers, i.e., is 0.x <= 0.y ?]
:(<xy) /.x 1
       /.y <x'(0)
       /=+x+y <-x-y
       +y

[List of all output of all n-bit programs p within time k.]
[ k is implicit argument.]
: (Bnp) /.n :v?k'!%p /.+v() +v
     [[ ^ B-n*0p B-n*1p [wrong order] ]]
        ^ B-n^p'(0) B-n^p'(1) [right order] [[[ELIMINATE DUPLICATES???]]]

:w !%             [Read and execute from remainder of tape
                   a program to compute an n-bit
                   initial piece of Omega.
                   IMPORTANT:  If Omega = .???1000000...
                   here must take Omega = .???0111111...
                   I.e., w = first n bits of the latter
                   binary representation.]
:(Lk)             [Main Loop]
  :x     Wk       [Compute the kth lower bound on Omega]
  /<wx   Bw()     [Are the first n bits OK?  I.e, is w <= x ? ]
                  [If so, form the list of all output of n-bit
                   programs within time k, output it, and halt.
                   This is bigger than anything of complexity
                   less than or equal to n!]
[I.e., this total output Bw() will be bigger than each individual output,
 and therefore must come from a program with more than n bits.
 Hence the size in bits of this prefix (= 1883) + the size in
 bits of any program for the first n bits of Omega must be
 greater than n.  Hence H(Omega_n) > n - 1883!
]
  L*1k            [If first n bits not OK, bump k and loop.]

L()           [Start main loop running with k initially zero.]

 )            [end literally]

#'            [Above prefix in binary is prepended to a
               program for U to compute the first n bits of Omega.]
"(            [begin literally]

[[[ Program to compute the first n bits of Omega: ]]]
[Cheat: test with the 7 bit approximation to Omega!]
[(Could also test this with 14 bit and 21 bit Omega approximations)]
[(Hence n = 7, and final output will have complexity greater than 7.)]
'(1001001)

 )            [end literally]
\end{verbatim}
}\chap{omega2.r}{\Size\begin{verbatim}
show.c

LISP Interpreter Run


[[[
 Show that H(Omega_n) > n - 1883.
 Omega_n is the first n bits of Omega.
]]]

[First test new stuff]

[<= for fractional binary numbers, i.e., is 0.x <= 0.y ?]
& (<xy) /.x 1
        /.y <x'(0)
        /=+x+y <-x-y
        +y

<:          (&(xy)(/(.x)1(/(.y)(<x('(0)))(/(=(+x)(+y))(<(-x)(-
            y))(+y)))))

<'(000)'(000)

expression  (<('(000))('(000)))
value       1

<'(000)'(001)

expression  (<('(000))('(001)))
value       1

<'(001)'(000)

expression  (<('(001))('(000)))
value       0

<'(001)'(001)

expression  (<('(001))('(001)))
value       1

<'(110)'(110)

expression  (<('(110))('(110)))
value       1

<'(110)'(111)

expression  (<('(110))('(111)))
value       1

<'(111)'(110)

expression  (<('(111))('(110)))
value       0

<'(111)'(111)

expression  (<('(111))('(111)))
value       1

<'()'(000)

expression  (<('())('(000)))
value       1

<'()'(001)

expression  (<('())('(001)))
value       1

<'(000)'()

expression  (<('(000))('()))
value       1

<'(001)'()

expression  (<('(001))('()))
value       0


[Now run it all!]

++?0'!%  [ Universal binary computer U ]
         [ Put together program for U :]

^#~'     [ Show prefix prepended to program for first n bits of Omega ]
         [ so that we can see how many bits long the prepended prefix is.]

"(       [ begin literally ]

[S = sum of three bits]
:(Sxyz) =x=yz
[C = carry of three bits]
:(Cxyz) /x/y1z/yz0
[A = addition of reversed binary x and y]
[c = carryin]
:(Axyc) /.x /.y /c'(1)() A'(0)yc
        /.y Ax'(0)c
        * S+x+yc A-x-yC+x+yc
[Pad x to length k on right and reverse]
:(Rxk) /.x/.k() *0Rx-k ^ R-x-k *+x()
[W = lower bound on Omega obtained by
 running all n-bit programs for time n]
:(Vkp) /.k /.+?n'!%p () '(1)
       A V-k*0p V-k*1p 0
:(Wn) [[[*0*".]]]R Vn() n  [No "0." in front]

[<= for fractional binary numbers, i.e., is 0.x <= 0.y ?]
:(<xy) /.x 1
       /.y <x'(0)
       /=+x+y <-x-y
       +y

[List of all output of all n-bit programs p within time k.]
[ k is implicit argument.]
: (Bnp) /.n :v?k'!%p /.+v() +v
     [[ ^ B-n*0p B-n*1p [wrong order] ]]
        ^ B-n^p'(0) B-n^p'(1) [right order] [[[ELIMINATE DUPLICATES???]]]

:w !%             [Read and execute from remainder of tape
                   a program to compute an n-bit
                   initial piece of Omega.
                   IMPORTANT:  If Omega = .???1000000...
                   here must take Omega = .???0111111...
                   I.e., w = first n bits of the latter
                   binary representation.]
:(Lk)             [Main Loop]
  :x     Wk       [Compute the kth lower bound on Omega]
  /<wx   Bw()     [Are the first n bits OK?  I.e, is w <= x ? ]
                  [If so, form the list of all output of n-bit
                   programs within time k, output it, and halt.
                   This is bigger than anything of complexity
                   less than or equal to n!]
[I.e., this total output Bw() will be bigger than each individual output,
 and therefore must come from a program with more than n bits.
 Hence the size in bits of this prefix (= 1883) + the size in
 bits of any program for the first n bits of Omega must be
 greater than n.  Hence H(Omega_n) > n - 1883!
]
  L*1k            [If first n bits not OK, bump k and loop.]

L()           [Start main loop running with k initially zero.]

 )            [end literally]

#'            [Above prefix in binary is prepended to a
               program for U to compute the first n bits of Omega.]
"(            [begin literally]

[[[ Program to compute the first n bits of Omega: ]]]
[Cheat: test with the 7 bit approximation to Omega!]
[(Could also test this with 14 bit and 21 bit Omega approximations)]
[(Hence n = 7, and final output will have complexity greater than 7.)]
'(1001001)

 )            [end literally]

expression  (+(+(?0('(!(%)))(^(#(~('(:(Sxyz)=x=yz:(Cxyz)/x/y1z
            /yz0:(Axyc)/.x/.y/c'(1)()A'(0)yc/.yAx'(0)c*S+x+ycA
            -x-yC+x+yc:(Rxk)/.x/.k()*0Rx-k^R-x-k*+x():(Vkp)/.k
            /.+?n'!%p()'(1)AV-k*0pV-k*1p0:(Wn)RVn()n:(<xy)/.x1
            /.y<x'(0)/=+x+y<-x-y+y:(Bnp)/.n:v?k'!%p/.+v()+v^B-
            n^p'(0)B-n^p'(1):w!%:(Lk):xWk/<wxBw()L*1kL()))))(#
            ('('(1001001))))))))
show        (:(Sxyz)=x=yz:(Cxyz)/x/y1z/yz0:(Axyc)/.x/.y/c'(1)(
            )A'(0)yc/.yAx'(0)c*S+x+ycA-x-yC+x+yc:(Rxk)/.x/.k()
            *0Rx-k^R-x-k*+x():(Vkp)/.k/.+?n'!%p()'(1)AV-k*0pV-
            k*1p0:(Wn)RVn()n:(<xy)/.x1/.y<x'(0)/=+x+y<-x-y+y:(
            Bnp)/.n:v?k'!%p/.+v()+v^B-n^p'(0)B-n^p'(1):w!%:(Lk
            ):xWk/<wxBw()L*1kL())
size        269(415)/1883(3533)
value       ($()0123456789;<>ABCDEFGHIJKLMNOPQRSTUVWXYZ\]_`abc
            defghijklmnopqrstuvwxyz|})

End of LISP Run

Elapsed time is 0 seconds.
\end{verbatim}
}\chap{sets0.l}{\Size\begin{verbatim}
[[[
 Test basic (finite) set functions.
]]]

[Set membership predicate; is e in set s?]
& (Mes) /.s0 /=e+s1 Me-s
Mx'(12345xabcde)
Mq'(12345xabcde)
[Eliminate duplicate elements from set s]
& (Ds) /.s() /M+s-s D-s *+sD-s
D'(1234512345abcde)
DD'(1234512345abcde)
[Set union]
& (Uxy) /.xy /M+xy U-xy *+xU-xy
U'(12345abcde)'(abcde67890)
[Set intersection]
& (Ixy) /.x() /M+xy *+xI-xy I-xy
I'(12345abcde)'(abcde67890)
[Subtract set y from set x]
& (Sxy) /.x() /M+xy S-xy *+xS-xy
S'(12345abcde)'(abcde67890)
[Identity function that displays a set of elements]
& (Os) /.s() *,+sO-s
O'(12345abcde)
[Combine two bit strings by interleaving them]
& (Cxy) /.xy /.yx *+x*+yC-x-y
C'(0000000000)'(11111111111111111111)
\end{verbatim}
}\chap{sets0.r}{\Size\begin{verbatim}
lisp.c

LISP Interpreter Run

[[[
 Test basic (finite) set functions.
]]]

[Set membership predicate; is e in set s?]
& (Mes) /.s0 /=e+s1 Me-s

M:          (&(es)(/(.s)0(/(=e(+s))1(Me(-s)))))

Mx'(12345xabcde)

expression  (Mx('(12345xabcde)))
value       1

Mq'(12345xabcde)

expression  (Mq('(12345xabcde)))
value       0

[Eliminate duplicate elements from set s]
& (Ds) /.s() /M+s-s D-s *+sD-s

D:          (&(s)(/(.s)()(/(M(+s)(-s))(D(-s))(*(+s)(D(-s))))))

D'(1234512345abcde)

expression  (D('(1234512345abcde)))
value       (12345abcde)

DD'(1234512345abcde)

expression  (D(D('(1234512345abcde))))
value       (12345abcde)

[Set union]
& (Uxy) /.xy /M+xy U-xy *+xU-xy

U:          (&(xy)(/(.x)y(/(M(+x)y)(U(-x)y)(*(+x)(U(-x)y)))))

U'(12345abcde)'(abcde67890)

expression  (U('(12345abcde))('(abcde67890)))
value       (12345abcde67890)

[Set intersection]
& (Ixy) /.x() /M+xy *+xI-xy I-xy

I:          (&(xy)(/(.x)()(/(M(+x)y)(*(+x)(I(-x)y))(I(-x)y))))

I'(12345abcde)'(abcde67890)

expression  (I('(12345abcde))('(abcde67890)))
value       (abcde)

[Subtract set y from set x]
& (Sxy) /.x() /M+xy S-xy *+xS-xy

S:          (&(xy)(/(.x)()(/(M(+x)y)(S(-x)y)(*(+x)(S(-x)y)))))

S'(12345abcde)'(abcde67890)

expression  (S('(12345abcde))('(abcde67890)))
value       (12345)

[Identity function that displays a set of elements]
& (Os) /.s() *,+sO-s

O:          (&(s)(/(.s)()(*(,(+s))(O(-s)))))

O'(12345abcde)

expression  (O('(12345abcde)))
display     1
display     2
display     3
display     4
display     5
display     a
display     b
display     c
display     d
display     e
value       (12345abcde)

[Combine two bit strings by interleaving them]
& (Cxy) /.xy /.yx *+x*+yC-x-y

C:          (&(xy)(/(.x)y(/(.y)x(*(+x)(*(+y)(C(-x)(-y)))))))

C'(0000000000)'(11111111111111111111)

expression  (C('(0000000000))('(11111111111111111111)))
value       (010101010101010101011111111111)

End of LISP Run

Elapsed time is 0 seconds.
\end{verbatim}
}\chap{sets1.l}{\Size\begin{verbatim}
[[[
 Show that
    H(X set-union Y) <= H(X) + H(Y) + 1799
 and that
    H(X set-intersection Y) <= H(X) + H(Y) + 1799.
 Here X and Y are INFINITE sets.
]]]

[Combine two bit strings by interleaving them]
& (Cxy) /.xy /.yx *+x*+yC-x-y

& (Dl) /.l"? ('&(xy)x D-l ,+l) [Display list in reverse]

D-
[[[++]]]?0'!%

^#~'"( [begin literally]

[Package of set functions from sets0.l]
: (Mes) /.s0 /=e+s1 Me-s
: (Ds) /.s() /M+s-s D-s *+sD-s
: (Uxy) /.xy /M+xy U-xy *+xU-xy
: (Ixy) /.x() /M+xy *+xI-xy I-xy
: (Sxy) /.x() /M+xy S-xy *+xS-xy
: (Os) /.s() *,+sO-s
[Main Loop:
 o is set of elements output so far.
 For first set,
 t is depth limit (time), b is bits read so far.
 For second set,
 u is depth limit (time), c is bits read so far.
]
: (Lotbuc)
 : v     ?t'!%b      [Run 1st computation again.]
 : w     ?u'!%c      [Run 2nd computation again.]
 : x     U-v-w       [Form UNION of sets so far]
 : y     OSxo        [Output all new elements]
 [This is an infinite loop. But to make debugging easier,
  halt if BOTH computations have halted.]
 / /=0.+v /=0.+w 100 "?
 [Bump everything before branching to head of loop]
 L x                 [Value of y is discarded, x is new o]
   /="?+v *1t    t   [Increment time for 1st computation]
   /="!+v ^b*@() b   [Increment tape for 1st computation]
   /="?+w *1u    u   [Increment time for 2nd computation]
   /="!+w ^c*@() c   [Increment tape for 2nd computation]

L()()()()()          [Initially all 5 induction variables
                      are empty.]
)                    [end literally]

C                    [Combine their bits in order needed]
                     [Wrong order if programs use @ or %]
#'"( *,a*,b*,c0 )    [Program to enumerate 1st FINITE set]
#'"( *,b*,c*,d0 )    [Program to enumerate 2nd FINITE set]
\end{verbatim}
}\chap{sets1.r}{\Size\begin{verbatim}
show.c

LISP Interpreter Run

[[[
 Show that
    H(X set-union Y) <= H(X) + H(Y) + 1799
 and that
    H(X set-intersection Y) <= H(X) + H(Y) + 1799.
 Here X and Y are INFINITE sets.
]]]

[Combine two bit strings by interleaving them]
& (Cxy) /.xy /.yx *+x*+yC-x-y

C:          (&(xy)(/(.x)y(/(.y)x(*(+x)(*(+y)(C(-x)(-y)))))))


& (Dl) /.l"? ('&(xy)x D-l ,+l) [Display list in reverse]

D:          (&(l)(/(.l)?(('(&(xy)x))(D(-l))(,(+l)))))


D-
[[[++]]]?0'!%

^#~'"( [begin literally]

[Package of set functions from sets0.l]
: (Mes) /.s0 /=e+s1 Me-s
: (Ds) /.s() /M+s-s D-s *+sD-s
: (Uxy) /.xy /M+xy U-xy *+xU-xy
: (Ixy) /.x() /M+xy *+xI-xy I-xy
: (Sxy) /.x() /M+xy S-xy *+xS-xy
: (Os) /.s() *,+sO-s
[Main Loop:
 o is set of elements output so far.
 For first set,
 t is depth limit (time), b is bits read so far.
 For second set,
 u is depth limit (time), c is bits read so far.
]
: (Lotbuc)
 : v     ?t'!%b      [Run 1st computation again.]
 : w     ?u'!%c      [Run 2nd computation again.]
 : x     U-v-w       [Form UNION of sets so far]
 : y     OSxo        [Output all new elements]
 [This is an infinite loop. But to make debugging easier,
  halt if BOTH computations have halted.]
 / /=0.+v /=0.+w 100 "?
 [Bump everything before branching to head of loop]
 L x                 [Value of y is discarded, x is new o]
   /="?+v *1t    t   [Increment time for 1st computation]
   /="!+v ^b*@() b   [Increment tape for 1st computation]
   /="?+w *1u    u   [Increment time for 2nd computation]
   /="!+w ^c*@() c   [Increment tape for 2nd computation]

L()()()()()          [Initially all 5 induction variables
                      are empty.]
)                    [end literally]

C                    [Combine their bits in order needed]
                     [Wrong order if programs use @ or %]
#'"( *,a*,b*,c0 )    [Program to enumerate 1st FINITE set]
#'"( *,b*,c*,d0 )    [Program to enumerate 2nd FINITE set]

expression  (D(-(?0('(!(%)))(^(#(~('(:(Mes)/.s0/=e+s1Me-s:(Ds)
            /.s()/M+s-sD-s*+sD-s:(Uxy)/.xy/M+xyU-xy*+xU-xy:(Ix
            y)/.x()/M+xy*+xI-xyI-xy:(Sxy)/.x()/M+xyS-xy*+xS-xy
            :(Os)/.s()*,+sO-s:(Lotbuc):v?t'!%b:w?u'!%c:xU-v-w:
            yOSxo//=0.+v/=0.+w100"?Lx/="?+v*1tt/="!+v^b*@()b/=
            "?+w*1uu/="!+w^c*@()cL()()()()()))))(C(#('(*,a*,b*
            ,c0)))(#('(*,b*,c*,d0))))))))
show        (:(Mes)/.s0/=e+s1Me-s:(Ds)/.s()/M+s-sD-s*+sD-s:(Ux
            y)/.xy/M+xyU-xy*+xU-xy:(Ixy)/.x()/M+xy*+xI-xyI-xy:
            (Sxy)/.x()/M+xyS-xy*+xS-xy:(Os)/.s()*,+sO-s:(Lotbu
            c):v?t'!%b:w?u'!%c:xU-v-w:yOSxo//=0.+v/=0.+w100"?L
            x/="?+v*1tt/="!+v^b*@()b/="?+w*1uu/="!+w^c*@()cL()
            ()()()())
size        257(401)/1799(3407)
display     a
display     d
display     c
display     b
value       ?

End of LISP Run

Elapsed time is 0 seconds.
\end{verbatim}
}\chap{sets2.l}{\Size\begin{verbatim}
[[[
 Show that
    H(X set-union Y) <= H(X) + H(Y) + 1799
 and that
    H(X set-intersection Y) <= H(X) + H(Y) + 1799.
 Here X and Y are INFINITE sets.
]]]

[Combine two bit strings by interleaving them]
& (Cxy) /.xy /.yx *+x*+yC-x-y

& (Dl) /.l"? ('&(xy)x D-l ,+l) [Display list in reverse]

D-
[[[++]]]?0'!%

^#~'"( [begin literally]

[Package of set functions from sets0.l]
: (Mes) /.s0 /=e+s1 Me-s
: (Ds) /.s() /M+s-s D-s *+sD-s
: (Uxy) /.xy /M+xy U-xy *+xU-xy
: (Ixy) /.x() /M+xy *+xI-xy I-xy
: (Sxy) /.x() /M+xy S-xy *+xS-xy
: (Os) /.s() *,+sO-s
[Main Loop:
 o is set of elements output so far.
 For first set,
 t is depth limit (time), b is bits read so far.
 For second set,
 u is depth limit (time), c is bits read so far.
]
: (Lotbuc)
 : v     ?t'!%b      [Run 1st computation again.]
 : w     ?u'!%c      [Run 2nd computation again.]
 : x     I-v-w       [Form INTERSECTION of sets so far]
 : y     OSxo        [Output all new elements]
 [This is an infinite loop. But to make debugging easier,
  halt if EITHER computation has halted.]
 / /=0.+v1 /=0.+w1 0 "?
 [Bump everything before branching to head of loop]
 L x                 [Value of y is discarded, x is new o]
   /="?+v *1t    t   [Increment time for 1st computation]
   /="!+v ^b*@() b   [Increment tape for 1st computation]
   /="?+w *1u    u   [Increment time for 2nd computation]
   /="!+w ^c*@() c   [Increment tape for 2nd computation]

L()()()()()          [Initially all 5 induction variables
                      are empty.]
)                    [end literally]

C                    [Combine their bits in order needed]
                     [Wrong order if programs use @ or %]
#'"( *,a*,b*,c0 )    [Program to enumerate 1st FINITE set]
#'"( *,b*,c*,d0 )    [Program to enumerate 2nd FINITE set]
\end{verbatim}
}\chap{sets2.r}{\Size\begin{verbatim}
show.c

LISP Interpreter Run

[[[
 Show that
    H(X set-union Y) <= H(X) + H(Y) + 1799
 and that
    H(X set-intersection Y) <= H(X) + H(Y) + 1799.
 Here X and Y are INFINITE sets.
]]]

[Combine two bit strings by interleaving them]
& (Cxy) /.xy /.yx *+x*+yC-x-y

C:          (&(xy)(/(.x)y(/(.y)x(*(+x)(*(+y)(C(-x)(-y)))))))


& (Dl) /.l"? ('&(xy)x D-l ,+l) [Display list in reverse]

D:          (&(l)(/(.l)?(('(&(xy)x))(D(-l))(,(+l)))))


D-
[[[++]]]?0'!%

^#~'"( [begin literally]

[Package of set functions from sets0.l]
: (Mes) /.s0 /=e+s1 Me-s
: (Ds) /.s() /M+s-s D-s *+sD-s
: (Uxy) /.xy /M+xy U-xy *+xU-xy
: (Ixy) /.x() /M+xy *+xI-xy I-xy
: (Sxy) /.x() /M+xy S-xy *+xS-xy
: (Os) /.s() *,+sO-s
[Main Loop:
 o is set of elements output so far.
 For first set,
 t is depth limit (time), b is bits read so far.
 For second set,
 u is depth limit (time), c is bits read so far.
]
: (Lotbuc)
 : v     ?t'!%b      [Run 1st computation again.]
 : w     ?u'!%c      [Run 2nd computation again.]
 : x     I-v-w       [Form INTERSECTION of sets so far]
 : y     OSxo        [Output all new elements]
 [This is an infinite loop. But to make debugging easier,
  halt if EITHER computation has halted.]
 / /=0.+v1 /=0.+w1 0 "?
 [Bump everything before branching to head of loop]
 L x                 [Value of y is discarded, x is new o]
   /="?+v *1t    t   [Increment time for 1st computation]
   /="!+v ^b*@() b   [Increment tape for 1st computation]
   /="?+w *1u    u   [Increment time for 2nd computation]
   /="!+w ^c*@() c   [Increment tape for 2nd computation]

L()()()()()          [Initially all 5 induction variables
                      are empty.]
)                    [end literally]

C                    [Combine their bits in order needed]
                     [Wrong order if programs use @ or %]
#'"( *,a*,b*,c0 )    [Program to enumerate 1st FINITE set]
#'"( *,b*,c*,d0 )    [Program to enumerate 2nd FINITE set]

expression  (D(-(?0('(!(%)))(^(#(~('(:(Mes)/.s0/=e+s1Me-s:(Ds)
            /.s()/M+s-sD-s*+sD-s:(Uxy)/.xy/M+xyU-xy*+xU-xy:(Ix
            y)/.x()/M+xy*+xI-xyI-xy:(Sxy)/.x()/M+xyS-xy*+xS-xy
            :(Os)/.s()*,+sO-s:(Lotbuc):v?t'!%b:w?u'!%c:xI-v-w:
            yOSxo//=0.+v1/=0.+w10"?Lx/="?+v*1tt/="!+v^b*@()b/=
            "?+w*1uu/="!+w^c*@()cL()()()()()))))(C(#('(*,a*,b*
            ,c0)))(#('(*,b*,c*,d0))))))))
show        (:(Mes)/.s0/=e+s1Me-s:(Ds)/.s()/M+s-sD-s*+sD-s:(Ux
            y)/.xy/M+xyU-xy*+xU-xy:(Ixy)/.x()/M+xy*+xI-xyI-xy:
            (Sxy)/.x()/M+xyS-xy*+xS-xy:(Os)/.s()*,+sO-s:(Lotbu
            c):v?t'!%b:w?u'!%c:xI-v-w:yOSxo//=0.+v1/=0.+w10"?L
            x/="?+v*1tt/="!+v^b*@()b/="?+w*1uu/="!+w^c*@()cL()
            ()()()())
size        257(401)/1799(3407)
display     c
display     b
value       ?

End of LISP Run

Elapsed time is 0 seconds.
\end{verbatim}
}\chap{sets3.l}{\Size\begin{verbatim}
[[[
 Show that
    H(X set-union Y) <= H(X) + H(Y) + 1799
 and that
    H(X set-intersection Y) <= H(X) + H(Y) + 1799.
 Here X and Y are INFINITE sets.
]]]

[IMPORTANT: This test case never halts, so] [<=====!!!]
[must run this with a depth limit on try/?]

[Combine two bit strings by interleaving them]
& (Cxy) /.xy /.yx *+x*+yC-x-y

& (2n) /.n() :k2-n /+n *1^kk ^kk [reverse binary to unary]

& (Dl) /.l"? ('&(xy)x D-l ,+l)   [display list in reverse]

D-
[[[++]]]?2'(000 000 001 1)'!% [ 768 base 10 = 1400 base 8 ]

^#'"( [begin literally]

[Package of set functions from sets0.l]
: (Mes) /.s0 /=e+s1 Me-s
: (Ds) /.s() /M+s-s D-s *+sD-s
: (Uxy) /.xy /M+xy U-xy *+xU-xy
: (Ixy) /.x() /M+xy *+xI-xy I-xy
: (Sxy) /.x() /M+xy S-xy *+xS-xy
: (Os) /.s() *,+sO-s
[Main Loop:
 o is set of elements output so far.
 For first set,
 t is depth limit (time), b is bits read so far.
 For second set,
 u is depth limit (time), c is bits read so far.
]
: (Lotbuc)
 : v     ?t'!%b      [Run 1st computation again.]
 : w     ?u'!%c      [Run 2nd computation again.]
 : x     I-v-w       [Form INTERSECTION of sets so far]
 : y     OSxo        [Output all new elements]
 [This is an infinite loop. But to make debugging easier,
  halt if EITHER computation has halted.]
 / /=0.+v1 /=0.+w1 0 "?
 [Bump everything before branching to head of loop]
 L x                 [Value of y is discarded, x is new o]
   /="?+v *1t    t   [Increment time for 1st computation]
   /="!+v ^b*@() b   [Increment tape for 1st computation]
   /="?+w *1u    u   [Increment time for 2nd computation]
   /="!+w ^c*@() c   [Increment tape for 2nd computation]

L()()()()()          [Initially all 5 induction variables
                      are empty.]
)                    [end literally]

C                    [Combine their bits in order needed]
                     [Wrong order if programs use @ or %]
                     [Program to enumerate 1st INFINITE set]
#'"( :(Lk)L,*1*1k L() )
                     [Program to enumerate 2nd INFINITE set]
#'"( :(Lk)L,*1*1*1k L() )
\end{verbatim}
}\chap{sets3.r}{\Size\begin{verbatim}
lisp.c

LISP Interpreter Run

[[[
 Show that
    H(X set-union Y) <= H(X) + H(Y) + 1799
 and that
    H(X set-intersection Y) <= H(X) + H(Y) + 1799.
 Here X and Y are INFINITE sets.
]]]

[IMPORTANT: This test case never halts, so] [<=====!!!]
[must run this with a depth limit on try/?]

[Combine two bit strings by interleaving them]
& (Cxy) /.xy /.yx *+x*+yC-x-y

C:          (&(xy)(/(.x)y(/(.y)x(*(+x)(*(+y)(C(-x)(-y)))))))


& (2n) /.n() :k2-n /+n *1^kk ^kk [reverse binary to unary]

2:          (&(n)(/(.n)()(('(&(k)(/(+n)(*1(^kk))(^kk))))(2(-n)
            ))))


& (Dl) /.l"? ('&(xy)x D-l ,+l)   [display list in reverse]

D:          (&(l)(/(.l)?(('(&(xy)x))(D(-l))(,(+l)))))


D-
[[[++]]]?2'(000 000 001 1)'!% [ 768 base 10 = 1400 base 8 ]

^#'"( [begin literally]

[Package of set functions from sets0.l]
: (Mes) /.s0 /=e+s1 Me-s
: (Ds) /.s() /M+s-s D-s *+sD-s
: (Uxy) /.xy /M+xy U-xy *+xU-xy
: (Ixy) /.x() /M+xy *+xI-xy I-xy
: (Sxy) /.x() /M+xy S-xy *+xS-xy
: (Os) /.s() *,+sO-s
[Main Loop:
 o is set of elements output so far.
 For first set,
 t is depth limit (time), b is bits read so far.
 For second set,
 u is depth limit (time), c is bits read so far.
]
: (Lotbuc)
 : v     ?t'!%b      [Run 1st computation again.]
 : w     ?u'!%c      [Run 2nd computation again.]
 : x     I-v-w       [Form INTERSECTION of sets so far]
 : y     OSxo        [Output all new elements]
 [This is an infinite loop. But to make debugging easier,
  halt if EITHER computation has halted.]
 / /=0.+v1 /=0.+w1 0 "?
 [Bump everything before branching to head of loop]
 L x                 [Value of y is discarded, x is new o]
   /="?+v *1t    t   [Increment time for 1st computation]
   /="!+v ^b*@() b   [Increment tape for 1st computation]
   /="?+w *1u    u   [Increment time for 2nd computation]
   /="!+w ^c*@() c   [Increment tape for 2nd computation]

L()()()()()          [Initially all 5 induction variables
                      are empty.]
)                    [end literally]

C                    [Combine their bits in order needed]
                     [Wrong order if programs use @ or %]
                     [Program to enumerate 1st INFINITE set]
#'"( :(Lk)L,*1*1k L() )
                     [Program to enumerate 2nd INFINITE set]
#'"( :(Lk)L,*1*1*1k L() )

expression  (D(-(?(2('(0000000011)))('(!(%)))(^(#('(:(Mes)/.s0
            /=e+s1Me-s:(Ds)/.s()/M+s-sD-s*+sD-s:(Uxy)/.xy/M+xy
            U-xy*+xU-xy:(Ixy)/.x()/M+xy*+xI-xyI-xy:(Sxy)/.x()/
            M+xyS-xy*+xS-xy:(Os)/.s()*,+sO-s:(Lotbuc):v?t'!%b:
            w?u'!%c:xI-v-w:yOSxo//=0.+v1/=0.+w10"?Lx/="?+v*1tt
            /="!+v^b*@()b/="?+w*1uu/="!+w^c*@()cL()()()()())))
            (C(#('(:(Lk)L,*1*1kL())))(#('(:(Lk)L,*1*1*1kL())))
            )))))
display     (111111)
display     (111111111111)
display     (111111111111111111)
display     (111111111111111111111111)
display     (111111111111111111111111111111)
display     (111111111111111111111111111111111111)
display     (111111111111111111111111111111111111111111)
display     (111111111111111111111111111111111111111111111111)
display     (1111111111111111111111111111111111111111111111111
            11111)
value       ?

End of LISP Run

Elapsed time is 0 seconds.
\end{verbatim}
}\chap{sets4.l}{\Size\begin{verbatim}
[[[
 Show that
    H(X set-union Y) <= H(X) + H(Y) + 1799
 and that
    H(X set-intersection Y) <= H(X) + H(Y) + 1799.
 Here X and Y are INFINITE sets.
]]]

[IMPORTANT: This test case never halts, so] [<=====!!!]
[must run this with a depth limit on try/?]

[Combine two bit strings by interleaving them]
& (Cxy) /.xy /.yx *+x*+yC-x-y

& (2n) /.n() :k2-n /+n *1^kk ^kk [reverse binary to unary]

& (Dl) /.l"? ('&(xy)x D-l ,+l)   [display list in reverse]

D-
[[[++]]]?2'(000 000 110 1)'!% [ 704 base 10 = 1300 base 8 ]

^#'"( [begin literally]

[Package of set functions from sets0.l]
: (Mes) /.s0 /=e+s1 Me-s
: (Ds) /.s() /M+s-s D-s *+sD-s
: (Uxy) /.xy /M+xy U-xy *+xU-xy
: (Ixy) /.x() /M+xy *+xI-xy I-xy
: (Sxy) /.x() /M+xy S-xy *+xS-xy
: (Os) /.s() *,+sO-s
[Main Loop:
 o is set of elements output so far.
 For first set,
 t is depth limit (time), b is bits read so far.
 For second set,
 u is depth limit (time), c is bits read so far.
]
: (Lotbuc)
 : v     ?t'!%b      [Run 1st computation again.]
 : w     ?u'!%c      [Run 2nd computation again.]
 : x     U-v-w       [Form UNION of sets so far]
 : y     OSxo        [Output all new elements]
 [This is an infinite loop. But to make debugging easier,
  halt if BOTH computations have halted.]
 / /=0.+v /=0.+w 100 "?
 [Bump everything before branching to head of loop]
 L x                 [Value of y is discarded, x is new o]
   /="?+v *1t    t   [Increment time for 1st computation]
   /="!+v ^b*@() b   [Increment tape for 1st computation]
   /="?+w *1u    u   [Increment time for 2nd computation]
   /="!+w ^c*@() c   [Increment tape for 2nd computation]

L()()()()()          [Initially all 5 induction variables
                      are empty.]
)                    [end literally]

C                    [Combine their bits in order needed]
                     [Wrong order if programs use @ or %]
                     [Program to enumerate 1st INFINITE set]
#'"( :(Lk)L,*1*1k L() )
                     [Program to enumerate 2nd INFINITE set]
#'"( :(Lk)L,*2*2*2k L() )
\end{verbatim}
}\chap{sets4.r}{\Size\begin{verbatim}
lisp.c

LISP Interpreter Run

[[[
 Show that
    H(X set-union Y) <= H(X) + H(Y) + 1799
 and that
    H(X set-intersection Y) <= H(X) + H(Y) + 1799.
 Here X and Y are INFINITE sets.
]]]

[IMPORTANT: This test case never halts, so] [<=====!!!]
[must run this with a depth limit on try/?]

[Combine two bit strings by interleaving them]
& (Cxy) /.xy /.yx *+x*+yC-x-y

C:          (&(xy)(/(.x)y(/(.y)x(*(+x)(*(+y)(C(-x)(-y)))))))


& (2n) /.n() :k2-n /+n *1^kk ^kk [reverse binary to unary]

2:          (&(n)(/(.n)()(('(&(k)(/(+n)(*1(^kk))(^kk))))(2(-n)
            ))))


& (Dl) /.l"? ('&(xy)x D-l ,+l)   [display list in reverse]

D:          (&(l)(/(.l)?(('(&(xy)x))(D(-l))(,(+l)))))


D-
[[[++]]]?2'(000 000 110 1)'!% [ 704 base 10 = 1300 base 8 ]

^#'"( [begin literally]

[Package of set functions from sets0.l]
: (Mes) /.s0 /=e+s1 Me-s
: (Ds) /.s() /M+s-s D-s *+sD-s
: (Uxy) /.xy /M+xy U-xy *+xU-xy
: (Ixy) /.x() /M+xy *+xI-xy I-xy
: (Sxy) /.x() /M+xy S-xy *+xS-xy
: (Os) /.s() *,+sO-s
[Main Loop:
 o is set of elements output so far.
 For first set,
 t is depth limit (time), b is bits read so far.
 For second set,
 u is depth limit (time), c is bits read so far.
]
: (Lotbuc)
 : v     ?t'!%b      [Run 1st computation again.]
 : w     ?u'!%c      [Run 2nd computation again.]
 : x     U-v-w       [Form UNION of sets so far]
 : y     OSxo        [Output all new elements]
 [This is an infinite loop. But to make debugging easier,
  halt if BOTH computations have halted.]
 / /=0.+v /=0.+w 100 "?
 [Bump everything before branching to head of loop]
 L x                 [Value of y is discarded, x is new o]
   /="?+v *1t    t   [Increment time for 1st computation]
   /="!+v ^b*@() b   [Increment tape for 1st computation]
   /="?+w *1u    u   [Increment time for 2nd computation]
   /="!+w ^c*@() c   [Increment tape for 2nd computation]

L()()()()()          [Initially all 5 induction variables
                      are empty.]
)                    [end literally]

C                    [Combine their bits in order needed]
                     [Wrong order if programs use @ or %]
                     [Program to enumerate 1st INFINITE set]
#'"( :(Lk)L,*1*1k L() )
                     [Program to enumerate 2nd INFINITE set]
#'"( :(Lk)L,*2*2*2k L() )

expression  (D(-(?(2('(0000001101)))('(!(%)))(^(#('(:(Mes)/.s0
            /=e+s1Me-s:(Ds)/.s()/M+s-sD-s*+sD-s:(Uxy)/.xy/M+xy
            U-xy*+xU-xy:(Ixy)/.x()/M+xy*+xI-xyI-xy:(Sxy)/.x()/
            M+xyS-xy*+xS-xy:(Os)/.s()*,+sO-s:(Lotbuc):v?t'!%b:
            w?u'!%c:xU-v-w:yOSxo//=0.+v/=0.+w100"?Lx/="?+v*1tt
            /="!+v^b*@()b/="?+w*1uu/="!+w^c*@()cL()()()()())))
            (C(#('(:(Lk)L,*1*1kL())))(#('(:(Lk)L,*2*2*2kL())))
            )))))
display     (11)
display     (1111)
display     (111111)
display     (11111111)
display     (1111111111)
display     (111111111111)
display     (11111111111111)
display     (1111111111111111)
display     (111111111111111111)
display     (11111111111111111111)
display     (1111111111111111111111)
display     (111111111111111111111111)
display     (11111111111111111111111111)
display     (1111111111111111111111111111)
display     (111111111111111111111111111111)
display     (222)
display     (11111111111111111111111111111111)
display     (222222)
display     (1111111111111111111111111111111111)
display     (222222222)
display     (111111111111111111111111111111111111)
display     (222222222222)
display     (11111111111111111111111111111111111111)
display     (222222222222222)
display     (1111111111111111111111111111111111111111)
display     (222222222222222222)
value       ?

End of LISP Run

Elapsed time is 0 seconds.
\end{verbatim}
}\chap{godel.l}{\Size\begin{verbatim}
[[[
 Show that a formal system of complexity N
 can't prove that a specific object has
 complexity > N + 994.
 Formal system is a never halting lisp expression
 that displays pairs (lisp object, lower bound
 on its complexity).  E.g., (x(1111)) means
 that x has complexity H(x) greater than or equal to 4.
]]]

[ (<xy) tells if x is less than y ]
& (<xy) /.y0 /.x1 <-x-y
<'(11)'(11)
<'(11)'(111)
<'(111)'(11)

[ Search list p for first s in pair (s,m) with H(s) >= m > n ]
[ (Returns 0/false to indicate not found.) ]
& (Epn) /.p 0 /<n+-+p ++p E-pn
E'((x(11))(y(111)))'()
E'((x(11))(y(111)))'(1)
E'((x(11))(y(111)))'(11)
E'((x(11))(y(111)))'(111)
E'((x(11))(y(111)))'(1111)

++?0'!%  [[Universal binary computer U]]
         [[Put together program for U:]]
^#~'     [[Show crucial prefix so can size it]]
"(       [[begin literally]]

[ (<xy) tells if x is less than y ]
: (<xy) /.y1 /.x1 <-x-y    [[ BAD DEFINITION!!! ALWAYS TRUE ]]

[ Search list p for first s in pair (s,m) with H(s) >= m > n ]
[ (Returns 0/false to indicate not found.) ]
: (Epn) /.p 0 /<n+-+p ++p E-pn

[ (2n) = convert reversed binary to unary ]
: (2n) /.n() :k2-n /+n *1^kk ^kk

[ k is the size in bits of this program for U, ]
[ excluding the bits in the formal axiomatic system. ]
: k 2'(010 001 111 1) [k = 994 base 10 = 1742 base 8]

[Main Loop - t is depth limit (time),
             b is bits of FAS read so far (buffer)]
: (Ltb)
 : v ?t'!%b [run FAS for time t]
 : s E-v^kb [look for s-exp s that is proved to have complexity H(s)
               > 994 + # of bits of FAS read]
 /s s       [Found it!  Output it and halt]
            [Contradiction: our value has complexity
             greater than our size!]
            [Our size is 994 bits + the number of bits of
             the FAS that were read at the point that the
             search succeeded (some bits of the FAS may be
             unread).]
 /="!+v  Lt^b*@()   [Read another bit of FAS]
 /="?+v  L*1tb      [Increase depth/time limit]
 "?     [Surprise, formal system halts, so we do too]

L()()      [Initially, 0 depth limit and no bits read]

 )      [[end literally]]

#'
"(      [[begin literally]]

[[This is the FAS]]
,'((xy)(11))  [FAS = {"H((xy)) >= 2"}]

 )      [[end literally]]
\end{verbatim}
}\chap{godel.r}{\Size\begin{verbatim}
show.c

LISP Interpreter Run

[[[
 Show that a formal system of complexity N
 can't prove that a specific object has
 complexity > N + 994.
 Formal system is a never halting lisp expression
 that displays pairs (lisp object, lower bound
 on its complexity).  E.g., (x(1111)) means
 that x has complexity H(x) greater than or equal to 4.
]]]

[ (<xy) tells if x is less than y ]
& (<xy) /.y0 /.x1 <-x-y

<:          (&(xy)(/(.y)0(/(.x)1(<(-x)(-y)))))

<'(11)'(11)

expression  (<('(11))('(11)))
value       0

<'(11)'(111)

expression  (<('(11))('(111)))
value       1

<'(111)'(11)

expression  (<('(111))('(11)))
value       0


[ Search list p for first s in pair (s,m) with H(s) >= m > n ]
[ (Returns 0/false to indicate not found.) ]
& (Epn) /.p 0 /<n+-+p ++p E-pn

E:          (&(pn)(/(.p)0(/(<n(+(-(+p))))(+(+p))(E(-p)n))))

E'((x(11))(y(111)))'()

expression  (E('((x(11))(y(111))))('()))
value       x

E'((x(11))(y(111)))'(1)

expression  (E('((x(11))(y(111))))('(1)))
value       x

E'((x(11))(y(111)))'(11)

expression  (E('((x(11))(y(111))))('(11)))
value       y

E'((x(11))(y(111)))'(111)

expression  (E('((x(11))(y(111))))('(111)))
value       0

E'((x(11))(y(111)))'(1111)

expression  (E('((x(11))(y(111))))('(1111)))
value       0


++?0'!%  [[Universal binary computer U]]
         [[Put together program for U:]]
^#~'     [[Show crucial prefix so can size it]]
"(       [[begin literally]]

[ (<xy) tells if x is less than y ]
: (<xy) /.y1 /.x1 <-x-y    [[ BAD DEFINITION!!! ALWAYS TRUE ]]

[ Search list p for first s in pair (s,m) with H(s) >= m > n ]
[ (Returns 0/false to indicate not found.) ]
: (Epn) /.p 0 /<n+-+p ++p E-pn

[ (2n) = convert reversed binary to unary ]
: (2n) /.n() :k2-n /+n *1^kk ^kk

[ k is the size in bits of this program for U, ]
[ excluding the bits in the formal axiomatic system. ]
: k 2'(010 001 111 1) [k = 994 base 10 = 1742 base 8]

[Main Loop - t is depth limit (time),
             b is bits of FAS read so far (buffer)]
: (Ltb)
 : v ?t'!%b [run FAS for time t]
 : s E-v^kb [look for s-exp s that is proved to have complexity H(s)
               > 994 + # of bits of FAS read]
 /s s       [Found it!  Output it and halt]
            [Contradiction: our value has complexity
             greater than our size!]
            [Our size is 994 bits + the number of bits of
             the FAS that were read at the point that the
             search succeeded (some bits of the FAS may be
             unread).]
 /="!+v  Lt^b*@()   [Read another bit of FAS]
 /="?+v  L*1tb      [Increase depth/time limit]
 "?     [Surprise, formal system halts, so we do too]

L()()      [Initially, 0 depth limit and no bits read]

 )      [[end literally]]

#'
"(      [[begin literally]]

[[This is the FAS]]
,'((xy)(11))  [FAS = {"H((xy)) >= 2"}]

 )      [[end literally]]

expression  (+(+(?0('(!(%)))(^(#(~('(:(<xy)/.y1/.x1<-x-y:(Epn)
            /.p0/<n+-+p++pE-pn:(2n)/.n():k2-n/+n*1^kk^kk:k2'(0
            100011111):(Ltb):v?t'!%b:sE-v^kb/ss/="!+vLt^b*@()/
            ="?+vL*1tb"?L()()))))(#('(,'((xy)(11)))))))))
show        (:(<xy)/.y1/.x1<-x-y:(Epn)/.p0/<n+-+p++pE-pn:(2n)/
            .n():k2-n/+n*1^kk^kk:k2'(0100011111):(Ltb):v?t'!%b
            :sE-v^kb/ss/="!+vLt^b*@()/="?+vL*1tb"?L()())
size        142(216)/994(1742)
value       (xy)

End of LISP Run

Elapsed time is 0 seconds.
\end{verbatim}
}\chap{godel2.l}{\Size\begin{verbatim}
[[[
 Show that a formal system of complexity N
 can't prove that a specific object has
 complexity > N + 994.
 Formal system is a never halting lisp expression
 that displays pairs (lisp object, lower bound
 on its complexity).  E.g., (x(1111)) means
 that x has complexity H(x) greater than or equal to 4.
]]]

[ (<xy) tells if x is less than y ]
& (<xy) /.y0 /.x1 <-x-y
<'(11)'(11)
<'(11)'(111)
<'(111)'(11)

[ Search list p for first s in pair (s,m) with H(s) >= m > n ]
[ (Returns 0/false to indicate not found.) ]
& (Epn) /.p 0 /<n+-+p ++p E-pn
E'((x(11))(y(111)))'()
E'((x(11))(y(111)))'(1)
E'((x(11))(y(111)))'(11)
E'((x(11))(y(111)))'(111)
E'((x(11))(y(111)))'(1111)

++?0'!%  [[Universal binary computer U]]
         [[Put together program for U:]]
^#~'     [[Show crucial prefix so can size it]]
"(       [[begin literally]]

[ (<xy) tells if x is less than y ]
: (<xy) /.y0 /.x1 <-x-y    [[ CORRECT DEFINITION!!! ]]

[ Search list p for first s in pair (s,m) with H(s) >= m > n ]
[ (Returns 0/false to indicate not found.) ]
: (Epn) /.p 0 /<n+-+p ++p E-pn

[ (2n) = convert reversed binary to unary ]
: (2n) /.n() :k2-n /+n *1^kk ^kk

[ k is the size in bits of this program for U, ]
[ excluding the bits in the formal axiomatic system. ]
: k 2'(010 001 111 1) [k = 994 base 10 = 1742 base 8]

[Main Loop - t is depth limit (time),
             b is bits of FAS read so far (buffer)]
: (Ltb)
 : v ?t'!%b [run FAS for time t]
 : s E-v^kb [look for s-exp s that is proved to have complexity H(s)
               > 994 + # of bits of FAS read]
 /s s       [Found it!  Output it and halt]
            [Contradiction: our value has complexity
             greater than our size!]
            [Our size is 994 bits + the number of bits of
             the FAS that were read at the point that the
             search succeeded (some bits of the FAS may be
             unread).]
 /="!+v  Lt^b*@()   [Read another bit of FAS]
 /="?+v  L*1tb      [Increase depth/time limit]
 "?     [Surprise, formal system halts, so we do too]

L()()      [Initially, 0 depth limit and no bits read]

 )      [[end literally]]

#'
"(      [[begin literally]]

[[This is the FAS]]
,'((xy)(11))  [FAS = {"H((xy)) >= 2"}]

 )      [[end literally]]
\end{verbatim}
}\chap{godel2.r}{\Size\begin{verbatim}
show.c

LISP Interpreter Run

[[[
 Show that a formal system of complexity N
 can't prove that a specific object has
 complexity > N + 994.
 Formal system is a never halting lisp expression
 that displays pairs (lisp object, lower bound
 on its complexity).  E.g., (x(1111)) means
 that x has complexity H(x) greater than or equal to 4.
]]]

[ (<xy) tells if x is less than y ]
& (<xy) /.y0 /.x1 <-x-y

<:          (&(xy)(/(.y)0(/(.x)1(<(-x)(-y)))))

<'(11)'(11)

expression  (<('(11))('(11)))
value       0

<'(11)'(111)

expression  (<('(11))('(111)))
value       1

<'(111)'(11)

expression  (<('(111))('(11)))
value       0


[ Search list p for first s in pair (s,m) with H(s) >= m > n ]
[ (Returns 0/false to indicate not found.) ]
& (Epn) /.p 0 /<n+-+p ++p E-pn

E:          (&(pn)(/(.p)0(/(<n(+(-(+p))))(+(+p))(E(-p)n))))

E'((x(11))(y(111)))'()

expression  (E('((x(11))(y(111))))('()))
value       x

E'((x(11))(y(111)))'(1)

expression  (E('((x(11))(y(111))))('(1)))
value       x

E'((x(11))(y(111)))'(11)

expression  (E('((x(11))(y(111))))('(11)))
value       y

E'((x(11))(y(111)))'(111)

expression  (E('((x(11))(y(111))))('(111)))
value       0

E'((x(11))(y(111)))'(1111)

expression  (E('((x(11))(y(111))))('(1111)))
value       0


++?0'!%  [[Universal binary computer U]]
         [[Put together program for U:]]
^#~'     [[Show crucial prefix so can size it]]
"(       [[begin literally]]

[ (<xy) tells if x is less than y ]
: (<xy) /.y0 /.x1 <-x-y    [[ CORRECT DEFINITION!!! ]]

[ Search list p for first s in pair (s,m) with H(s) >= m > n ]
[ (Returns 0/false to indicate not found.) ]
: (Epn) /.p 0 /<n+-+p ++p E-pn

[ (2n) = convert reversed binary to unary ]
: (2n) /.n() :k2-n /+n *1^kk ^kk

[ k is the size in bits of this program for U, ]
[ excluding the bits in the formal axiomatic system. ]
: k 2'(010 001 111 1) [k = 994 base 10 = 1742 base 8]

[Main Loop - t is depth limit (time),
             b is bits of FAS read so far (buffer)]
: (Ltb)
 : v ?t'!%b [run FAS for time t]
 : s E-v^kb [look for s-exp s that is proved to have complexity H(s)
               > 994 + # of bits of FAS read]
 /s s       [Found it!  Output it and halt]
            [Contradiction: our value has complexity
             greater than our size!]
            [Our size is 994 bits + the number of bits of
             the FAS that were read at the point that the
             search succeeded (some bits of the FAS may be
             unread).]
 /="!+v  Lt^b*@()   [Read another bit of FAS]
 /="?+v  L*1tb      [Increase depth/time limit]
 "?     [Surprise, formal system halts, so we do too]

L()()      [Initially, 0 depth limit and no bits read]

 )      [[end literally]]

#'
"(      [[begin literally]]

[[This is the FAS]]
,'((xy)(11))  [FAS = {"H((xy)) >= 2"}]

 )      [[end literally]]

expression  (+(+(?0('(!(%)))(^(#(~('(:(<xy)/.y0/.x1<-x-y:(Epn)
            /.p0/<n+-+p++pE-pn:(2n)/.n():k2-n/+n*1^kk^kk:k2'(0
            100011111):(Ltb):v?t'!%b:sE-v^kb/ss/="!+vLt^b*@()/
            ="?+vL*1tb"?L()()))))(#('(,'((xy)(11)))))))))
show        (:(<xy)/.y0/.x1<-x-y:(Epn)/.p0/<n+-+p++pE-pn:(2n)/
            .n():k2-n/+n*1^kk^kk:k2'(0100011111):(Ltb):v?t'!%b
            :sE-v^kb/ss/="!+vLt^b*@()/="?+vL*1tb"?L()())
size        142(216)/994(1742)
value       ?

End of LISP Run

Elapsed time is 1 seconds.
\end{verbatim}
}\chap{godel3.l}{\Size\begin{verbatim}
[[[
 Show that a formal system of complexity N
 can't determine more than N + 3192 bits of Omega.
 Formal system is a never halting lisp expression
 that displays lists of the form (10X0XXXX10).
 This stands for the fractional part of Omega,
 and means that these 0,1 bits of Omega are known.
 X stands for an unknown bit.
]]]

[Count number of bits in an omega that are determined.]
& (Cw) /.w() /=X+w C-w *1C-w
C'(XXX)
C'(1XX)
C'(1X0)
C'(110)

[Merge bits of data into unknown bits of an omega.]
& (Mw) /.w() * /=X+w@+w M-w
[Test it.]
++?0 ':(Mw)/.w()*/=X+w@+wM-w M'(00X00X00X) '(111)
++?0 ':(Mw)/.w()*/=X+w@+wM-w M'(11X11X111) '(00)

[(<xy) tells if x is less than y.]
& (<xy) /.y0 /.x1 <-x-y
<'(11)'(11)
<'(11)'(111)
<'(111)'(11)

[
 Examine omegas in list w to see if in any one of them
 the number of bits that are determined is greater than n.
 Returns 0 to indicate not found, or what it found.
]
& (Ewn) /.w 0 /<nC+w +w E-wn
E'((00)(000))'()
E'((00)(000))'(1)
E'((00)(000))'(11)
E'((00)(000))'(111)
E'((00)(000))'(1111)

++?0'!%  [ The universal computer U ]
         [ Put together program for U : ]

^#~'     [ Show crucial prefix so that we can size it ]
"(       [ begin literally ]

[Count number of bits in an omega that are determined.]
: (Cw) /.w() /=X+w C-w *1C-w

[Merge bits of data into unknown bits of an omega.]
: (Mw) /.w() * /=X+w@+w M-w

[(<xy) tells if x is less than y.]
: (<xy) /.y1 /.x1 <-x-y    [[FORCED TO "TRUE" FOR TEST]]

[
 Examine omegas in list w to see if in any one of them
 the number of bits that are determined is greater than n.
 Returns 0 to indicate not found, or what it found.
]
: (Ewn) /.w 0 /<nC+w +w E-wn

[ (2n) = convert reverse binary to unary ]
: (2n) /.n() :k2-n /+n *1^kk ^kk

[We know that H(Omega_n) > n - 1883]
[Let k be 1883 plus the size of this program in bits = 1883 + 1309 = 3192]
: k 2'(000 111 100 11)  [ k = 3192 base 10 = 6170 base 8 ]
[
 Size of this program is 1309 + bits of FAS read + missing bits of Omega.
 Program tries to output this many + 1883 bits of Omega.
 But that would give us a program whose output is more complex
 than the size of the program.  Contradiction!
 Thus this program won't find what it is looking for.
 So FAS of complexity N cannot determine > N + 1883 + 1309 bits of Omega,
 i.e., > N + 3192 bits of Omega.
]
[Main Loop: t is depth limit (time),
            b is bits of FAS read so far (buffer).]
:(Ltb)
 :v      ?t'!%b     [Run FAS again.]
 :s      E-v^kb     [Look for an omega with >
                     (size of this program + 1883) bits determined.]
 /s      Ms         [Found it!  Merge in undetermined bits,
                     output result, and halt.]
 /="!+v  Lt^b*@()   [Read another bit of FAS.]
 /="?+v  L*1tb      [Increase depth/time limit.]
 "?                 [Surprise, formal system halts,
                     so we do too.]

L()()               [Initially, 0 depth limit
                     and no bits read.]
 )                  [ end literally ]

^#'"(
,'(1X0) [Toy formal system with only one theorem.]
    )

'(0) [Missing bit of omega that is needed.]
\end{verbatim}
}\chap{godel3.r}{\Size\begin{verbatim}
show.c

LISP Interpreter Run

[[[
 Show that a formal system of complexity N
 can't determine more than N + 3192 bits of Omega.
 Formal system is a never halting lisp expression
 that displays lists of the form (10X0XXXX10).
 This stands for the fractional part of Omega,
 and means that these 0,1 bits of Omega are known.
 X stands for an unknown bit.
]]]

[Count number of bits in an omega that are determined.]
& (Cw) /.w() /=X+w C-w *1C-w

C:          (&(w)(/(.w)()(/(=X(+w))(C(-w))(*1(C(-w))))))

C'(XXX)

expression  (C('(XXX)))
value       ()

C'(1XX)

expression  (C('(1XX)))
value       (1)

C'(1X0)

expression  (C('(1X0)))
value       (11)

C'(110)

expression  (C('(110)))
value       (111)


[Merge bits of data into unknown bits of an omega.]
& (Mw) /.w() * /=X+w@+w M-w

M:          (&(w)(/(.w)()(*(/(=X(+w))(@)(+w))(M(-w)))))

[Test it.]
++?0 ':(Mw)/.w()*/=X+w@+wM-w M'(00X00X00X) '(111)

expression  (+(+(?0('(('(&((Mw))(M('(00X00X00X)))))('(&()(/(.w
            )()(*(/(=X(+w))(@)(+w))(M(-w))))))))('(111)))))
value       M

++?0 ':(Mw)/.w()*/=X+w@+wM-w M'(11X11X111) '(00)

expression  (+(+(?0('(('(&((Mw))(M('(11X11X111)))))('(&()(/(.w
            )()(*(/(=X(+w))(@)(+w))(M(-w))))))))('(00)))))
value       M


[(<xy) tells if x is less than y.]
& (<xy) /.y0 /.x1 <-x-y

<:          (&(xy)(/(.y)0(/(.x)1(<(-x)(-y)))))

<'(11)'(11)

expression  (<('(11))('(11)))
value       0

<'(11)'(111)

expression  (<('(11))('(111)))
value       1

<'(111)'(11)

expression  (<('(111))('(11)))
value       0


[
 Examine omegas in list w to see if in any one of them
 the number of bits that are determined is greater than n.
 Returns 0 to indicate not found, or what it found.
]
& (Ewn) /.w 0 /<nC+w +w E-wn

E:          (&(wn)(/(.w)0(/(<n(C(+w)))(+w)(E(-w)n))))

E'((00)(000))'()

expression  (E('((00)(000)))('()))
value       (00)

E'((00)(000))'(1)

expression  (E('((00)(000)))('(1)))
value       (00)

E'((00)(000))'(11)

expression  (E('((00)(000)))('(11)))
value       (000)

E'((00)(000))'(111)

expression  (E('((00)(000)))('(111)))
value       0

E'((00)(000))'(1111)

expression  (E('((00)(000)))('(1111)))
value       0


++?0'!%  [ The universal computer U ]
         [ Put together program for U : ]

^#~'     [ Show crucial prefix so that we can size it ]
"(       [ begin literally ]

[Count number of bits in an omega that are determined.]
: (Cw) /.w() /=X+w C-w *1C-w

[Merge bits of data into unknown bits of an omega.]
: (Mw) /.w() * /=X+w@+w M-w

[(<xy) tells if x is less than y.]
: (<xy) /.y1 /.x1 <-x-y    [[FORCED TO "TRUE" FOR TEST]]

[
 Examine omegas in list w to see if in any one of them
 the number of bits that are determined is greater than n.
 Returns 0 to indicate not found, or what it found.
]
: (Ewn) /.w 0 /<nC+w +w E-wn

[ (2n) = convert reverse binary to unary ]
: (2n) /.n() :k2-n /+n *1^kk ^kk

[We know that H(Omega_n) > n - 1883]
[Let k be 1883 plus the size of this program in bits = 1883 + 1309 = 3192]
: k 2'(000 111 100 11)  [ k = 3192 base 10 = 6170 base 8 ]
[
 Size of this program is 1309 + bits of FAS read + missing bits of Omega.
 Program tries to output this many + 1883 bits of Omega.
 But that would give us a program whose output is more complex
 than the size of the program.  Contradiction!
 Thus this program won't find what it is looking for.
 So FAS of complexity N cannot determine > N + 1883 + 1309 bits of Omega,
 i.e., > N + 3192 bits of Omega.
]
[Main Loop: t is depth limit (time),
            b is bits of FAS read so far (buffer).]
:(Ltb)
 :v      ?t'!%b     [Run FAS again.]
 :s      E-v^kb     [Look for an omega with >
                     (size of this program + 1883) bits determined.]
 /s      Ms         [Found it!  Merge in undetermined bits,
                     output result, and halt.]
 /="!+v  Lt^b*@()   [Read another bit of FAS.]
 /="?+v  L*1tb      [Increase depth/time limit.]
 "?                 [Surprise, formal system halts,
                     so we do too.]

L()()               [Initially, 0 depth limit
                     and no bits read.]
 )                  [ end literally ]

^#'"(
,'(1X0) [Toy formal system with only one theorem.]
    )

'(0) [Missing bit of omega that is needed.]

expression  (+(+(?0('(!(%)))(^(#(~('(:(Cw)/.w()/=X+wC-w*1C-w:(
            Mw)/.w()*/=X+w@+wM-w:(<xy)/.y1/.x1<-x-y:(Ewn)/.w0/
            <nC+w+wE-wn:(2n)/.n():k2-n/+n*1^kk^kk:k2'(00011110
            011):(Ltb):v?t'!%b:sE-v^kb/sMs/="!+vLt^b*@()/="?+v
            L*1tb"?L()()))))(^(#('(,'(1X0))))('(0)))))))
show        (:(Cw)/.w()/=X+wC-w*1C-w:(Mw)/.w()*/=X+w@+wM-w:(<x
            y)/.y1/.x1<-x-y:(Ewn)/.w0/<nC+w+wE-wn:(2n)/.n():k2
            -n/+n*1^kk^kk:k2'(00011110011):(Ltb):v?t'!%b:sE-v^
            kb/sMs/="!+vLt^b*@()/="?+vL*1tb"?L()())
size        187(273)/1309(2435)
value       (100)

End of LISP Run

Elapsed time is 0 seconds.
\end{verbatim}
}\chap{godel4.l}{\Size\begin{verbatim}
[[[
 Show that a formal system of complexity N
 can't determine more than N + 3192 bits of Omega.
 Formal system is a never halting lisp expression
 that displays lists of the form (10X0XXXX10).
 This stands for the fractional part of Omega,
 and means that these 0,1 bits of Omega are known.
 X stands for an unknown bit.
]]]

[Count number of bits in an omega that are determined.]
& (Cw) /.w() /=X+w C-w *1C-w
C'(XXX)
C'(1XX)
C'(1X0)
C'(110)

[Merge bits of data into unknown bits of an omega.]
& (Mw) /.w() * /=X+w@+w M-w
[Test it.]
++?0 ':(Mw)/.w()*/=X+w@+wM-w M'(00X00X00X) '(111)
++?0 ':(Mw)/.w()*/=X+w@+wM-w M'(11X11X111) '(00)

[(<xy) tells if x is less than y.]
& (<xy) /.y0 /.x1 <-x-y
<'(11)'(11)
<'(11)'(111)
<'(111)'(11)

[
 Examine omegas in list w to see if in any one of them
 the number of bits that are determined is greater than n.
 Returns 0 to indicate not found, or what it found.
]
& (Ewn) /.w 0 /<nC+w +w E-wn
E'((00)(000))'()
E'((00)(000))'(1)
E'((00)(000))'(11)
E'((00)(000))'(111)
E'((00)(000))'(1111)

++?0'!%  [ The universal computer U ]
         [ Put together program for U : ]

^#~'     [ Show crucial prefix so that we can size it ]
"(       [ begin literally ]

[Count number of bits in an omega that are determined.]
: (Cw) /.w() /=X+w C-w *1C-w

[Merge bits of data into unknown bits of an omega.]
: (Mw) /.w() * /=X+w@+w M-w

[(<xy) tells if x is less than y.]
: (<xy) /.y0 /.x1 <-x-y    [[CORRECT DEFINITION OF <]]

[
 Examine omegas in list w to see if in any one of them
 the number of bits that are determined is greater than n.
 Returns 0 to indicate not found, or what it found.
]
: (Ewn) /.w 0 /<nC+w +w E-wn

[ (2n) = convert reverse binary to unary ]
: (2n) /.n() :k2-n /+n *1^kk ^kk

[We know that H(Omega_n) > n - 1883]
[Let k be 1883 plus the size of this program in bits = 1883 + 1309 = 3192]
: k 2'(000 111 100 11)  [ k = 3192 base 10 = 6170 base 8 ]
[
 Size of this program is 1309 + bits of FAS read + missing bits of Omega.
 Program tries to output this many + 1883 bits of Omega.
 But that would give us a program whose output is more complex
 than the size of the program.  Contradiction!
 Thus this program won't find what it is looking for.
 So FAS of complexity N cannot determine > N + 1883 + 1309 bits of Omega,
 i.e., > N + 3192 bits of Omega.
]
[Main Loop: t is depth limit (time),
            b is bits of FAS read so far (buffer).]
:(Ltb)
 :v      ?t'!%b     [Run FAS again.]
 :s      E-v^kb     [Look for an omega with >
                     (size of this program + 1883) bits determined.]
 /s      Ms         [Found it!  Merge in undetermined bits,
                     output result, and halt.]
 /="!+v  Lt^b*@()   [Read another bit of FAS.]
 /="?+v  L*1tb      [Increase depth/time limit.]
 "?                 [Surprise, formal system halts,
                     so we do too.]

L()()               [Initially, 0 depth limit
                     and no bits read.]
 )                  [ end literally ]

^#'"(
,'(1X0) [Toy formal system with only one theorem.]
    )

'(0) [Missing bit of omega that is needed.]
\end{verbatim}
}\chap{godel4.r}{\Size\begin{verbatim}
show.c

LISP Interpreter Run

[[[
 Show that a formal system of complexity N
 can't determine more than N + 3192 bits of Omega.
 Formal system is a never halting lisp expression
 that displays lists of the form (10X0XXXX10).
 This stands for the fractional part of Omega,
 and means that these 0,1 bits of Omega are known.
 X stands for an unknown bit.
]]]

[Count number of bits in an omega that are determined.]
& (Cw) /.w() /=X+w C-w *1C-w

C:          (&(w)(/(.w)()(/(=X(+w))(C(-w))(*1(C(-w))))))

C'(XXX)

expression  (C('(XXX)))
value       ()

C'(1XX)

expression  (C('(1XX)))
value       (1)

C'(1X0)

expression  (C('(1X0)))
value       (11)

C'(110)

expression  (C('(110)))
value       (111)


[Merge bits of data into unknown bits of an omega.]
& (Mw) /.w() * /=X+w@+w M-w

M:          (&(w)(/(.w)()(*(/(=X(+w))(@)(+w))(M(-w)))))

[Test it.]
++?0 ':(Mw)/.w()*/=X+w@+wM-w M'(00X00X00X) '(111)

expression  (+(+(?0('(('(&((Mw))(M('(00X00X00X)))))('(&()(/(.w
            )()(*(/(=X(+w))(@)(+w))(M(-w))))))))('(111)))))
value       M

++?0 ':(Mw)/.w()*/=X+w@+wM-w M'(11X11X111) '(00)

expression  (+(+(?0('(('(&((Mw))(M('(11X11X111)))))('(&()(/(.w
            )()(*(/(=X(+w))(@)(+w))(M(-w))))))))('(00)))))
value       M


[(<xy) tells if x is less than y.]
& (<xy) /.y0 /.x1 <-x-y

<:          (&(xy)(/(.y)0(/(.x)1(<(-x)(-y)))))

<'(11)'(11)

expression  (<('(11))('(11)))
value       0

<'(11)'(111)

expression  (<('(11))('(111)))
value       1

<'(111)'(11)

expression  (<('(111))('(11)))
value       0


[
 Examine omegas in list w to see if in any one of them
 the number of bits that are determined is greater than n.
 Returns 0 to indicate not found, or what it found.
]
& (Ewn) /.w 0 /<nC+w +w E-wn

E:          (&(wn)(/(.w)0(/(<n(C(+w)))(+w)(E(-w)n))))

E'((00)(000))'()

expression  (E('((00)(000)))('()))
value       (00)

E'((00)(000))'(1)

expression  (E('((00)(000)))('(1)))
value       (00)

E'((00)(000))'(11)

expression  (E('((00)(000)))('(11)))
value       (000)

E'((00)(000))'(111)

expression  (E('((00)(000)))('(111)))
value       0

E'((00)(000))'(1111)

expression  (E('((00)(000)))('(1111)))
value       0


++?0'!%  [ The universal computer U ]
         [ Put together program for U : ]

^#~'     [ Show crucial prefix so that we can size it ]
"(       [ begin literally ]

[Count number of bits in an omega that are determined.]
: (Cw) /.w() /=X+w C-w *1C-w

[Merge bits of data into unknown bits of an omega.]
: (Mw) /.w() * /=X+w@+w M-w

[(<xy) tells if x is less than y.]
: (<xy) /.y0 /.x1 <-x-y    [[CORRECT DEFINITION OF <]]

[
 Examine omegas in list w to see if in any one of them
 the number of bits that are determined is greater than n.
 Returns 0 to indicate not found, or what it found.
]
: (Ewn) /.w 0 /<nC+w +w E-wn

[ (2n) = convert reverse binary to unary ]
: (2n) /.n() :k2-n /+n *1^kk ^kk

[We know that H(Omega_n) > n - 1883]
[Let k be 1883 plus the size of this program in bits = 1883 + 1309 = 3192]
: k 2'(000 111 100 11)  [ k = 3192 base 10 = 6170 base 8 ]
[
 Size of this program is 1309 + bits of FAS read + missing bits of Omega.
 Program tries to output this many + 1883 bits of Omega.
 But that would give us a program whose output is more complex
 than the size of the program.  Contradiction!
 Thus this program won't find what it is looking for.
 So FAS of complexity N cannot determine > N + 1883 + 1309 bits of Omega,
 i.e., > N + 3192 bits of Omega.
]
[Main Loop: t is depth limit (time),
            b is bits of FAS read so far (buffer).]
:(Ltb)
 :v      ?t'!%b     [Run FAS again.]
 :s      E-v^kb     [Look for an omega with >
                     (size of this program + 1883) bits determined.]
 /s      Ms         [Found it!  Merge in undetermined bits,
                     output result, and halt.]
 /="!+v  Lt^b*@()   [Read another bit of FAS.]
 /="?+v  L*1tb      [Increase depth/time limit.]
 "?                 [Surprise, formal system halts,
                     so we do too.]

L()()               [Initially, 0 depth limit
                     and no bits read.]
 )                  [ end literally ]

^#'"(
,'(1X0) [Toy formal system with only one theorem.]
    )

'(0) [Missing bit of omega that is needed.]

expression  (+(+(?0('(!(%)))(^(#(~('(:(Cw)/.w()/=X+wC-w*1C-w:(
            Mw)/.w()*/=X+w@+wM-w:(<xy)/.y0/.x1<-x-y:(Ewn)/.w0/
            <nC+w+wE-wn:(2n)/.n():k2-n/+n*1^kk^kk:k2'(00011110
            011):(Ltb):v?t'!%b:sE-v^kb/sMs/="!+vLt^b*@()/="?+v
            L*1tb"?L()()))))(^(#('(,'(1X0))))('(0)))))))
show        (:(Cw)/.w()/=X+wC-w*1C-w:(Mw)/.w()*/=X+w@+wM-w:(<x
            y)/.y0/.x1<-x-y:(Ewn)/.w0/<nC+w+wE-wn:(2n)/.n():k2
            -n/+n*1^kk^kk:k2'(00011110011):(Ltb):v?t'!%b:sE-v^
            kb/sMs/="!+vLt^b*@()/="?+vL*1tb"?L()())
size        187(273)/1309(2435)
value       ?

End of LISP Run

Elapsed time is 0 seconds.
\end{verbatim}
}\chap{godel5.l}{\Size\begin{verbatim}
[[[
 COROLLARY EXTRACTED FROM GODEL3/4 & OMEGA2:
 Consider a partial determination of Omega, e.g.,
 a lisp expression of the form (10X0XXXX10).
 This stands for the fractional part of Omega,
 and means that these 0,1 bits of Omega are known.
 X stands for an unknown bit.
 Then the complexity H of a partial determination
 of Omega is greater than the number of bits that
 are determined minus 1883 + 175 = 2058.
 H((10X0XXXX10)) > number of bits determined - 2058.
]]]

++?0'!%  [ The universal computer U ]
         [ Put together program for U : ]

^#~'     [ Show crucial prefix so that we can size it ]
"(       [ begin literally ]

[Merge bits of data into unknown bits of an omega.]
: (Mw) /.w() * /=X+w@+w M-w

       M!%        [Merge in undetermined bits,
                    output result, and halt.]
 )       [ end literally ]

^#'"(
'(10X0XXXX10) [Partial determination of Omega.]
    )

'(11100) [Missing bits of omega that are needed.]
\end{verbatim}
}\chap{godel5.r}{\Size\begin{verbatim}
show.c

LISP Interpreter Run

[[[
 COROLLARY EXTRACTED FROM GODEL3/4 & OMEGA2:
 Consider a partial determination of Omega, e.g.,
 a lisp expression of the form (10X0XXXX10).
 This stands for the fractional part of Omega,
 and means that these 0,1 bits of Omega are known.
 X stands for an unknown bit.
 Then the complexity H of a partial determination
 of Omega is greater than the number of bits that
 are determined minus 1883 + 175 = 2058.
 H((10X0XXXX10)) > number of bits determined - 2058.
]]]

++?0'!%  [ The universal computer U ]
         [ Put together program for U : ]

^#~'     [ Show crucial prefix so that we can size it ]
"(       [ begin literally ]

[Merge bits of data into unknown bits of an omega.]
: (Mw) /.w() * /=X+w@+w M-w

       M!%        [Merge in undetermined bits,
                    output result, and halt.]
 )       [ end literally ]

^#'"(
'(10X0XXXX10) [Partial determination of Omega.]
    )

'(11100) [Missing bits of omega that are needed.]

expression  (+(+(?0('(!(%)))(^(#(~('(:(Mw)/.w()*/=X+w@+wM-wM!%
            ))))(^(#('('(10X0XXXX10))))('(11100)))))))
show        (:(Mw)/.w()*/=X+w@+wM-wM!%)
size        25(31)/175(257)
value       (1010110010)

End of LISP Run

Elapsed time is 0 seconds.
\end{verbatim}
}\chap{lisp.c}{\Size\begin{verbatim}
/* lisp.c: high-speed LISP interpreter */

/*
   The storage required by this interpreter is 8 bytes times
   the symbolic constant SIZE, which is 8 * 16,000,000 =
   128 megabytes.  To run this interpreter in small machines,
   reduce the #define SIZE 16000000 below.

   To compile, type
      cc -O -olisp lisp.c
   To run interactively, type
      lisp
   To run with output on screen, type
      lisp <test.l
   To run with output in file, type
      lisp <test.l >test.r

   Reference:  Kernighan & Ritchie,
   The C Programming Language, Second Edition,
   Prentice-Hall, 1988.
*/

#include <stdio.h>
#include <time.h>

#define SIZE 16000000 /* numbers of nodes of tree storage */
#define LAST_ATOM 128 /* highest integer value of character */
#define nil 128 /* null pointer in tree storage */
#define question -1 /* error pointer in tree storage */
#define exclamation -2 /* error pointer in tree storage */
#define infinity 999999999 /* "infinite" depth limit */

/* For very small PC's, change following line to
   make hd & tl unsigned short instead of long:
   (If so, SIZE must be less than 64K.) */
long hd[SIZE+1], tl[SIZE+1]; /* tree storage */
long next = nil; /* list of free nodes */
long low = LAST_ATOM+1; /* first never-used node */
long vlst[LAST_ATOM+1]; /* bindings of each atom */
long tape; /* Turing machine tapes */
long display; /* display indicators */
long outputs; /* output stacks */
long q; /* for converting expressions to binary */
long col; /* column in each 50 character chunk of output
            (preceeded by 12 char prefix) */
long cc; /* character count */
time_t time1; /* clock at start of execution */
time_t time2; /* clock at end of execution */

long ev(long e); /* initialize and evaluate expression */
void initialize_atoms(void); /* initialize atoms */
void clean_env(void); /* clean environment */
void restore_env(void); /* restore dirty environment */
long eval(long e, long d); /* evaluate expression */
/* evaluate list of expressions */
long evalst(long e, long d);
/* bind values of arguments to formal parameters */
void bind(long vars, long args);
long at(long x); /* atomic predicate */
long jn(long x, long y); /* join head to tail */
long pop(long x); /* return tl & free node */
void fr(long x); /* free list of nodes */
long eq(long x, long y); /* equal predicate */
long cardinality(long x); /* number of elements in list */
long append(long x, long y); /* append two lists */
/* read one square of Turing machine tape */
long getbit(void);
/* read one character from Turing machine tape */
long getchr(void);
/* read expression from Turing machine tape */
void putchr(long x); /* convert character to binary */
void putexp(long x); /* convert expression to binary */
void putexp2(long x); /* convert expression to binary */
long out(char *x, long y); /* output expression */
void out2(long x); /* really output expression */
void out3(long x); /* really really output expression */
long chr2(void); /* read character - skip blanks,
                    tabs and new line characters */
long chr(void); /* read character - skip comments */
long in(long mexp, long rparenokay); /* input m-exp */
long (*pchr)(void); /* pointer to chr2 or getchr */
long p; /* parens associated with each operator for in() */
long p0; /* parens associated with each primitive function */
long p1; /* parens associated with each operator */

main() /* lisp main program */
{
char name_colon[] = "X:"; /* for printing name: def pairs */

time1 = time(NULL); /* start timer */
printf("lisp.c\n\nLISP Interpreter Run\n");
initialize_atoms();
p = nil;
p = jn(jn('@',nil),p);
p = jn(jn('%',nil),p);
p = jn(jn('+',jn('1',nil)),p);
p = jn(jn('-',jn('1',nil)),p);
p = jn(jn('.',jn('1',nil)),p);
p = jn(jn('\'',jn('1',nil)),p);
p = jn(jn(',',jn('1',nil)),p);
p = jn(jn('!',jn('1',nil)),p);
p = jn(jn('#',jn('1',nil)),p);
p = jn(jn('~',jn('1',nil)),p);
p = jn(jn('*',jn('1',jn('2',nil))),p);
p = jn(jn('=',jn('1',jn('2',nil))),p);
p = jn(jn('&',jn('1',jn('2',nil))),p);
p = jn(jn('^',jn('1',jn('2',nil))),p);
p = jn(jn('/',jn('1',jn('2',jn('3',nil)))),p);
p = jn(jn(':',jn('1',jn('2',jn('3',nil)))),p);
p = jn(jn('?',jn('1',jn('2',jn('3',nil)))),p);
p1 = p0 = p;

while (1) {
      long e, f, name, def;
      printf("\n");
      /* read lisp meta-expression from stdin, ) not okay */
      pchr = chr2; cc = 0; p = p1; e = in(1,0); p1 = p;
      /* flush rest of input line */
      while (putchar(getchar()) != '\n');
      printf("\n");
      f = hd[e];
      name = hd[tl[e]];
      def = hd[tl[tl[e]]];
      if (f == '&') {
      /* definition */
         if (at(name)) {
         /* variable definition, e.g., & x '(abc) */
            def = out("expression",def);
            def = ev(def);
         } /* end of variable definition */
         else          {
         /* function definition, e.g., & (Fxy) *x*y() */
            long var_list = tl[name];
            name = hd[name];
            def = jn('&',jn(var_list,jn(def,nil)));
         } /* end of function definition */
         name_colon[0] = name;
         out(name_colon,def);
         /* new binding replaces old */
         vlst[name] = jn(def,nil);
         continue;
      } /* end of definition */
      /* write corresponding s-expression */
      e = out("expression",e);
      /* evaluate expression */
      e = out("value",ev(e));
   }
}

long ev(long e) /* initialize and evaluate expression */
{
 long d = infinity; /* "infinite" depth limit */
 long v;
 tape = jn(nil,nil);
 display = jn('Y',nil);
 outputs = jn(nil,nil);
 v = eval(e,d);
 if (v == question) v = '?';
 if (v == exclamation) v = '!';
 return v;
}

void initialize_atoms(void) /* initialize atoms */
{
 long i;
 for (i = 0; i <= LAST_ATOM; ++i) {
 hd[i] = tl[i] = i; /* so that hd & tl of atom = atom */
 /* initially each atom evaluates to self */
 vlst[i] = jn(i,nil);
 }
}

long jn(long x, long y) /* join two lists */
{
 long z;
 /* if y is not a list, then jn is x */
 if ( y != nil && at(y) ) return x;

 if (next == nil) {
  if (low > SIZE) {
  printf("Storage overflow!\n");
  exit(0);
  }
 next = low++;
 tl[next] = nil;
 }

 z = next;
 next = tl[next];
 hd[z] = x;
 tl[z] = y;

 return z;
}

long pop(long x) /* return tl & free node */
{
 long y;
 y = tl[x];
 tl[x] = next;
 next = x;
 return y;
}

void fr(long x) /* free list of nodes */
{
 while (x != nil) x = pop(x);
}

long at(long x) /* atom predicate */
{
 return ( x <= LAST_ATOM );
}

long eq(long x, long y) /* equal predicate */
{
 if (x == y) return 1;
 if (at(x)) return 0;
 if (at(y)) return 0;
 if (eq(hd[x],hd[y])) return eq(tl[x],tl[y]);
 return 0;
}

long eval(long e, long d) /* evaluate expression */
{
/*
 e is expression to be evaluated
 d is permitted depth - integer, not pointer to tree storage
*/
 long f, v, args, x, y, z, vars, body;

 /* find current binding of atomic expression */
 if (at(e)) return hd[vlst[e]];

 f = eval(hd[e],d); /* evaluate function */
 e = tl[e]; /* remove function from list of arguments */
 if (f < 0) return f; /* function = error value? */

 if (f == '\'') return hd[e]; /* quote */

 if (f == '/') { /* if then else */
 v = eval(hd[e],d);
 e = tl[e];
 if (v < 0) return v; /* error? */
 if (v == '0') e = tl[e];
 return eval(hd[e],d);
 }

 args = evalst(e,d); /* evaluate list of arguments */
 if (args < 0) return args; /* error? */

 x = hd[args]; /* pick up first argument */
 y = hd[tl[args]]; /* pick up second argument */
 z = hd[tl[tl[args]]]; /* pick up third argument */

 switch (f) {
 case '@': {fr(args); return getbit();}
         /* read lisp meta-expression from TM tape, ) not okay */
 case '%': {fr(args); pchr = getchr; cc = 1; p = p0; return in(1,0);}
 case '#': {fr(args);
           v = q = jn(nil,nil); putexp(x); return pop(v);}
 case '+': {fr(args); return hd[x];}
 case '-': {fr(args); return tl[x];}
 case '.': {fr(args); return (at(x) ? '1' : '0');}
 case ',': {fr(args); hd[outputs] = jn(x,hd[outputs]);
           return (hd[display] == 'Y' ? out("display",x): x);}
 case '~': {fr(args); return /* out("show", */ x /* ) */ ;}
 case '=': {fr(args); return (eq(x,y) ? '1' : '0');}
 case '*': {fr(args); return jn(x,y);}
 case '^': {fr(args);
           return append((at(x)?nil:x),(at(y)?nil:y));}
 }

 if (d == 0) {fr(args); return question;} /* depth exceeded
                                             -> error! */
 d--; /* decrement depth */

 if (f == '!') {
 fr(args);
 clean_env(); /* clean environment */
 v = eval(x,d);
 restore_env(); /* restore unclean environment */
 return v;
 }

 if (f == '?') {
 fr(args);
 x = cardinality(x); /* convert s-exp into number */
 clean_env();
 tape = jn(z,tape);
 display = jn('N',display);
 outputs = jn(nil,outputs);
 v = eval(y,(d <= x ? d : x));
 restore_env();
 z = hd[outputs];
 tape = pop(tape);
 display = pop(display);
 outputs = pop(outputs);
 if (v == question) return (d <= x ? question : jn('?',z));
 if (v == exclamation) return jn('!',z);
 return jn(jn(v,nil),z);
 }

 f = tl[f];
 vars = hd[f];
 f = tl[f];
 body = hd[f];

 bind(vars,args);
 fr(args);

 v = eval(body,d);

 /* unbind */
 while (!at(vars)) {
 if (at(hd[vars]))
 vlst[hd[vars]] = pop(vlst[hd[vars]]);
 vars = tl[vars];
 }

 return v;
}

void clean_env(void) /* clean environment */
{
 long i;
 for (i = 0; i <= LAST_ATOM; ++i)
 vlst[i] = jn(i,vlst[i]); /* clean environment */
}

void restore_env(void) /* restore unclean environment */
{
 long i;
 for (i = 0; i <= LAST_ATOM; ++i)
 vlst[i] = pop(vlst[i]); /* restore unclean environment */
}

long cardinality(long x) /* number of elements in list */
{
 if (at(x)) return (x == nil ? 0 : infinity);
 return 1+cardinality(tl[x]);
}

/* bind values of arguments to formal parameters */
void bind(long vars, long args)
{
 if (at(vars)) return;
 bind(tl[vars],tl[args]);
 if (at(hd[vars]))
 vlst[hd[vars]] = jn(hd[args],vlst[hd[vars]]);
}

long evalst(long e, long d) /* evaluate list of expressions */
{
 long x, y;
 if (at(e)) return nil;
 x = eval(hd[e],d);
 if (x < 0) return x; /* error? */
 y = evalst(tl[e],d);
 if (y < 0) return y; /* error? */
 return jn(x,y);
}

long append(long x, long y) /* append two lists */
{
 if (at(x)) return y;
 return jn(hd[x],append(tl[x],y));
}

/* read one square of Turing machine tape */
long getbit(void)
{
 long x;
 if (at(hd[tape])) return exclamation; /* tape finished ! */
 x = hd[hd[tape]];
 hd[tape] = tl[hd[tape]];
 return (x == '0' ? '0' : '1');
}

/* read one character from Turing machine tape */
long getchr(void)
{
 long c, b, i;
 c = 0;
 do {
    for (i = 0; i < 7; ++i) {
    b = getbit();
    if (b < 0) return b; /* error? */
    c = c + c + b - '0';
    }
 }
/* keep only non-blank printable ASCII codes */
 while (c >= 127 || c <= 32) ;
 return c;
}

void putchr(long x) /* convert character to binary */
{
 q = tl[q] = jn(( x &  64 ? '1' : '0' ), nil);
 q = tl[q] = jn(( x &  32 ? '1' : '0' ), nil);
 q = tl[q] = jn(( x &  16 ? '1' : '0' ), nil);
 q = tl[q] = jn(( x &   8 ? '1' : '0' ), nil);
 q = tl[q] = jn(( x &   4 ? '1' : '0' ), nil);
 q = tl[q] = jn(( x &   2 ? '1' : '0' ), nil);
 q = tl[q] = jn(( x &   1 ? '1' : '0' ), nil);
}

void putexp(long x) /* convert expression to binary */
{
 /* remove containing parens at top level! */
 while (!at(x)) {
 putexp2(hd[x]);
 x = tl[x];
 }
}

void putexp2(long x) /* convert expression to binary */
{
 if ( at(x) && x != nil ) {putchr(x); return;}
 putchr('(');

 while (!at(x)) {
 putexp2(hd[x]);
 x = tl[x];
 }

 putchr(')');
}

long out(char *x, long y) /* output expression */
{
   printf("%-12s",x);
   col = 0; /* so can insert \n and 12 blanks
               every 50 characters of output */
   cc = -2; /* count characters in m-expression */
   out2(y);
   printf("\n");
   if (*x == 's' && cc > 0)
   printf("%-12s%d(%o)/%d(%o)\n","size",cc,cc,7*cc,7*cc);
   return y;
}

void out2(long x) /* really output expression */
{
   if ( at(x) && x != nil ) {out3(x); return;}
   out3('(');
   while (!at(x)) {
      out2(hd[x]);
      x = tl[x];
      }
   out3(')');
}

void out3(long x) /* really really output expression */
{
   if (col++ == 50) {printf("\n%-12s"," ");  col = 1;}
   putchar(x);
   cc++;
}

long chr2(void) /* read character - skip blanks,
                   tabs and new line characters */
{
   long c;
   do {
      c = getchar();
      if (c == EOF) {
         time2 = time(NULL);
         printf(
         "End of LISP Run\n\nElapsed time is %.0f seconds.\n",
         difftime(time2,time1)
      /* on some systems, above line should instead be: */
      /* time2 - time1 */
         );
         exit(0); /* terminate execution */
         }
      putchar(c);
   }
/* keep only non-blank printable ASCII codes */
   while (c >= 127 || c <= 32) ;
   return c;
}

long chr(void) /* read character - skip comments */
{              /* here pchr -> chr2, getchr */
   long c;
   while (1) {
      c = (*pchr)();
      if (c < 0) return c; /* error? */
      if (c != '[') return c;
   /* comments may be nested */
      while ((c = chr()) != ']')
      if (c < 0) return c; /* error? */
   }
}

/* read expression from Turing machine tape */
long in(long mexp, long rparenokay) /* input m-exp */
{
   long c = chr();
   if (c < 0) return c; /* error? */
   cc++; /* bump character count */
   if (c == ')') if (rparenokay) return ')'; else return nil;
   if (c == '(') { /* explicit list */
      long first, last, next;
      first = last = jn(nil,nil);
      while ((next = in(mexp,1)) != ')')
      {
      if (next < 0) return next; /* error? */
      last = tl[last] = jn(next,nil);
      }
      return pop(first);
      }
   if (!mexp) return c; /* atom */
   if (c == '{') { /* number */
      long n = 0, u;
      while ((c = chr()) != '}') {
      if (c < 0) return c; /* error? */
      c = c - '0';
      if (c >= 0 && c <= 9) n = 10 * n + c;
      }
      for (u = nil; n > 0; n--) u = jn('1',u);
      return u;
      }
   if (c == '"') return in(0,0); /* s-exp */
   if (c == '&' && cc == 1) {
   /* expand "define" only if & is first character */
      long name, def;
      name = in(0,0); if (name < 0) return name; /* error? */
      p = jn(name,p);
      def  = in(1,0); if (def  < 0) return def ; /* error? */
      return jn('&',jn(name,jn(def,nil)));
      }
   if (c == ':') { /* expand "let" */
      long name, def, body;
      name = in(0,0); if (name < 0) return name; /* error? */
      p = jn(name,p);
      def  = in(1,0); if (def  < 0) return def ; /* error? */
      body = in(1,0); if (body < 0) return body; /* error? */
      p = pop(p);
      if (!at(name)) {
         long var_list;
         var_list = tl[name];
         name = hd[name];
         def =
         jn('\'',jn(jn('&',jn(var_list,jn(def,nil))),nil));
         }
      return
      jn(
       jn('\'',jn(jn('&',jn(jn(name,nil),jn(body,nil))),nil)),
       jn(def,nil)
      );
      }
  {long p2;
   for (p2 = p; !at(p2); p2 = tl[p2]) {
      long p3 = hd[p2];
      if (p3 == c) return c;
      if (hd[p3] == c) {
         long first, last, next;
         first = last = jn(c,nil);
         for (p3 = tl[p3]; !at(p3); p3 = tl[p3]) {
            next = in(1,0);
            if (next < 0) return next; /* error? */
            last = tl[last] = jn(next,nil);
            }
         return first;
         }
      }
   return c;
   }
}
\end{verbatim}
}\chap{show.c}{\Size\begin{verbatim}
/* show.c: high-speed LISP interpreter */

/*
   The storage required by this interpreter is 8 bytes times
   the symbolic constant SIZE, which is 8 * 16,000,000 =
   128 megabytes.  To run this interpreter in small machines,
   reduce the #define SIZE 16000000 below.

   To compile, type
      cc -O -oshow show.c
   To run interactively, type
      show
   To run with output on screen, type
      show <test.l
   To run with output in file, type
      show <test.l >test.r

   Reference:  Kernighan & Ritchie,
   The C Programming Language, Second Edition,
   Prentice-Hall, 1988.
*/

#include <stdio.h>
#include <time.h>

#define SIZE 16000000 /* numbers of nodes of tree storage */
#define LAST_ATOM 128 /* highest integer value of character */
#define nil 128 /* null pointer in tree storage */
#define question -1 /* error pointer in tree storage */
#define exclamation -2 /* error pointer in tree storage */
#define infinity 999999999 /* "infinite" depth limit */

/* For very small PC's, change following line to
   make hd & tl unsigned short instead of long:
   (If so, SIZE must be less than 64K.) */
long hd[SIZE+1], tl[SIZE+1]; /* tree storage */
long next = nil; /* list of free nodes */
long low = LAST_ATOM+1; /* first never-used node */
long vlst[LAST_ATOM+1]; /* bindings of each atom */
long tape; /* Turing machine tapes */
long display; /* display indicators */
long outputs; /* output stacks */
long q; /* for converting expressions to binary */
long col; /* column in each 50 character chunk of output
            (preceeded by 12 char prefix) */
long cc; /* character count */
time_t time1; /* clock at start of execution */
time_t time2; /* clock at end of execution */

long ev(long e); /* initialize and evaluate expression */
void initialize_atoms(void); /* initialize atoms */
void clean_env(void); /* clean environment */
void restore_env(void); /* restore dirty environment */
long eval(long e, long d); /* evaluate expression */
/* evaluate list of expressions */
long evalst(long e, long d);
/* bind values of arguments to formal parameters */
void bind(long vars, long args);
long at(long x); /* atomic predicate */
long jn(long x, long y); /* join head to tail */
long pop(long x); /* return tl & free node */
void fr(long x); /* free list of nodes */
long eq(long x, long y); /* equal predicate */
long cardinality(long x); /* number of elements in list */
long append(long x, long y); /* append two lists */
/* read one square of Turing machine tape */
long getbit(void);
/* read one character from Turing machine tape */
long getchr(void);
/* read expression from Turing machine tape */
void putchr(long x); /* convert character to binary */
void putexp(long x); /* convert expression to binary */
void putexp2(long x); /* convert expression to binary */
long out(char *x, long y); /* output expression */
void out2(long x); /* really output expression */
void out3(long x); /* really really output expression */
long chr2(void); /* read character - skip blanks,
                    tabs and new line characters */
long chr(void); /* read character - skip comments */
long in(long mexp, long rparenokay); /* input m-exp */
long (*pchr)(void); /* pointer to chr2 or getchr */
long p; /* parens associated with each operator for in() */
long p0; /* parens associated with each primitive function */
long p1; /* parens associated with each operator */

main() /* lisp main program */
{
char name_colon[] = "X:"; /* for printing name: def pairs */

time1 = time(NULL); /* start timer */
printf("show.c\n\nLISP Interpreter Run\n");
initialize_atoms();
p = nil;
p = jn(jn('@',nil),p);
p = jn(jn('%',nil),p);
p = jn(jn('+',jn('1',nil)),p);
p = jn(jn('-',jn('1',nil)),p);
p = jn(jn('.',jn('1',nil)),p);
p = jn(jn('\'',jn('1',nil)),p);
p = jn(jn(',',jn('1',nil)),p);
p = jn(jn('!',jn('1',nil)),p);
p = jn(jn('#',jn('1',nil)),p);
p = jn(jn('~',jn('1',nil)),p);
p = jn(jn('*',jn('1',jn('2',nil))),p);
p = jn(jn('=',jn('1',jn('2',nil))),p);
p = jn(jn('&',jn('1',jn('2',nil))),p);
p = jn(jn('^',jn('1',jn('2',nil))),p);
p = jn(jn('/',jn('1',jn('2',jn('3',nil)))),p);
p = jn(jn(':',jn('1',jn('2',jn('3',nil)))),p);
p = jn(jn('?',jn('1',jn('2',jn('3',nil)))),p);
p1 = p0 = p;

while (1) {
      long e, f, name, def;
      printf("\n");
      /* read lisp meta-expression from stdin, ) not okay */
      pchr = chr2; cc = 0; p = p1; e = in(1,0); p1 = p;
      /* flush rest of input line */
      while (putchar(getchar()) != '\n');
      printf("\n");
      f = hd[e];
      name = hd[tl[e]];
      def = hd[tl[tl[e]]];
      if (f == '&') {
      /* definition */
         if (at(name)) {
         /* variable definition, e.g., & x '(abc) */
            def = out("expression",def);
            def = ev(def);
         } /* end of variable definition */
         else          {
         /* function definition, e.g., & (Fxy) *x*y() */
            long var_list = tl[name];
            name = hd[name];
            def = jn('&',jn(var_list,jn(def,nil)));
         } /* end of function definition */
         name_colon[0] = name;
         out(name_colon,def);
         /* new binding replaces old */
         vlst[name] = jn(def,nil);
         continue;
      } /* end of definition */
      /* write corresponding s-expression */
      e = out("expression",e);
      /* evaluate expression */
      e = out("value",ev(e));
   }
}

long ev(long e) /* initialize and evaluate expression */
{
 long d = infinity; /* "infinite" depth limit */
 long v;
 tape = jn(nil,nil);
 display = jn('Y',nil);
 outputs = jn(nil,nil);
 v = eval(e,d);
 if (v == question) v = '?';
 if (v == exclamation) v = '!';
 return v;
}

void initialize_atoms(void) /* initialize atoms */
{
 long i;
 for (i = 0; i <= LAST_ATOM; ++i) {
 hd[i] = tl[i] = i; /* so that hd & tl of atom = atom */
 /* initially each atom evaluates to self */
 vlst[i] = jn(i,nil);
 }
}

long jn(long x, long y) /* join two lists */
{
 long z;
 /* if y is not a list, then jn is x */
 if ( y != nil && at(y) ) return x;

 if (next == nil) {
  if (low > SIZE) {
  printf("Storage overflow!\n");
  exit(0);
  }
 next = low++;
 tl[next] = nil;
 }

 z = next;
 next = tl[next];
 hd[z] = x;
 tl[z] = y;

 return z;
}

long pop(long x) /* return tl & free node */
{
 long y;
 y = tl[x];
 tl[x] = next;
 next = x;
 return y;
}

void fr(long x) /* free list of nodes */
{
 while (x != nil) x = pop(x);
}

long at(long x) /* atom predicate */
{
 return ( x <= LAST_ATOM );
}

long eq(long x, long y) /* equal predicate */
{
 if (x == y) return 1;
 if (at(x)) return 0;
 if (at(y)) return 0;
 if (eq(hd[x],hd[y])) return eq(tl[x],tl[y]);
 return 0;
}

long eval(long e, long d) /* evaluate expression */
{
/*
 e is expression to be evaluated
 d is permitted depth - integer, not pointer to tree storage
*/
 long f, v, args, x, y, z, vars, body;

 /* find current binding of atomic expression */
 if (at(e)) return hd[vlst[e]];

 f = eval(hd[e],d); /* evaluate function */
 e = tl[e]; /* remove function from list of arguments */
 if (f < 0) return f; /* function = error value? */

 if (f == '\'') return hd[e]; /* quote */

 if (f == '/') { /* if then else */
 v = eval(hd[e],d);
 e = tl[e];
 if (v < 0) return v; /* error? */
 if (v == '0') e = tl[e];
 return eval(hd[e],d);
 }

 args = evalst(e,d); /* evaluate list of arguments */
 if (args < 0) return args; /* error? */

 x = hd[args]; /* pick up first argument */
 y = hd[tl[args]]; /* pick up second argument */
 z = hd[tl[tl[args]]]; /* pick up third argument */

 switch (f) {
 case '@': {fr(args); return getbit();}
         /* read lisp meta-expression from TM tape, ) not okay */
 case '%': {fr(args); pchr = getchr; cc = 1; p = p0; return in(1,0);}
 case '#': {fr(args);
           v = q = jn(nil,nil); putexp(x); return pop(v);}
 case '+': {fr(args); return hd[x];}
 case '-': {fr(args); return tl[x];}
 case '.': {fr(args); return (at(x) ? '1' : '0');}
 case ',': {fr(args); hd[outputs] = jn(x,hd[outputs]);
           return (hd[display] == 'Y' ? out("display",x): x);}
 case '~': {fr(args); return /* */ out("show", /* */ x /* */ ) /* */ ;}
 case '=': {fr(args); return (eq(x,y) ? '1' : '0');}
 case '*': {fr(args); return jn(x,y);}
 case '^': {fr(args);
           return append((at(x)?nil:x),(at(y)?nil:y));}
 }

 if (d == 0) {fr(args); return question;} /* depth exceeded
                                             -> error! */
 d--; /* decrement depth */

 if (f == '!') {
 fr(args);
 clean_env(); /* clean environment */
 v = eval(x,d);
 restore_env(); /* restore unclean environment */
 return v;
 }

 if (f == '?') {
 fr(args);
 x = cardinality(x); /* convert s-exp into number */
 clean_env();
 tape = jn(z,tape);
 display = jn('N',display);
 outputs = jn(nil,outputs);
 v = eval(y,(d <= x ? d : x));
 restore_env();
 z = hd[outputs];
 tape = pop(tape);
 display = pop(display);
 outputs = pop(outputs);
 if (v == question) return (d <= x ? question : jn('?',z));
 if (v == exclamation) return jn('!',z);
 return jn(jn(v,nil),z);
 }

 f = tl[f];
 vars = hd[f];
 f = tl[f];
 body = hd[f];

 bind(vars,args);
 fr(args);

 v = eval(body,d);

 /* unbind */
 while (!at(vars)) {
 if (at(hd[vars]))
 vlst[hd[vars]] = pop(vlst[hd[vars]]);
 vars = tl[vars];
 }

 return v;
}

void clean_env(void) /* clean environment */
{
 long i;
 for (i = 0; i <= LAST_ATOM; ++i)
 vlst[i] = jn(i,vlst[i]); /* clean environment */
}

void restore_env(void) /* restore unclean environment */
{
 long i;
 for (i = 0; i <= LAST_ATOM; ++i)
 vlst[i] = pop(vlst[i]); /* restore unclean environment */
}

long cardinality(long x) /* number of elements in list */
{
 if (at(x)) return (x == nil ? 0 : infinity);
 return 1+cardinality(tl[x]);
}

/* bind values of arguments to formal parameters */
void bind(long vars, long args)
{
 if (at(vars)) return;
 bind(tl[vars],tl[args]);
 if (at(hd[vars]))
 vlst[hd[vars]] = jn(hd[args],vlst[hd[vars]]);
}

long evalst(long e, long d) /* evaluate list of expressions */
{
 long x, y;
 if (at(e)) return nil;
 x = eval(hd[e],d);
 if (x < 0) return x; /* error? */
 y = evalst(tl[e],d);
 if (y < 0) return y; /* error? */
 return jn(x,y);
}

long append(long x, long y) /* append two lists */
{
 if (at(x)) return y;
 return jn(hd[x],append(tl[x],y));
}

/* read one square of Turing machine tape */
long getbit(void)
{
 long x;
 if (at(hd[tape])) return exclamation; /* tape finished ! */
 x = hd[hd[tape]];
 hd[tape] = tl[hd[tape]];
 return (x == '0' ? '0' : '1');
}

/* read one character from Turing machine tape */
long getchr(void)
{
 long c, b, i;
 c = 0;
 do {
    for (i = 0; i < 7; ++i) {
    b = getbit();
    if (b < 0) return b; /* error? */
    c = c + c + b - '0';
    }
 }
/* keep only non-blank printable ASCII codes */
 while (c >= 127 || c <= 32) ;
 return c;
}

void putchr(long x) /* convert character to binary */
{
 q = tl[q] = jn(( x &  64 ? '1' : '0' ), nil);
 q = tl[q] = jn(( x &  32 ? '1' : '0' ), nil);
 q = tl[q] = jn(( x &  16 ? '1' : '0' ), nil);
 q = tl[q] = jn(( x &   8 ? '1' : '0' ), nil);
 q = tl[q] = jn(( x &   4 ? '1' : '0' ), nil);
 q = tl[q] = jn(( x &   2 ? '1' : '0' ), nil);
 q = tl[q] = jn(( x &   1 ? '1' : '0' ), nil);
}

void putexp(long x) /* convert expression to binary */
{
 /* remove containing parens at top level! */
 while (!at(x)) {
 putexp2(hd[x]);
 x = tl[x];
 }
}

void putexp2(long x) /* convert expression to binary */
{
 if ( at(x) && x != nil ) {putchr(x); return;}
 putchr('(');

 while (!at(x)) {
 putexp2(hd[x]);
 x = tl[x];
 }

 putchr(')');
}

long out(char *x, long y) /* output expression */
{
   printf("%-12s",x);
   col = 0; /* so can insert \n and 12 blanks
               every 50 characters of output */
   cc = -2; /* count characters in m-expression */
   out2(y);
   printf("\n");
   if (*x == 's' && cc > 0)
   printf("%-12s%d(%o)/%d(%o)\n","size",cc,cc,7*cc,7*cc);
   return y;
}

void out2(long x) /* really output expression */
{
   if ( at(x) && x != nil ) {out3(x); return;}
   out3('(');
   while (!at(x)) {
      out2(hd[x]);
      x = tl[x];
      }
   out3(')');
}

void out3(long x) /* really really output expression */
{
   if (col++ == 50) {printf("\n%-12s"," ");  col = 1;}
   putchar(x);
   cc++;
}

long chr2(void) /* read character - skip blanks,
                   tabs and new line characters */
{
   long c;
   do {
      c = getchar();
      if (c == EOF) {
         time2 = time(NULL);
         printf(
         "End of LISP Run\n\nElapsed time is %.0f seconds.\n",
         difftime(time2,time1)
      /* on some systems, above line should instead be: */
      /* time2 - time1 */
         );
         exit(0); /* terminate execution */
         }
      putchar(c);
   }
/* keep only non-blank printable ASCII codes */
   while (c >= 127 || c <= 32) ;
   return c;
}

long chr(void) /* read character - skip comments */
{              /* here pchr -> chr2, getchr */
   long c;
   while (1) {
      c = (*pchr)();
      if (c < 0) return c; /* error? */
      if (c != '[') return c;
   /* comments may be nested */
      while ((c = chr()) != ']')
      if (c < 0) return c; /* error? */
   }
}

/* read expression from Turing machine tape */
long in(long mexp, long rparenokay) /* input m-exp */
{
   long c = chr();
   if (c < 0) return c; /* error? */
   cc++; /* bump character count */
   if (c == ')') if (rparenokay) return ')'; else return nil;
   if (c == '(') { /* explicit list */
      long first, last, next;
      first = last = jn(nil,nil);
      while ((next = in(mexp,1)) != ')')
      {
      if (next < 0) return next; /* error? */
      last = tl[last] = jn(next,nil);
      }
      return pop(first);
      }
   if (!mexp) return c; /* atom */
   if (c == '{') { /* number */
      long n = 0, u;
      while ((c = chr()) != '}') {
      if (c < 0) return c; /* error? */
      c = c - '0';
      if (c >= 0 && c <= 9) n = 10 * n + c;
      }
      for (u = nil; n > 0; n--) u = jn('1',u);
      return u;
      }
   if (c == '"') return in(0,0); /* s-exp */
   if (c == '&' && cc == 1) {
   /* expand "define" only if & is first character */
      long name, def;
      name = in(0,0); if (name < 0) return name; /* error? */
      p = jn(name,p);
      def  = in(1,0); if (def  < 0) return def ; /* error? */
      return jn('&',jn(name,jn(def,nil)));
      }
   if (c == ':') { /* expand "let" */
      long name, def, body;
      name = in(0,0); if (name < 0) return name; /* error? */
      p = jn(name,p);
      def  = in(1,0); if (def  < 0) return def ; /* error? */
      body = in(1,0); if (body < 0) return body; /* error? */
      p = pop(p);
      if (!at(name)) {
         long var_list;
         var_list = tl[name];
         name = hd[name];
         def =
         jn('\'',jn(jn('&',jn(var_list,jn(def,nil))),nil));
         }
      return
      jn(
       jn('\'',jn(jn('&',jn(jn(name,nil),jn(body,nil))),nil)),
       jn(def,nil)
      );
      }
  {long p2;
   for (p2 = p; !at(p2); p2 = tl[p2]) {
      long p3 = hd[p2];
      if (p3 == c) return c;
      if (hd[p3] == c) {
         long first, last, next;
         first = last = jn(c,nil);
         for (p3 = tl[p3]; !at(p3); p3 = tl[p3]) {
            next = in(1,0);
            if (next < 0) return next; /* error? */
            last = tl[last] = jn(next,nil);
            }
         return first;
         }
      }
   return c;
   }
}
\end{verbatim}
}\chap{big.c}{\Size\begin{verbatim}
/* big.c: high-speed LISP interpreter */

/*
   The storage required by this interpreter is 8 bytes times
   the symbolic constant SIZE, which is 8 * 64,000,000 =
   512 megabytes.  To run this interpreter in small machines,
   reduce the #define SIZE 64000000 below.

   To compile, type
      cc -O -obig -bmaxdata:0x40000000 big.c
   To run interactively, type
      big
   To run with output on screen, type
      big <test.l
   To run with output in file, type
      big <test.l >test.r

   Reference:  Kernighan & Ritchie,
   The C Programming Language, Second Edition,
   Prentice-Hall, 1988.
*/

#include <stdio.h>
#include <time.h>

#define SIZE 64000000 /* numbers of nodes of tree storage */
#define LAST_ATOM 128 /* highest integer value of character */
#define nil 128 /* null pointer in tree storage */
#define question -1 /* error pointer in tree storage */
#define exclamation -2 /* error pointer in tree storage */
#define infinity 999999999 /* "infinite" depth limit */

/* For very small PC's, change following line to
   make hd & tl unsigned short instead of long:
   (If so, SIZE must be less than 64K.) */
long hd[SIZE+1], tl[SIZE+1]; /* tree storage */
long next = nil; /* list of free nodes */
long low = LAST_ATOM+1; /* first never-used node */
long vlst[LAST_ATOM+1]; /* bindings of each atom */
long tape; /* Turing machine tapes */
long display; /* display indicators */
long outputs; /* output stacks */
long q; /* for converting expressions to binary */
long col; /* column in each 50 character chunk of output
            (preceeded by 12 char prefix) */
long cc; /* character count */
time_t time1; /* clock at start of execution */
time_t time2; /* clock at end of execution */

long ev(long e); /* initialize and evaluate expression */
void initialize_atoms(void); /* initialize atoms */
void clean_env(void); /* clean environment */
void restore_env(void); /* restore dirty environment */
long eval(long e, long d); /* evaluate expression */
/* evaluate list of expressions */
long evalst(long e, long d);
/* bind values of arguments to formal parameters */
void bind(long vars, long args);
long at(long x); /* atomic predicate */
long jn(long x, long y); /* join head to tail */
long pop(long x); /* return tl & free node */
void fr(long x); /* free list of nodes */
long eq(long x, long y); /* equal predicate */
long cardinality(long x); /* number of elements in list */
long append(long x, long y); /* append two lists */
/* read one square of Turing machine tape */
long getbit(void);
/* read one character from Turing machine tape */
long getchr(void);
/* read expression from Turing machine tape */
void putchr(long x); /* convert character to binary */
void putexp(long x); /* convert expression to binary */
void putexp2(long x); /* convert expression to binary */
long out(char *x, long y); /* output expression */
void out2(long x); /* really output expression */
void out3(long x); /* really really output expression */
long chr2(void); /* read character - skip blanks,
                    tabs and new line characters */
long chr(void); /* read character - skip comments */
long in(long mexp, long rparenokay); /* input m-exp */
long (*pchr)(void); /* pointer to chr2 or getchr */
long p; /* parens associated with each operator for in() */
long p0; /* parens associated with each primitive function */
long p1; /* parens associated with each operator */

main() /* lisp main program */
{
char name_colon[] = "X:"; /* for printing name: def pairs */

time1 = time(NULL); /* start timer */
printf("big.c\n\nLISP Interpreter Run\n");
initialize_atoms();
p = nil;
p = jn(jn('@',nil),p);
p = jn(jn('%',nil),p);
p = jn(jn('+',jn('1',nil)),p);
p = jn(jn('-',jn('1',nil)),p);
p = jn(jn('.',jn('1',nil)),p);
p = jn(jn('\'',jn('1',nil)),p);
p = jn(jn(',',jn('1',nil)),p);
p = jn(jn('!',jn('1',nil)),p);
p = jn(jn('#',jn('1',nil)),p);
p = jn(jn('~',jn('1',nil)),p);
p = jn(jn('*',jn('1',jn('2',nil))),p);
p = jn(jn('=',jn('1',jn('2',nil))),p);
p = jn(jn('&',jn('1',jn('2',nil))),p);
p = jn(jn('^',jn('1',jn('2',nil))),p);
p = jn(jn('/',jn('1',jn('2',jn('3',nil)))),p);
p = jn(jn(':',jn('1',jn('2',jn('3',nil)))),p);
p = jn(jn('?',jn('1',jn('2',jn('3',nil)))),p);
p1 = p0 = p;

while (1) {
      long e, f, name, def;
      printf("\n");
      /* read lisp meta-expression from stdin, ) not okay */
      pchr = chr2; cc = 0; p = p1; e = in(1,0); p1 = p;
      /* flush rest of input line */
      while (putchar(getchar()) != '\n');
      printf("\n");
      f = hd[e];
      name = hd[tl[e]];
      def = hd[tl[tl[e]]];
      if (f == '&') {
      /* definition */
         if (at(name)) {
         /* variable definition, e.g., & x '(abc) */
            def = out("expression",def);
            def = ev(def);
         } /* end of variable definition */
         else          {
         /* function definition, e.g., & (Fxy) *x*y() */
            long var_list = tl[name];
            name = hd[name];
            def = jn('&',jn(var_list,jn(def,nil)));
         } /* end of function definition */
         name_colon[0] = name;
         out(name_colon,def);
         /* new binding replaces old */
         vlst[name] = jn(def,nil);
         continue;
      } /* end of definition */
      /* write corresponding s-expression */
      e = out("expression",e);
      /* evaluate expression */
      e = out("value",ev(e));
   }
}

long ev(long e) /* initialize and evaluate expression */
{
 long d = infinity; /* "infinite" depth limit */
 long v;
 tape = jn(nil,nil);
 display = jn('Y',nil);
 outputs = jn(nil,nil);
 v = eval(e,d);
 if (v == question) v = '?';
 if (v == exclamation) v = '!';
 return v;
}

void initialize_atoms(void) /* initialize atoms */
{
 long i;
 for (i = 0; i <= LAST_ATOM; ++i) {
 hd[i] = tl[i] = i; /* so that hd & tl of atom = atom */
 /* initially each atom evaluates to self */
 vlst[i] = jn(i,nil);
 }
}

long jn(long x, long y) /* join two lists */
{
 long z;
 /* if y is not a list, then jn is x */
 if ( y != nil && at(y) ) return x;

 if (next == nil) {
  if (low > SIZE) {
  printf("Storage overflow!\n");
  exit(0);
  }
 next = low++;
 tl[next] = nil;
 }

 z = next;
 next = tl[next];
 hd[z] = x;
 tl[z] = y;

 return z;
}

long pop(long x) /* return tl & free node */
{
 long y;
 y = tl[x];
 tl[x] = next;
 next = x;
 return y;
}

void fr(long x) /* free list of nodes */
{
 while (x != nil) x = pop(x);
}

long at(long x) /* atom predicate */
{
 return ( x <= LAST_ATOM );
}

long eq(long x, long y) /* equal predicate */
{
 if (x == y) return 1;
 if (at(x)) return 0;
 if (at(y)) return 0;
 if (eq(hd[x],hd[y])) return eq(tl[x],tl[y]);
 return 0;
}

long eval(long e, long d) /* evaluate expression */
{
/*
 e is expression to be evaluated
 d is permitted depth - integer, not pointer to tree storage
*/
 long f, v, args, x, y, z, vars, body;

 /* find current binding of atomic expression */
 if (at(e)) return hd[vlst[e]];

 f = eval(hd[e],d); /* evaluate function */
 e = tl[e]; /* remove function from list of arguments */
 if (f < 0) return f; /* function = error value? */

 if (f == '\'') return hd[e]; /* quote */

 if (f == '/') { /* if then else */
 v = eval(hd[e],d);
 e = tl[e];
 if (v < 0) return v; /* error? */
 if (v == '0') e = tl[e];
 return eval(hd[e],d);
 }

 args = evalst(e,d); /* evaluate list of arguments */
 if (args < 0) return args; /* error? */

 x = hd[args]; /* pick up first argument */
 y = hd[tl[args]]; /* pick up second argument */
 z = hd[tl[tl[args]]]; /* pick up third argument */

 switch (f) {
 case '@': {fr(args); return getbit();}
         /* read lisp meta-expression from TM tape, ) not okay */
 case '%': {fr(args); pchr = getchr; cc = 1; p = p0; return in(1,0);}
 case '#': {fr(args);
           v = q = jn(nil,nil); putexp(x); return pop(v);}
 case '+': {fr(args); return hd[x];}
 case '-': {fr(args); return tl[x];}
 case '.': {fr(args); return (at(x) ? '1' : '0');}
 case ',': {fr(args); hd[outputs] = jn(x,hd[outputs]);
           return (hd[display] == 'Y' ? out("display",x): x);}
 case '~': {fr(args); return /* out("show", */ x /* ) */ ;}
 case '=': {fr(args); return (eq(x,y) ? '1' : '0');}
 case '*': {fr(args); return jn(x,y);}
 case '^': {fr(args);
           return append((at(x)?nil:x),(at(y)?nil:y));}
 }

 if (d == 0) {fr(args); return question;} /* depth exceeded
                                             -> error! */
 d--; /* decrement depth */

 if (f == '!') {
 fr(args);
 clean_env(); /* clean environment */
 v = eval(x,d);
 restore_env(); /* restore unclean environment */
 return v;
 }

 if (f == '?') {
 fr(args);
 x = cardinality(x); /* convert s-exp into number */
 clean_env();
 tape = jn(z,tape);
 display = jn('N',display);
 outputs = jn(nil,outputs);
 v = eval(y,(d <= x ? d : x));
 restore_env();
 z = hd[outputs];
 tape = pop(tape);
 display = pop(display);
 outputs = pop(outputs);
 if (v == question) return (d <= x ? question : jn('?',z));
 if (v == exclamation) return jn('!',z);
 return jn(jn(v,nil),z);
 }

 f = tl[f];
 vars = hd[f];
 f = tl[f];
 body = hd[f];

 bind(vars,args);
 fr(args);

 v = eval(body,d);

 /* unbind */
 while (!at(vars)) {
 if (at(hd[vars]))
 vlst[hd[vars]] = pop(vlst[hd[vars]]);
 vars = tl[vars];
 }

 return v;
}

void clean_env(void) /* clean environment */
{
 long i;
 for (i = 0; i <= LAST_ATOM; ++i)
 vlst[i] = jn(i,vlst[i]); /* clean environment */
}

void restore_env(void) /* restore unclean environment */
{
 long i;
 for (i = 0; i <= LAST_ATOM; ++i)
 vlst[i] = pop(vlst[i]); /* restore unclean environment */
}

long cardinality(long x) /* number of elements in list */
{
 if (at(x)) return (x == nil ? 0 : infinity);
 return 1+cardinality(tl[x]);
}

/* bind values of arguments to formal parameters */
void bind(long vars, long args)
{
 if (at(vars)) return;
 bind(tl[vars],tl[args]);
 if (at(hd[vars]))
 vlst[hd[vars]] = jn(hd[args],vlst[hd[vars]]);
}

long evalst(long e, long d) /* evaluate list of expressions */
{
 long x, y;
 if (at(e)) return nil;
 x = eval(hd[e],d);
 if (x < 0) return x; /* error? */
 y = evalst(tl[e],d);
 if (y < 0) return y; /* error? */
 return jn(x,y);
}

long append(long x, long y) /* append two lists */
{
 if (at(x)) return y;
 return jn(hd[x],append(tl[x],y));
}

/* read one square of Turing machine tape */
long getbit(void)
{
 long x;
 if (at(hd[tape])) return exclamation; /* tape finished ! */
 x = hd[hd[tape]];
 hd[tape] = tl[hd[tape]];
 return (x == '0' ? '0' : '1');
}

/* read one character from Turing machine tape */
long getchr(void)
{
 long c, b, i;
 c = 0;
 do {
    for (i = 0; i < 7; ++i) {
    b = getbit();
    if (b < 0) return b; /* error? */
    c = c + c + b - '0';
    }
 }
/* keep only non-blank printable ASCII codes */
 while (c >= 127 || c <= 32) ;
 return c;
}

void putchr(long x) /* convert character to binary */
{
 q = tl[q] = jn(( x &  64 ? '1' : '0' ), nil);
 q = tl[q] = jn(( x &  32 ? '1' : '0' ), nil);
 q = tl[q] = jn(( x &  16 ? '1' : '0' ), nil);
 q = tl[q] = jn(( x &   8 ? '1' : '0' ), nil);
 q = tl[q] = jn(( x &   4 ? '1' : '0' ), nil);
 q = tl[q] = jn(( x &   2 ? '1' : '0' ), nil);
 q = tl[q] = jn(( x &   1 ? '1' : '0' ), nil);
}

void putexp(long x) /* convert expression to binary */
{
 /* remove containing parens at top level! */
 while (!at(x)) {
 putexp2(hd[x]);
 x = tl[x];
 }
}

void putexp2(long x) /* convert expression to binary */
{
 if ( at(x) && x != nil ) {putchr(x); return;}
 putchr('(');

 while (!at(x)) {
 putexp2(hd[x]);
 x = tl[x];
 }

 putchr(')');
}

long out(char *x, long y) /* output expression */
{
   printf("%-12s",x);
   col = 0; /* so can insert \n and 12 blanks
               every 50 characters of output */
   cc = -2; /* count characters in m-expression */
   out2(y);
   printf("\n");
   if (*x == 's' && cc > 0)
   printf("%-12s%d(%o)/%d(%o)\n","size",cc,cc,7*cc,7*cc);
   return y;
}

void out2(long x) /* really output expression */
{
   if ( at(x) && x != nil ) {out3(x); return;}
   out3('(');
   while (!at(x)) {
      out2(hd[x]);
      x = tl[x];
      }
   out3(')');
}

void out3(long x) /* really really output expression */
{
   if (col++ == 50) {printf("\n%-12s"," ");  col = 1;}
   putchar(x);
   cc++;
}

long chr2(void) /* read character - skip blanks,
                   tabs and new line characters */
{
   long c;
   do {
      c = getchar();
      if (c == EOF) {
         time2 = time(NULL);
         printf(
         "End of LISP Run\n\nElapsed time is %.0f seconds.\n",
         difftime(time2,time1)
      /* on some systems, above line should instead be: */
      /* time2 - time1 */
         );
         exit(0); /* terminate execution */
         }
      putchar(c);
   }
/* keep only non-blank printable ASCII codes */
   while (c >= 127 || c <= 32) ;
   return c;
}

long chr(void) /* read character - skip comments */
{              /* here pchr -> chr2, getchr */
   long c;
   while (1) {
      c = (*pchr)();
      if (c < 0) return c; /* error? */
      if (c != '[') return c;
   /* comments may be nested */
      while ((c = chr()) != ']')
      if (c < 0) return c; /* error? */
   }
}

/* read expression from Turing machine tape */
long in(long mexp, long rparenokay) /* input m-exp */
{
   long c = chr();
   if (c < 0) return c; /* error? */
   cc++; /* bump character count */
   if (c == ')') if (rparenokay) return ')'; else return nil;
   if (c == '(') { /* explicit list */
      long first, last, next;
      first = last = jn(nil,nil);
      while ((next = in(mexp,1)) != ')')
      {
      if (next < 0) return next; /* error? */
      last = tl[last] = jn(next,nil);
      }
      return pop(first);
      }
   if (!mexp) return c; /* atom */
   if (c == '{') { /* number */
      long n = 0, u;
      while ((c = chr()) != '}') {
      if (c < 0) return c; /* error? */
      c = c - '0';
      if (c >= 0 && c <= 9) n = 10 * n + c;
      }
      for (u = nil; n > 0; n--) u = jn('1',u);
      return u;
      }
   if (c == '"') return in(0,0); /* s-exp */
   if (c == '&' && cc == 1) {
   /* expand "define" only if & is first character */
      long name, def;
      name = in(0,0); if (name < 0) return name; /* error? */
      p = jn(name,p);
      def  = in(1,0); if (def  < 0) return def ; /* error? */
      return jn('&',jn(name,jn(def,nil)));
      }
   if (c == ':') { /* expand "let" */
      long name, def, body;
      name = in(0,0); if (name < 0) return name; /* error? */
      p = jn(name,p);
      def  = in(1,0); if (def  < 0) return def ; /* error? */
      body = in(1,0); if (body < 0) return body; /* error? */
      p = pop(p);
      if (!at(name)) {
         long var_list;
         var_list = tl[name];
         name = hd[name];
         def =
         jn('\'',jn(jn('&',jn(var_list,jn(def,nil))),nil));
         }
      return
      jn(
       jn('\'',jn(jn('&',jn(jn(name,nil),jn(body,nil))),nil)),
       jn(def,nil)
      );
      }
  {long p2;
   for (p2 = p; !at(p2); p2 = tl[p2]) {
      long p3 = hd[p2];
      if (p3 == c) return c;
      if (hd[p3] == c) {
         long first, last, next;
         first = last = jn(c,nil);
         for (p3 = tl[p3]; !at(p3); p3 = tl[p3]) {
            next = in(1,0);
            if (next < 0) return next; /* error? */
            last = tl[last] = jn(next,nil);
            }
         return first;
         }
      }
   return c;
   }
}
\end{verbatim}
}

\end{document}